\documentclass[fleqn,usenatbib]{mnras}

\usepackage[T1]{fontenc}
\usepackage{ae,aecompl}

\usepackage{appendix}

\usepackage{amsmath}
\usepackage{longtable}
\usepackage{color}
\usepackage{float}
\usepackage{setspace}
\usepackage{amsfonts}
\usepackage{times}
\usepackage{graphicx}
\usepackage{framed}
\usepackage{rotating}

\def\be{\begin{equation}}
\def\ee{\end{equation}}
\def\ba{\begin{eqnarray}}
\def\ea{\end{eqnarry}}
\def\bal#1\eal{\begin{align}#1\end{align}}

\title[Actinide opacities]{Actinide opacities for modeling the spectra and\\
light curves of kilonovae}

\author[C.~J.~Fontes et al.]{
C.~J.~Fontes$^{1,2}$\thanks{E-mail: cjf@lanl.gov},
C.~L.~Fryer$^{1,3,4,5,6}$,
R.~T.~Wollaeger$^{1,3}$,
M.~R.~Mumpower$^{1,7}$, and
T.~M.~Sprouse$^{7}$
\\
$^1$Center for Theoretical Astrophysics, Los Alamos National Laboratory,
Los Alamos, NM 87545, USA\\
$^2$Computational Physics Division, Los Alamos National Laboratory,
Los Alamos, NM 87545, USA\\
$^3$Computer, Computational, and Statistical Sciences Division,
Los Alamos National Laboratory, Los Alamos, NM 87545, USA\\
$^4$Physics Department, University of Arizona, Tucson, AZ 85721, USA\\
$^5$Physics and Astronomy Department, University of New Mexico, Albuquerque,
NM 87131, USA\\
$^6$The George Washington University, Washington, DC 20052, USA\\
$^7$Theoretical Division, Los Alamos National Laboratory,
Los Alamos, NM 87545, USA
}

\date{Accepted XXX. Received YYY; in original form ZZZ}

\pubyear{2021}

\hypersetup{
        colorlinks=true,
        linkcolor=blue,
        citecolor=blue,
        urlcolor=cyan
}
\begin{document}
\label{firstpage}
\pagerange{\pageref{firstpage}--\pageref{lastpage}}
\maketitle

%%%%%%%%%%%%%%%%%%%%%%%%%%%%%%%%%%%%%%%%%%%%%%%%%%%%%%%%%%%%%%%%%


\begin{abstract}

We extend previous ab initio calculations of lanthanide opacities
\hbox{(Fontes et al., 2020, MNRAS, 493, 4143)}
to include a complete set of actinide opacities
for use in the modeling of kilonova light curves and spectra.
Detailed, fine-structure line features are generated using
the configuration-interaction approach.
These actinide opacities display similar trends to those observed
for lanthanide opacities, such as the lighter actinides producing higher
opacity than the heavier ones for relevant conditions in the dynamical ejecta.
A line-binned treatment is employed to pre-compute opacity tables
for 14 actinide elements $(89 \le Z \le 102)$
over a grid of relevant temperatures and densities.
These tabular opacities will be made publicly available for general
usage in kilonova modeling.
We demonstrate the usefulness of these opacities in kilonova simulations
by exploring the sensitivity of light curves and spectra to different
actinide abundance distributions that are predicted by different nuclear
theories, as well as to different choices of ejecta mass and velocity.
We find very little sensitivity to the two considered distributions,
indicating that opacities for actinides with $Z \ge 99$ do not contribute
strongly. On the other hand, a single actinide element, protactinium,
is found to produce faint spectral features in the far infrared at late times
(5--7 days post merger). More generally, we find that the choice of ejecta
mass and velocity have the most significant effect on KN emission for
this study.

\end{abstract}


%%%%%%%%%%%%%%%%%%%%%%%%%%%%%%%%%%%%%%%%%%%%%%%%%%%%%%%%%%%%%%%%%%%

%%%%%%%%%%%%%%%%% BODY OF PAPER %%%%%%%%%%%%%%%%%%

\section{Introduction}
\label{sec:intro}

The effect of lanthanide opacities on the electromagnetic
spectra produced during a kilonova (KN) event has been
the focus of intense research over the past several years, as
catalyzed by the recent KN observation associated with the gravitational
wave detection known as GW170817 \citep{abbott17h,abbott17a}.
The term kilonova \citep{li_nsm98,kulkarni_nsm05,metzger10} refers
to the electromagnetic radiation that can be
produced after a neutron star merger (NSM).
The emission from such events is powered by the radioactive decay
of $r$-process elements that are created in this neutron-rich environment.
High-energy photons, along with alpha particles, beta particles
and fission fragments, produced during these decay events eventually thermalize,
which leads to the thermal emission of radiation.
The passage of thermal photons through, and possible escape from,
the dynamical ejecta
is expected to be controlled by the opacity of heavy elements that possess
a large number (greater than 50) of bound electrons.
These bound electrons are characterized by complex energy level diagrams
and a corresponding, densely packed forest of bound-bound absorption (line)
features that can greatly inhibit the flow of photons through the dynamical
ejecta. The atomic level populations, which are required to calculate
the opaciy, are typically assumed to be
in local thermodynamic equilibrium (LTE), at least during the early
phase of a KN, i.e. less than about 7~days post merger.

While lanthanide opacities have been studied in some detail
in this context, actinide opacities, which are the focus of this work,
have been relatively ignored. The opacities of low-charge ion stages of actinide
elements are expected to possess complex absorption features that
are similar to those exhibited by the corresponding lanthanide elements.
Furthermore, the production of actinides in NSMs remains an open topic
of pursuit in nuclear physics \citep{wanajo14,wu19,vassh20}.
For example, the synthesis of actinides may yield distinct
gamma-ray emission from nascent fission fragments \citep{wang20} or show
unique imprints from the production of specific elements
\citep{zhu18,korobkin20},
but large uncertainties in nuclear theory leave unresolved
just how heavy an element can be created during these events
\citep{mumpower18,cote18,moller19}.

As far as lanthanide elements are concerned, detailed line opacities
were initially considered by \cite{kasen13}, \cite{tanaka13},
\cite{barnes13} and \cite{kasen15} for use in KN simulations.
An independent study \citep{fontes15,fontes17} was carried out
to investigate alternative methods of including detailed line opacities
in KN modeling, i.e. the line-binned treatment considered in the present
work.
Subsequently, each of these three research groups produced a complete set
of lanthanide opacities \citep{kasen17,fontes20,tanaka20}. Recent lanthanide
opacity efforts have also focused on the calculation of precise atomic data
(energies and radiative decay rates) that are used in the generation
of opacities for KN modeling
\citep{quinet20,gaigalas19,radiute20,carvajal21,carvajal22}.

As the analysis and understanding of GW170817 (and KNe in general) evolve,
the need for detailed opacities has expanded beyond the lanthanide elements
to include the fourth- through sixth-row elements of the periodic table.
KN modelers now include opacities for these elements,
as well as the lanthanides,
to investigate an ever-growing list of applications,
such as morphology and composition effects,
specific line signatures for the purpose
of elemental identification, non-LTE
effects on the atomic level populations, and KN detectability studies,
e.g. \cite{kasen17}, \cite{tanvir17},
\cite{troja17}, \cite{tanaka18}, \cite{wollaeger18}, \cite{kawaguchi18},
\cite{watson19}, \cite{wollaeger19},
\cite{kawaguchi20}, \cite{tanaka20}, \cite{even20},
\cite{banerjee20}, \cite{bulla21}, \cite{zhu21}, \cite{korobkin21},
\cite{oconnor21}, \cite{gillanders21}, \cite{domoto21}, \cite{kawaguchi21},
\cite{hotokezaka21}, \cite{ristic21}, \cite{chase21}, \cite{wollaeger21}.

The purpose of the present work is to extend our earlier calculations
of lanthanide opacites \citep[hereafter referred to as Paper~I]{fontes20}
to include the actinide elements. As noted in Fig.~1 of Paper~I,
the abundance of actinide elements is predicted to be similar to that of the
lanthanides within the dynamical ejecta of a KN.
To our knowledge, detailed actinide opacities
have not been generated for KN modeling, with the exception of uranium,
which we considered as the sole, representative actinide element in Paper~I.
(Those uranium opacity data are reproduced in the present work for
completeness.) Here, we provide a complete set 
of frequency-dependent opacities for the actinide elements, and also
demonstrate their usefulness in testing the sensitivity of
KN spectra and light curves to different actinide abundance distributions
that may arise due to variations in conditions or from large uncertainties
predicted by current nuclear theory modeling \citep{sprouse20,cote21}.

\section{Atomic Physics Considerations}
\label{sec:atomic}

\subsection{Basic considerations}
\label{subsec:basic_atomic}

The computational framework that is used to generate the actinide opacities
in this work is the same as that described in Paper~I, and so we provide
a summary of our approach. We use
the Los Alamos suite of atomic physics and plasma modeling
codes~(see \citealt{LANL_suite} and references therein)
to generate the fundamental atomic data, e.g. level energies,
radiative decay rates, photoionization cross sections, as well
as the LTE atomic level populations and resultant frequency-dependent
opacities of interest.

The calculation of accurate, ab initio actinide energies and oscillator
strengths for such a broad set of atomic levels and elements is very
challenging, and is made more difficult by the paucity of benchmark values that
are available compared to the lanthanides. So this dataset should be
considered a first attempt to calculate a complete set of actinide opacities.
For this work, we chose the semi-relativistic Hatree-Fock capability
within the Los Alamos suite because it provides a more complete description
of the (non-local) electron-electron exchange potential compared
to the (local-exchange) potential in the fully relativistic approach.
The semi-relativistic approach also produces more
physically reasonable results across the entire range of actinide elements.
On the other hand, this approach lacks the relativistic physics
that can be important to describe subtle, indirect effects concerning
the $5f$ electrons that exist
in neutral and low-charged actinide ions \citep{tatewaki17,rose78}.

From a computational perspective,
the semi-relativistic calculations begin with the CATS atomic
structure code \citep{cats_man}, which employs the Hartree-Fock method
of Cowan~\citep{cowan}.
These calculations produce detailed, fine-structure energy levels
that include configuration interaction for the specified list of configurations
(see Table~\ref{tab:configs}).
Oscillator strengths are also generated in this step, and are eventually used
to produce the bound-bound contribution to the opacity.
After the atomic structure calculations are complete,
the GIPPER ionization code is used
to obtain the relevant photoionization cross sections
in the distorted-wave approximation. The photoionization data are used
to generate the bound-free contribution to the opacity and are not
expected to be too important for the present application, due to
the range of relevant photon energies, but are included
for completeness. Therefore, they are calculated in the configuration-average
approximation, rather than fine-structure detail, in order to minimize the
computational time.

The atomic level populations are calculated with the ATOMIC code
from these fundamental atomic data. This code can be used in either
LTE or non-LTE mode \citep{atomic1,atomic2,hakel06,colgan_oplib,fontes_cr16}.
The LTE approach was chosen for the present
application, which requires only the atomic structure data,
along with the temperature and density, to calculate
the populations. The populations are then combined with the oscillator
strengths and photoionization cross sections in ATOMIC to obtain
the monochromatic opacities, which are constructed from the standard
four contributions:
bound-bound (b-b), bound-free (b-f), free-free (f-f) and scattering.
Specific formulae for these contributions are readily available in various
textbooks, such as \citet{huebner}.

To obtain the bound-bound contribution, we use the line-binned treatment
described in Paper~I. As this line-binned approach differs from the
traditional expansion-opacity method \citep{sobolev60,castor74,karp77},
we provide a bit more detail on this topic.
The expression for the (monochromatic) bound-bound contribution is given by
\begin{equation}
\kappa^{\rm b-b}_\nu = \frac{\pi e^2}{\rho m_e c} \sum_i N_i
\, |f_{i}| \, L_{i,\nu} \,,
\label{opac_bb}
\end{equation}
where $\nu$ is the photon energy,
$\rho$ is the mass density, $N_i$ is the number density of the initial
level in transition $i$,
$f_{i}$ is the oscillator strength describing the photo-excitation
of transition $i$, and $L_{i,\nu}$ is the associated line profile function.
The corresponding line-binned, bound-bound opacities are
comprised of discrete frequency (or wavelength) bins
that contain a sum over all of the lines contained within a bin.
An expression for this discrete opacity is obtained
from the continuous opacity displayed in equation~(\ref{opac_bb})
by replacing the line profile with $1/\Delta \nu_j$, i.e.
\begin{equation}
\kappa^{\rm bin}_{\nu,j} = \frac{1}{\Delta \nu_j}\frac{\pi e^2}{\rho m_e c}
\sum_{i \in \Delta \nu_j} N_i
\, |f_{i}| \,,
\label{opac_bb_bin}
\end{equation}
where $\Delta \nu_j$ represents the frequency width of a bin denoted
by integer index $j$. So, the summation encompasses all lines $i$ with
centers that reside in bin $j$.
As in our previous work on lanthanides, we include
only E1 dipole radiatiave transitions in the present models.
Under LTE conditions, these allowed transitions are expected to provide the
dominant contribution to the line absorption over a significant fraction
of the energy range of interest. Forbidden transitions can
produce weaker spectral features at relatively low photon energies,
and a feasibility study of whether such features could be identified
in observed KN spectra is a potential topic of future research.

As demonstrated in Paper~I, simulated KN light curves produced
with line-binned opacities are similar to those produced with the more commonly
used expansion opacities, but specific spectral features can differ
between the two methods, e.g. we observed that the expansion-opacity approach
produced shallower absorption troughs.
Furthermore, line-binned, bound-bound opacities have the advantage of being
independent of the particular type of hydrodynamic expansion. Thus, one can
pre-compute tabulated opacities on a grid of relevant temperatures
and densities, thereby eliminating the need to explicitly calculate
a large number of opacities during the radiation-transport portion
of KN simulations.

\subsection{Detailed considerations and comparisons}
\label{subsec:detailed_atomic}

In this section, we discuss some convergence concepts associated with the
calculation of atomic data and opacities.
We also provide some comparisons to assess the accuracy of our atomic data. 

From a general perspective, the ability to calculate large amounts of atomic
data for actinide elements, with any significant accuracy, is a daunting
theoretical challenge. It is well known that many ``workhorse''
atomic structure codes, such as the one currently employed, as well as others,
are not expected to produce accurate transition energies
and radiative rates for near-neutral ion stages of actinide elements
without some sort of tuning,
i.e. without the incorporation of some experimental data in order
to improve ab initio calculations.
As mentioned in the Introduction, the available benchmark data
that can be used to evaluate the accuracy of our actinide calculations
is rather scarce, which precludes the tuning approach. 
For the present application of KN modeling, this lack of accuracy is somewhat
mitigated because the transport of radiation depends on the
density of lines per energy (or wavelength) range in the opacity.
From a theoretical perspective, many lines are summed within a given energy
(or wavelength) bin to obtain their contribution to the opacity and then
this quasi-continuum
of lines is significantly smeared within the radiation-transport calculations
due to the large velocity gradients, i.e.
$\Delta v/c \sim 0.01$, that are predicted to occur in the dynamical ejecta of
a KN. These considerations provide some justification
for prioritizing completeness of the line contributions over accuracy
of individual lines when calculating opacities for KN modeling.
Thus, our strategy in the present, first attempt, is to produce
atomic models that attempt to preserve the statistical nature of the
myriad absorption features in actinide opacities.

\subsubsection{Convergence discussion}
\label{subsubsec:conv_discussion}

Our method for generating lists of atomic configurations is similar to
that employed in our previous work in Paper~I on lanthanides. 
For a given ion stage, one starts with the ground configuration and
then systematically considers single- and double-electron promotions
to higher lying $nl$ orbitals in order to form excited configurations.
The relevant temperatures are sufficiently low for the KN modeling of interest
in this work (see Fig.~\ref{fig:ionfrac_u} and the associated
discussion in the next section) that one only needs to consider promotions from
the valence orbitals, i.e. $5f$, $6d$, $7s$ subshells. The electron promotions
are performed up to some maximum principal quantum number, $n_{\rm max}$,
in order to ensure that that converged partition functions and level
populations will be obtained from thermal considerations, i.e.
based on Saha-Boltzmann statistics. The excited configurations must
also permit all of the relevant dipole excitations  from lower lying
configurations whose fine-structure levels will contain significant
populations. Additionally, these excited configurations must also be chosen
such that enough configuration-interaction is included to produce
reasonably converged level energies and radiative transition rates in the atomic
structure calculations. Through trial and error, we found that a value of
$n_{\rm max} = 7$ or 8 provides a reasonable balance between convergence
and computational-resource considerations. Once a list of configurations
is chosen, we perform fine-structure
calculations based on those lists of configurations to obtain energy
levels, wavefunctions and radiative decay rates. When then carry out
some numerical convergence checks by extending the configuration lists
using $n_{\rm max} + 1$, instead of $n_{\rm max}$. These tests indicate
that the populations of low-lying atomic levels, which are
responsible for producing the majority of lines, changed by a maximum
of a few percent. So convergence with respect to atomic level populations
is good. The spectral details of the opacities sometimes
change in that additional lines appear at the very high and
low ends in photon energy range of interest. However, these lines are
not expected to contribute significantly to the KN modeling because:
(1) At the high-energy end, the opacity is so large that the ejecta is already
optically thick and adding additional lines would not change that condition.
(2) At the low-energy end, the opacity is so small that the corresponding
absorption and emission is typically too weak to contribute significantly to
the KN simulations. These tests also sometimes reveal shifting in the energies
and increases/decreases in the strengths of the lines that appear within
the main photon energy range of interest, indicating that convergence with
respect to configuration interaction is not complete. However, we mention
that the qualitative behavior of the frequency-dependent opacities is very
similar in these comparisons and the Planck mean opacities typically differ
by less than 5--10\%, which provides some confidence that the bulk, statistical
properties of the forests of lines are preserved, even if the individual
lines are not fully converged. A more extensive investigation of the
configuration-list convergence is desirable for future work,
in accord with similar statements made by~\cite{kasen13}
in the context of lanthanide opacity calculations.

\subsubsection{Accuracy discussion}
\label{subsubsec:acc_discussion}

\begin{figure}
\centering
\includegraphics[clip=true,angle=0,width=0.9\columnwidth]
{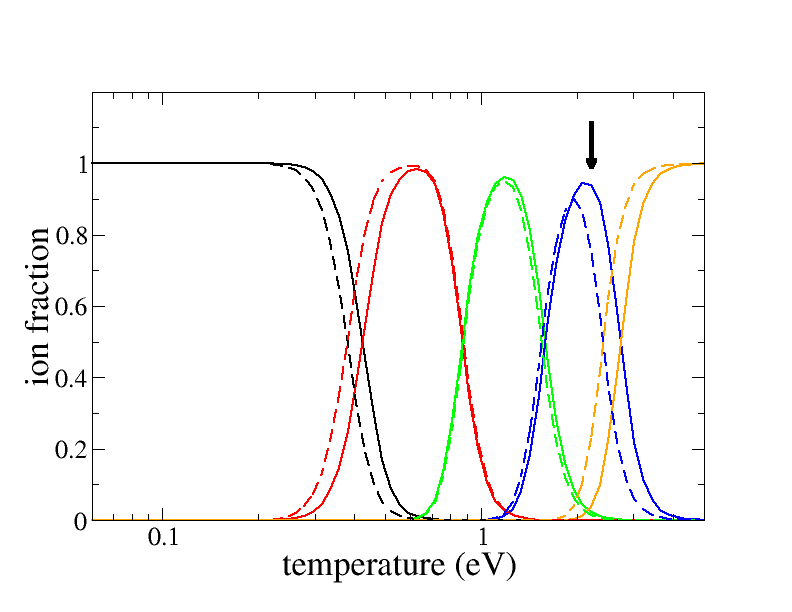}
\caption{
Ionization-stage fraction versus temperature for U at a typical mass
density of $\rho = 10^{-6}$~g/cm$^3$.
The black curves refer to U~{\sc i}, the
red ones to U~{\sc ii}, the green ones to U~{\sc iii}, and
the blue ones to U~{\sc iv}. The solid curves use NIST-corrected
ionization energies~\citep{nist}, while the dashed curves use
uncorrected values. The vertical arrow, located at a temperature of about
2~eV, indicates the temperature above which the fraction of U~{\sc v}
starts to become significant.
}
\label{fig:ionfrac_u}
\end{figure}

As in Paper~I, for improved accuracy, we replace our calculated ionization
potentials for the actinide elements with the values provided in the NIST
Atomic Spectra Database (ASD)~\citep{nist}. An illustrative example
for this type of correction, is provied in Fig.~\ref{fig:ionfrac_u}
for uranium at a mass density of $\rho = 10^{-6}$~g/cm$^3$.
We see that using NIST-corrected energies shifts the onset of the next higher
ion stage to slightly higher temperatures. Similar figures (not shown)
are obtained for the other actinides, due to the similarities in their
ionization potentials for each ion stage.
As mentioned in the previous section, this type of figure is also useful
to estimate the maximum temperature at which each ion stage exists,
which aids in the selection of appropriate list of configurations.
This figure can also be used to estimate the maximum temperature
of validity for the atomic models that are considered in this work.
Since we only consider the first four ion stages of each element,
the opacity calculations are expected to be valid up to a maximum temperature
of $T \sim 2$~eV (23~kK), as indicated by the vertical arrow
in Fig.~\ref{fig:ionfrac_u}. Going to higher temperatures would require
the inclusion of additional ion stages in the models.

While the actinide data in the NIST database is not plentiful,
there is one quantity that is readily available for each ion stage of
every actinide element, the level label of the predicted ground state.
We display in Table~\ref{tab:ground_label} a simple comparison of this
information. 
\begin{table}
\centering
\caption{ A tabular display of whether the present calculations predict
the same ground state as that listed in the NIST 
Atomic Spectra Database~\citep{nist}. Each row contains assessment codes
for the four ion stages associated with a particular actinide element, which
is labeled by its $Z$ value in the first column. A checkmark indicates
that our calculations match the NIST result.
An integer indicates a mismatch in the predicted level and denotes
the level number in our list of energy-ordered levels (for that ion stage)
that corresponds to the NIST label.
}
\vspace*{0.5\baselineskip}
\begin{tabular}{rcccc}
\hline $Z$ & \multicolumn{4}{c}{assessement code}  \\
\hline
 89 & \checkmark& \checkmark&  2 & \checkmark \\
 90 & \checkmark& 3&  3 & \checkmark \\
 91 & 2& 10& 12& \checkmark \\
 92 & \checkmark& 6& 7& \checkmark \\
 93 & \checkmark& \checkmark& 3& \checkmark \\
 94 & \checkmark& 3& \checkmark& \checkmark \\
 95 & \checkmark& \checkmark& \checkmark& \checkmark \\
 96 & \checkmark& 2& 3& \checkmark \\
 97 & \checkmark& 2& \checkmark& \checkmark \\
 98 & \checkmark& \checkmark& \checkmark& \checkmark \\
 99 & \checkmark& \checkmark& \checkmark& \checkmark \\
100 & \checkmark& \checkmark& \checkmark& \checkmark \\
101 & \checkmark& \checkmark& \checkmark& \checkmark \\
102 & \checkmark& \checkmark& \checkmark& \checkmark \\
\hline
\end{tabular}
\label{tab:ground_label}
\end{table}
For a given ion stage, a checkmark indicates
that our calculations predict the same ground level as the one in the NIST
database. An integer indicates a mismatch in the predicted level and denotes
the level number in our list of energy-ordered levels (for that ion stage)
that corresponds to the NIST label. We see that our ab initio calculations
are not perfect, but they match the NIST label 77\% of the time. When there
is a mismatch, the NIST ground level is often a relatively low-lying 
level in our calculations, i.e. the second or third level.
However, our code displays stronger mismatches for the second and third
ion stages of $Z=91$, protactinium, and moderate mismatches for
the same ion stages of $Z=92$, uranium, illustrating the challenges
of calculating the entire range of actinide elements within a single,
self-consistent framework.
These mismatches could likely be improved by tuning our atomic structure
calculations, but such manipulations do not necessarily result in more accurate
values of the excited-state energies, and so we prefer to use a consistent,
ab initio approach in the present work.

\begin{figure}
\centering
\includegraphics[clip=true,angle=0,width=0.9\columnwidth]
{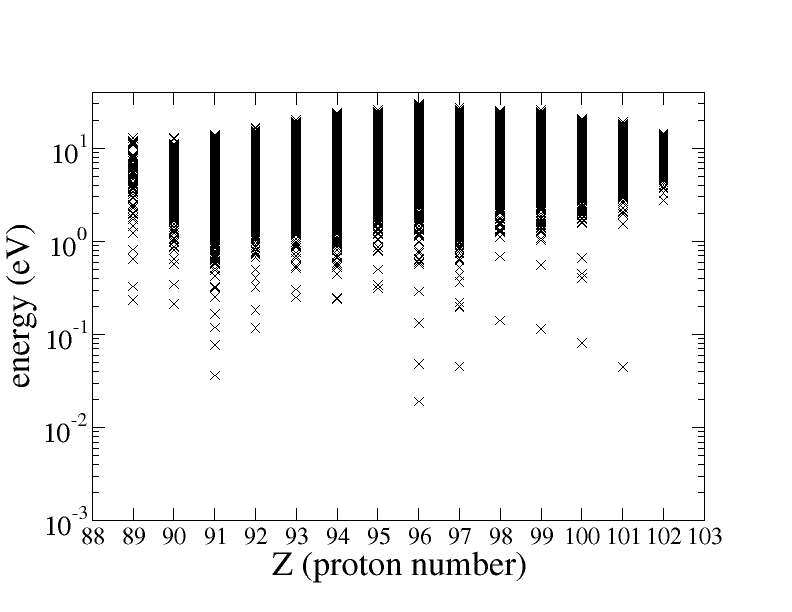}
\caption{
Calculated energy levels for the singly ionized stage of each actinide element.
Each energy level is represented by an $\times$ symbol.
}
\label{fig:elev02}
\end{figure}

Before providing a comparison of more refined atomic data,
we offer a simple visual inspection of the trends in our level energies.
In Fig.~\ref{fig:elev02}, we display our excited-state energies
for the singly charged ion stages, in a schematic format.
As expected from basic atomic theory,
there is some visible left-right symmetry about the $Z=95$ column,
which is characterized by a ground state that contains a half-filled
$(5f^7)$ $f$ shell. From energy-minimization and angular-momentum-coupling
considerations, ion stages that contain ground states with half-filled
subshells are expected to (1) be semi-stable, i.e. a larger energy gap between
the ground and first excited states compared to adjacent ion stages, and
(2) contain more excited states due to the coupling of 7 electrons
in an $f$ subshell.
This symmetry is not as strong as that exhibited in the corresponding
lanthanide diagram, which might be an indication of a difference in
electron correlation effects in actinides compared to lanthanides. Or it
could indicate a lack of accuracy in the atomic structure calculations
of the singly ionized actinides.

\begin{figure}
\centering
\includegraphics[clip=true,angle=0,width=0.9\columnwidth]
{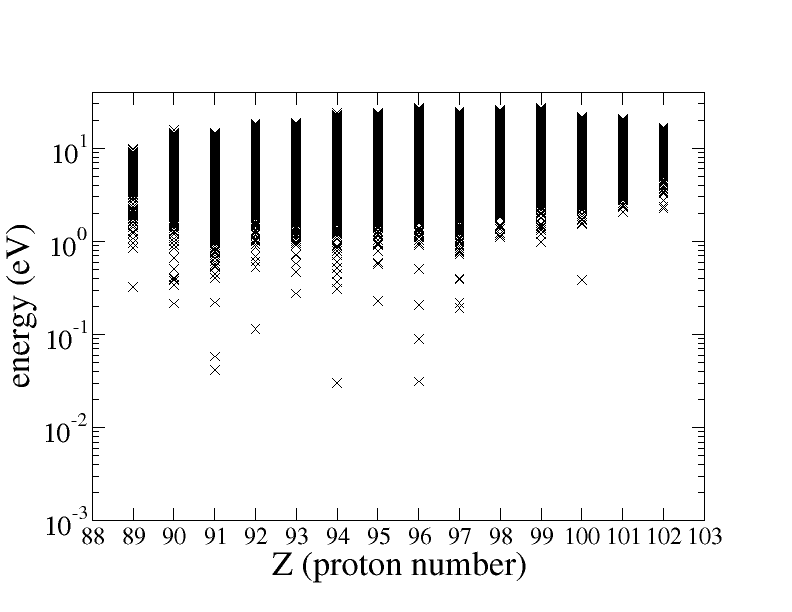}
\caption{
Calculated energy levels for the neutral stage of each actinide element.
Each energy level is represented by an $\times$ symbol.
}
\label{fig:elev01}
\end{figure}

Continuing with the discussion of excited-state energies,
in Fig.~\ref{fig:elev01} we present the energies for the neutral
stage of each actinide element.
There is less symmetry in this figure compared to that displayed
in the previous Fig.~\ref{fig:elev02} for the singly ionized ions,
with a noticeable lack of lower-lying excited
states for the rightmost five elements. Since there is no obvious
reason for this pattern, it is likely an
indication of a lack of accuracy in the calculations for these neutral cases,
which is not surprising from a theoretical perspective.
In general, the neutral stage of an element is the most difficult to calculate
in an accurate manner due to the existence of valence electrons that are
only lightly bound to the nucleus, and so it becomes important to
calculate the electron-electron interaction with higher than usual precision.
These neutral calculations for lanthanides can pose significant numerical,
even for the accurate multiconfiguration Dirac-Hartree-Fock (MCDHF)
method~\citep{grant07,gaigalas19}.

Next, we provide comparisons between our calculated results
and the line data that are available in the NIST database.
Some line information is available for the first 11 actinides.
Of those 11, only the first one, actinium $(Z=89)$, contains
both transition rates and level designations, which is the most
useful type of information for providing a meaningful accuracy assessment. 
NIST data are available for the first three ion stages of Ac, and
so we provide a comparison of those results in Table~\ref{tab:ac_data}.
\begin{table*}
\centering
\caption{A comparison of our calculated energy levels and
radiative transition rates with data available in the NIST database.
Comparisons are provided for the first three ion stages of Ac $(Z=89)$.
The NIST accuracy estimate code is listed in the final
column. (See text for details.)
}
\vspace*{0.5\baselineskip}
\begin{tabular}{cccccccl}
\hline
   &   & \multicolumn{2}{c}{transition energy (eV)}&
 \multicolumn{2}{c}{radiative decay rate (s$^{-1}$)} & ratio of decay rates &
  accuracy \\
lower level & upper level & NIST & Present & NIST & present & (present/NIST) &
  estimate \\
\hline
\multicolumn{8}{c}{Ac \sc{i}} \\
\hline
$6d 7s^2\, ^2\!D_{5/2}$  & $6d^2 7p\, ^4\!D_{5/2}$  &  4.0204406 & 4.00 &
    3.5E+07  & 5.57E+06 & 0.159 & E \\
$6d 7s^2\, ^2\!D_{5/2}$  & $7s^2 5f\, ^2\!F_{7/2}$  &  3.9834653 & 4.19 &
    1.1E+08  & 2.25E+07 & 0.205 & D+ \\
$6d 7s^2\, ^2\!D_{5/2}$  & $6d^2 7p\, ^2\!F_{7/2}$  &  3.9086043 & 3.96 &
    4.9E+07  & 4.43E+07 & 0.904 & D+ \\
$6d 7s^2\, ^2\!D_{5/2}$  & $7s^2 5f\, ^2\!F_{5/2}$  &  3.8983382 & 4.19 &
    1.5E+07  & 2.08E+06 & 0.139 & D+ \\
$6d 7s^2\, ^2\!D_{5/2}$  & $6d^2 7p\, ^4\!S_{3/2}$  &  3.8047011 & 4.06 &
    ---      & 8.54E+06 & --- & --- \\
$6d 7s^2\, ^2\!D_{5/2}$  & $6d^2 7p\, ^4\!D_{3/2}$  &  3.7522887 & 3.70 &
    ---      & 2.23E+06 & --- & --- \\
$6d 7s^2\, ^2\!D_{5/2}$  & $6d^2 7p\, ^2\!F_{5/2}$  &  3.6660786 & 3.84 &
    ---      & 1.20E+06 & --- & --- \\
$6d 7s^2\, ^2\!D_{5/2}$  & $6d^2 7p\, ^2\!D_{5/2}$  &  3.6281802 & 4.00 &
    6.6E+07  & 6.57E+06 & 0.0995 & D+ \\
$6d 7s^2\, ^2\!D_{5/2}$  & $6d^2 7p\, ^4\!D_{3/2}$  &  3.4920317 & 3.82 &
    1.3E+07  & 4.21E+06 & 0.324 & E \\
$6d 7s^2\, ^2\!D_{5/2}$  & $6d^2 7p\, ^4\!F_{7/2}$  &  3.265365 & 3.45 &
    8.E+06  & 9.78E+05 & 0.122 & D+ \\
$6d 7s^2\, ^2\!D_{5/2}$  & $6d 6s 7p\, ^2\!F_{7/2}$  &  3.1899927 & 2.79 &
    1.1E+08  & 1.90E+08 & 1.723 & D+ \\
$6d 7s^2\, ^2\!D_{5/2}$  & $6d^2 7p\, ^4\!F_{5/2}$  &  3.0721332 & 3.30 &
    3.7E+07  & 3.49E+06 & 0.0943 & D+ \\
$6d 7s^2\, ^2\!D_{5/2}$  & $6d 7s 7p\, ^2\!F_{5/2}$  &  3.0506018 & 2.60 &
    ---      & 1.99E+07 & --- & --- \\
$6d 7s^2\, ^2\!D_{5/2}$  & $6d^2 7p\, ^4\!F_{3/2}$  &  3.0130314 & 3.17  &
    ---      & 9.74E+05 & --- & --- \\
$6d 7s^2\, ^2\!D_{5/2}$  & $6d^2 7p\, ^2\!D_{5/2}$  &  2.9551157 & 2.81 &
    1.0E+08  & 1.71E+08 & 1.71 & D+ \\
$6d 7s^2\, ^2\!D_{5/2}$  & $6d 7s 7p\, ^2\!P_{3/2}$  &  2.8191355 & 2.61 &
    7.2E+07  & 7.64E+07 & 1.06 & D+ \\
$6d 7s^2\, ^2\!D_{5/2}$  & $6d 7s 7p\, ^2\!F_{7/2}$  &  2.7774347& 2.46 &
    4.1E+07  & 3.00E+07 & 0.732 & D+ \\
$6d 7s^2\, ^2\!D_{5/2}$  & $6d^2 7p\, ^2\!D_{3/2}$  &  2.6886477 & 2.44 &
    ---      & 1.46E+07 & --- & --- \\
$6d 7s^2\, ^2\!D_{5/2}$  & $6d^2 7p\, ^4\!G_{5/2}$  &  2.6864196 & 2.90 &
    1.1E+07  & 8.63E+06 & 0.785 & E \\
$6d 7s^2\, ^2\!D_{5/2}$  & $6d^2 7p\, ^4\!G_{7/2}$  &  2.6339837 & 3.09 &
    5.E+06  & 1.42E+07 & 2.84 & E \\
$6d 7s^2\, ^2\!D_{5/2}$  & $6d 7s 7p\, ^4\!P_{5/2}$  &  2.5503147 & 2.06 &
    ---     & 2.23E+05 & --- & --- \\
$6d 7s^2\, ^2\!D_{5/2}$  & $6d 7s 7p\, ^2\!F_{5/2}$  &  2.3512907 & 2.21 &
    1.6E+07 & 2.25E+07 & 1.41 & D+ \\
$6d 7s^2\, ^2\!D_{5/2}$  & $6d 7s 7p\, ^2\!D_{3/2}$  &  2.0805811 & 1.75 &
    ---     & 2.08E+04 & --- & --- \\
$6d 7s^2\, ^2\!D_{5/2}$  & $6d 7s 7p\, ^2\!D_{5/2}$ or $^4\!D_{5/2}$  &  1.948942  & 1.62 &
    7.6E+06 & 2.77E+06 & 0.364 & B \\
$6d 7s^2\, ^2\!D_{5/2}$  & $6d 7s 7p\, ^4\!D_{3/2}$  &  1.9223543 & 1.53 &
    7.7E+05 & 9.97E+04 & 0.129 & C \\
$6d 7s^2\, ^2\!D_{5/2}$  & $6d 7s 7p\, ^4\!F_{7/2}$  &  1.9158584 & 1.58 &
    1.18E+06 & 7.01E+05 & 0.594 & C+ \\
$6d 7s^2\, ^2\!D_{5/2}$  & $6d 7s 7p\, ^4\!F_{5/2}$  &  1.5757503 & 1.30 &
    1.56E+06 & 9.36E+05 & 0.600 & C+ \\
$6d 7s^2\, ^2\!D_{5/2}$  & $6d 7s 7p\, ^4\!F_{3/2}$  &  1.4235204 & 1.17 &
    1.7E+05 & 4.35E+04 & 0.256 & C \\
$6d 7s^2\, ^2\!D_{5/2}$  & $7s^2 7p\, ^2\!P_{3/2}$  &  1.2454607 & 0.958 &
    3.56E+06 & 2.54E+06 & 0.713 & B+ \\
\hline
\multicolumn{8}{c}{Ac \sc{ii}} \\
\hline
$7s^2\, ^1\!S_{0}$  & $6d 7p\, ^1\!P_{1}$  &  5.480092 & 5.26 &
    ---      & 5.72E+08 & --- & --- \\
$7s^2\, ^1\!S_{0}$  & $6d 7p\, ^3\!P_{1}$  &  4.5694975 & 4.03 &
    2.5E+07  & 5.48E+07 & 2.19 & E \\
$7s^2\, ^1\!S_{0}$  & $6d 7p\, ^3\!D_{1}$  &  4.139654  & 3.67 &
    1.6E+08  & 1.69E+08 & 1.056 & E \\
$7s^2\, ^1\!S_{0}$  & $6d 7p$ or $7s 7p\, ^1\!P_{1}$  &  3.6265838 & 3.19 &
    1.2E+08  & 1.09E+08 & 0.908 & E \\
$7s^2\, ^1\!S_{0}$  & $7s 7p\, ^3\!P_{1}$  &  2.750035  & 2.42 &
    2.13E+07 & 1.26E+07 & 0.591 & B \\
$6d 7s\, ^3\!D_{1}$  & $5f 6d\, ^1\!D_{2}$  & 4.9551105 & 5.84 &
    ---      & 8.79E+06 & --- & --- \\
$6d 7s\, ^3\!D_{1}$  & $5f 6d\, ^3\!F_{2}$  & 4.5674445 & 5.27 &
    3.4E+07  & 5.70E+07 & 1.68 & E \\
$6d 7s\, ^3\!D_{1}$  & $6d 7p\, ^3\!P_{2}$  & 4.1698345 & 4.02 &
    ---      & 1.54E+07 & --- & --- \\
$6d 7s\, ^3\!D_{1}$  & $6d 7p\, ^3\!P_{1}$  & 3.981858  & 3.79 &
    2.4E+08  & 1.26E+08 & 0.525 & D+ \\
$6d 7s\, ^3\!D_{1}$  & $6d 7p\, ^3\!P_{0}$  & 3.9725015 & 3.74 &
    4.E+08  & 3.79E+08 & 0.948 & E \\
$6d 7s\, ^3\!D_{1}$  & $6d 7p\, ^3\!F_{2}$  & 3.8010436 & 4.46  &
    1.7E+08  & 3.24E+08 & 1.91 & E \\
$6d 7s\, ^3\!D_{1}$  & $6d 7p\, ^3\!D_{1}$  & 3.5520145 & 3.43 &
    1.0E+08  & 1.49E+08 & 1.49 & D+ \\
$6d 7s\, ^3\!D_{1}$  & $6d 7p\, ^3\!D_{2}$  & 3.5416468 & 3.49 &
    ---      & 3.31E+07 & --- & --- \\
$6d 7s\, ^3\!D_{1}$  & $6d 7p\, ^1\!D_{2}$  & 3.3648353 & 3.36 &
    1.0E+07  & 3.17E+07 & 3.17 & E \\
$6d 7s\, ^3\!D_{1}$  & $6d 7p\, ^1\!P_{1}$  & 3.0389443 & 2.95 &
    ---      & 1.21E+07 & --- & --- \\
$6d 7s\, ^3\!D_{1}$  & $7s 7p\, ^3\!P_{2}$  & 2.9088535 & 2.69 &
    ---      & 6.93E+06 & --- & --- \\
$6d 7s\, ^3\!D_{1}$  & $6d 7p\, ^3\!F_{2}$  & 2.6913618 & 2.75 &
    2.6E+07  & 3.99E+07 & 1.53 & D+ \\
$6d 7s\, ^3\!D_{1}$  & $7s 7p\, ^3\!P_{1}$  & 2.1623955 & 2.18 &
    1.04E+07 & 1.13E+07 & 1.09 & B \\
$6d 7s\, ^3\!D_{1}$  & $7s 7p\, ^3\!P_{0}$  & 2.0106255 & 2.05 &
    2.80E+07 & 3.59E+07 & 1.28 & B \\
\hline
\multicolumn{8}{c}{Ac \sc{iii}} \\
\hline
$6p^6 7s\, ^2\!S_{1/2}$  & $6p^6 7p\, ^2\!P_{3/2}$  &  4.719214 & 4.51 &
    3.97e+08 & 5.04E+08 & 1.27 & B+ \\
$6p^6 6d\, ^2\!D_{3/2}$  & $6p^6 7p\, ^2\!P_{3/2}$  &  4.619904 & 4.60 &
    2.89e+07 & 4.52E+07 & 1.56 & B+ \\
$6p^6 6d\, ^2\!D_{5/2}$  & $6p^6 7p\, ^2\!P_{3/2}$  &  4.197993 & 4.13 &
    2.30e+08 & 2.93E+08 & 1.27 & B+ \\
$6p^6 7s\, ^2\!S_{1/2}$  & $6p^6 7p\, ^2\!P_{1/2}$  &  3.653305 & 3.67 &
    1.90e+08 & 2.74E+08 & 1.44 & B+ \\
$6p^6 6d\, ^2\!D_{3/2}$  & $6p^6 7p\, ^2\!P_{1/2}$  &  3.553995 & 3.77 &
    1.58e+08 & 2.49E+08 & 1.57 & B+ \\
$6p^6 6d\, ^2\!D_{3/2}$  & $6p^6 5f\, ^2\!F_{5/2}$  &  2.808675 & 3.62 &
    1.85e+07 & 8.81E+07 & 4.76 & B+ \\
$6p^6 6d\, ^2\!D_{5/2}$  & $6p^6 5f\, ^2\!F_{7/2}$  &  2.712320 & 3.51 &
    2.11e+07 & 8.56E+07 & 4.06 & B+ \\
$6p^6 6d\, ^2\!D_{5/2}$  & $6p^6 5f\, ^2\!F_{5/2}$  &  2.386764 & 3.14 &
    8.79e+05 & 4.12E+06 & 4.69 & B+ \\
\hline
\end{tabular}
\label{tab:ac_data}
\end{table*}
In these comparisons, we consider all of the available NIST data
for Ac~{\sc iii}, but limit the comparisons for Ac~{\sc i}
and Ac~{\sc ii} in the interest of brevity and also due to the labor-intenstive
process required to match the level designations between the present
and NIST data sets. Not surprisingly, the labels do not appear in the same 
energy order, and the appearance of duplicate labels (due to the complexity
of the angular momentum coupling) makes it challenging to make clear
matches. Sometimes, an inspection of the mixing purity 
(``Leading percentages'') quantity is also required
in order to make a definitive identification, if such information is provided.
Thus, for Ac~{\sc i} we considered only those radiative transitions
with a lower level of the first excited state, $6d\, 7s^2\, ^2\!D_{5/2}$,
which is predicted to lie only 0.277~eV above the $6d\, 7s^2\, ^2\!D_{3/2}$
ground state, according to the NIST database. We chose the first excited
state, rather than the ground state, because the NIST database includes
a $6d - 5f$ transition for the excited state, while no such transitions
are provided for the ground state. For Ac~{\sc ii} we chose two lower levels,
the ground and first excited states, in order to highlight the various
types of orbital transitions. Despite these constraints,
the comparisons in Table~\ref{tab:ac_data} provide 
a representative sample of the accuracy for the various types
of radiative transitions in our Ac model.

Starting with the highest charged ion stage, Ac~{\sc iii}, which is
expected to be the most accurately calculated, we see that the
transition energies agree well (within 6\%) for the
the first five transitions, which are of the type $6d - 7p$ and $7s - 7p$.
On the other hand, the energies for the remaining three transitions,
which are of the type $6d -5f$, differ by about 30\%, indicating
the difficulty in accurately calculating the $5f$ wavefunction
with the semi-relativistic approach. The radiative decay rates follow a
similar trend, with the first five transitions showing differences
of 27--57\%, while the last three transitions display significantly
larger differences, i.e. greater than a factor of four. For convenience,
we have also provided the letter codes for the NIST accuracy estimate
in the final column of the table. These codes range from AAA $(< 0.3\%)$
to E $(> 50\%)$. (See the NIST website for a full listing of accuracy codes.)
All eight transitions for Ac~{\sc iii} are rated B+,
which indicates an accuracy of $\le 7$\%. Thus, all of the calculated
rates fall outside of the NIST estimated accuracy for this ion stage.

Moving on to Ac~{\sc ii}, the five transitions with the ground state
as the lowest level display differences in the transition energies
ranging from 4--14\%. The fourteen transitions with the first excited
state as the lowest level display similar differences, ranging from
0.1--18\%. The decay rates show a broad range of differences, ranging
from 5.6\% to a factor of 3.17. A number of the calculated rates
fall within the NIST accuracy estimate, but not all of them.
We also note that the calculated energies and decay rates are
sometimes higher, and sometimes lower, than the NIST values,
indicating a lack of a systematic shift in the calculated data.

Finally, we consider the 29 transitions within the neutral stage, Ac~{\sc i},
with the first excited state as the lower level. The comparisons for
this ion stage display the largest disagreements. The transition energy
differences range from 0.5--30\%.
The radiative decay rates differ by as little as 6\%, while a factor
of about 10 is obtained for two of the transitions.
A few of the calculated rates fall within the NIST accuracy estimate,
but most do not.
As in the case of the singly ionized stage, there does not appear
to be a systematic (upward or downward) shift in the calculated data
relative to the NIST data.

\section{Sample Opacities and Tables}
\label{sec:opac}

In this section, we provide some example opacities and information
concerning the tabular opacities that are generated for KN simulations.
The content presented here is similar to that provided in Section~3 of Paper I
and is, therefore, presented in a more concentrated, abbreviated form.

\begin{figure}
\includegraphics[clip=true,angle=0,width=1.0\columnwidth]
{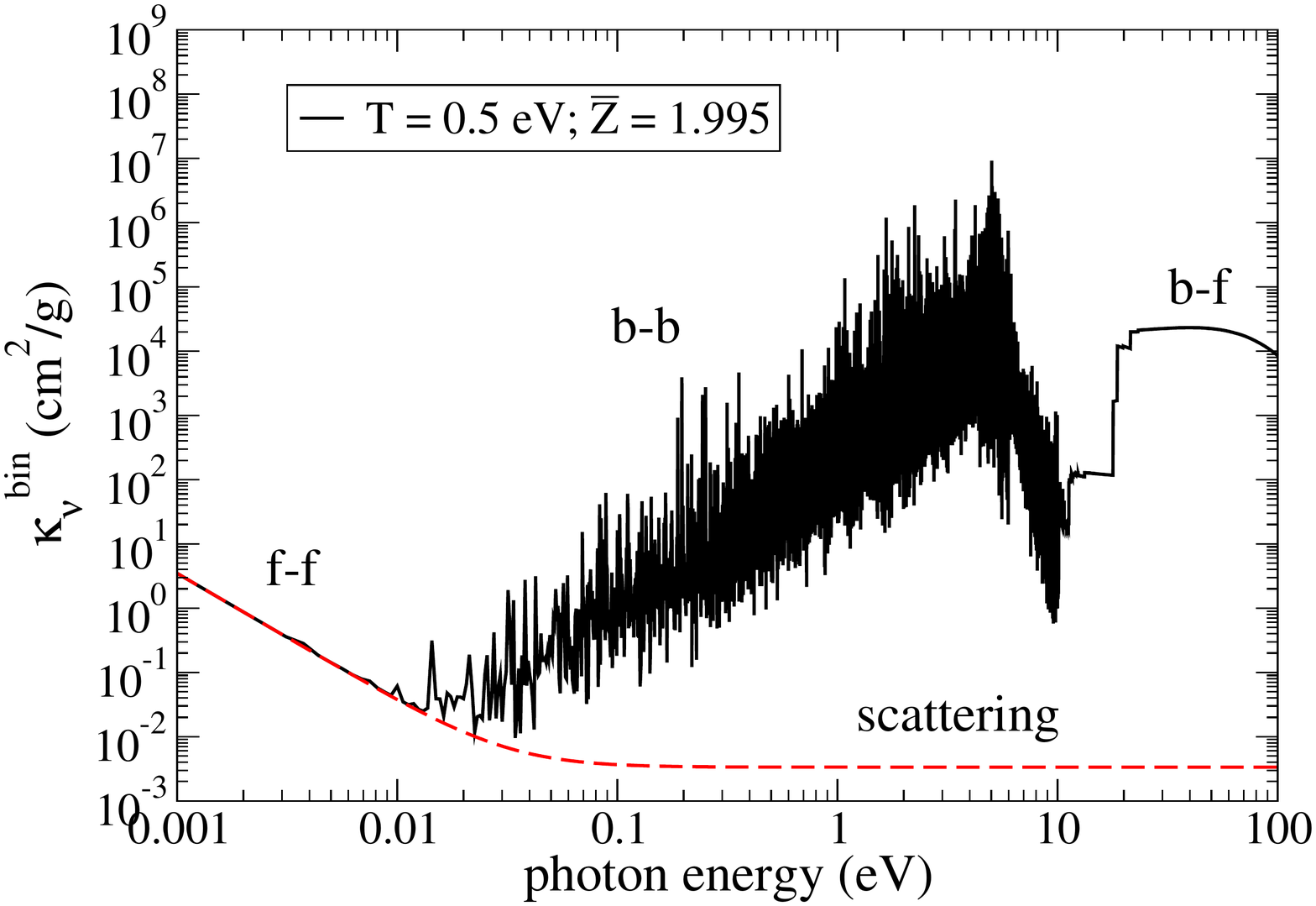}
\caption{
The LTE line-binned opacity for uranium at $T = 0.5$~eV and
$\rho = 10^{-13}$~g/cm$^3$. The solid black curve displays the complete
(or total) opacity, which includes the bound-bound, bound-free,
free-free and scattering contributions.
The dashed red curve displays only the contributions due to free electrons,
i.e. the free-free and scattering contributions.
The average charge state, $\overline{Z}$, for these conditions is listed
in the legend.
}
\label{fig:opac_u_1}
\end{figure}

In order to illustrate the basic characteristics of the opacities used
in this study, we present in Fig.~\ref{fig:opac_u_1}
the monochromatic opacity of U for typical ejecta conditions
of $T = 0.5$~eV and $\rho = 10^{-13}$~g/cm$^3$.
The solid black curve displays the complete (or total) opacity, with all four
contributions (b-b, b-f, f-f and scattering), while the dashed red curve
shows the contributions that arise only from free electrons (f-f and
scattering) in order to highlight the massive
differences that can occur when the bound electrons are taken into
account. The b-b contribution was calculated via the line-binned
expression in equation~(\ref{opac_bb_bin}).
The f-f and scattering contributions were obtained from the
simple, analytic formulae \citep{huebner} associated with Kramers and
Thomson, respectively. The gap between the b-b features and the onset
of the b-f edges occurring at $\sim 10$--20~eV is due to missing lines that
would be present if more excited configurations had been included in
the model.
(Our transport calculations have minimal intensity at these energies.)
We note that a mean charge state of $\overline{Z} = 1.995$ is
obtained for these conditions, indicating that the opacity is
dominated by U~{\sc iii}, or doubly ionized uranium.

\subsection{Examples of line-binned opacities}
\label{sub:opac_lb}

\begin{figure*}
\hbox{
\includegraphics[clip=true,angle=0,width=0.666\columnwidth]
{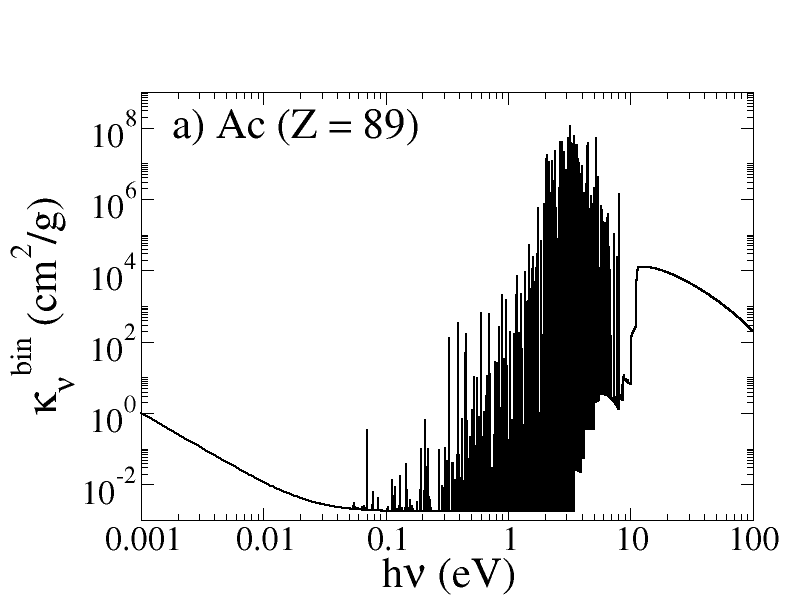}
%\hfill
\includegraphics[clip=true,angle=0,width=0.666\columnwidth]
{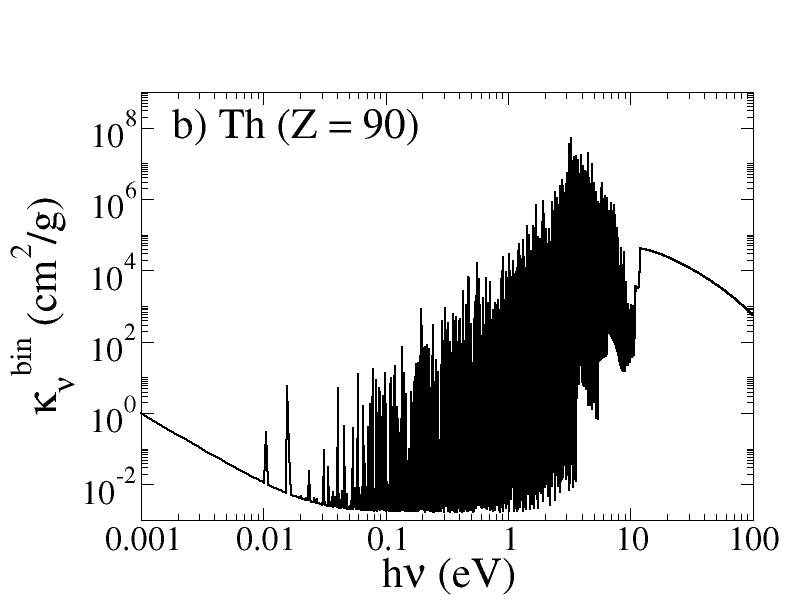}
%\hfill
\includegraphics[clip=true,angle=0,width=0.666\columnwidth]
{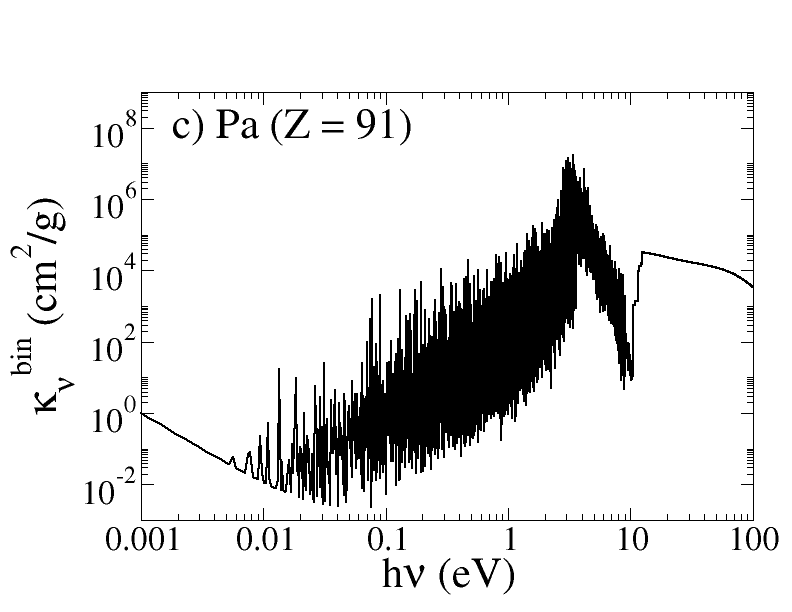}
}
\hbox{
\includegraphics[clip=true,angle=0,width=0.666\columnwidth]
{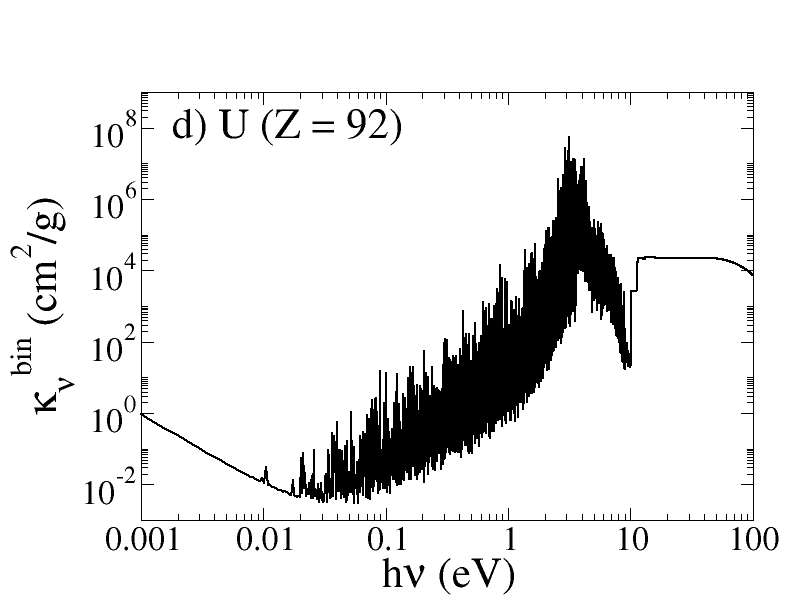}
%\hfill
\includegraphics[clip=true,angle=0,width=0.666\columnwidth]
{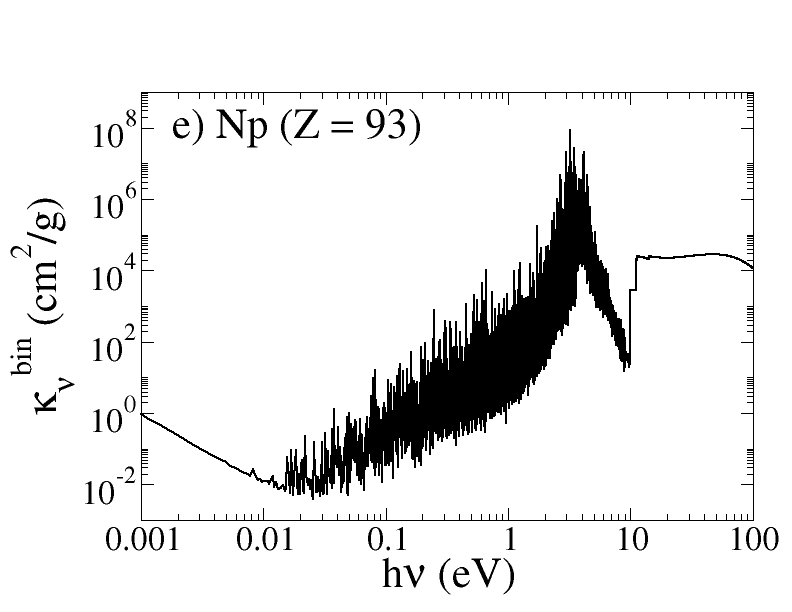}
%\hfill
\includegraphics[clip=true,angle=0,width=0.666\columnwidth]
{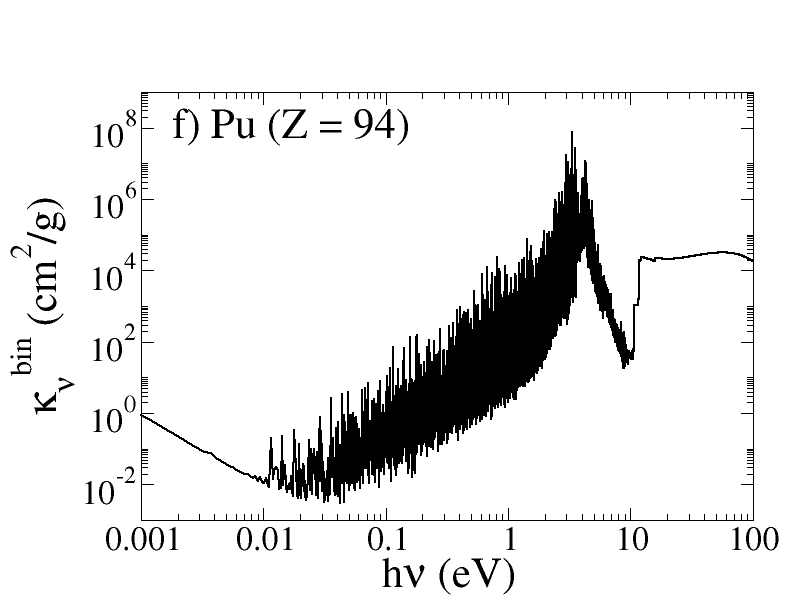}
}
\hbox{
\includegraphics[clip=true,angle=0,width=0.666\columnwidth]
{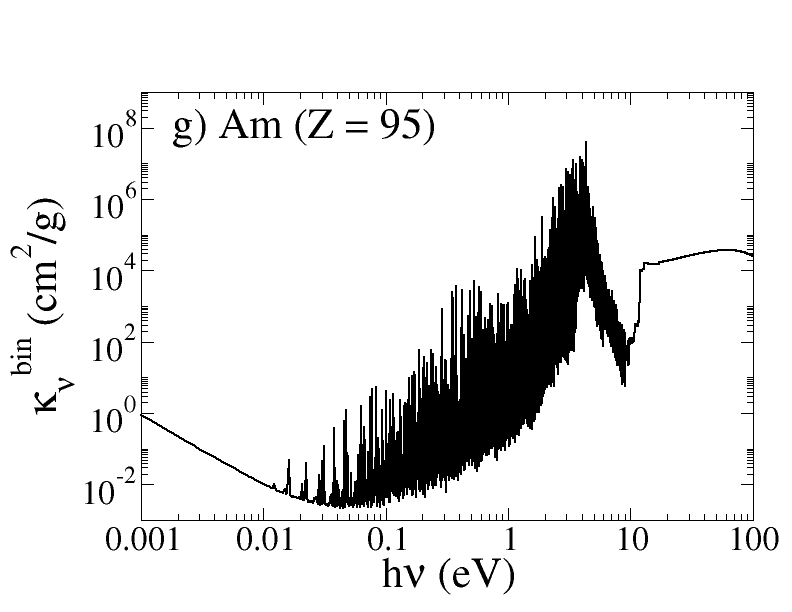}
%\hfill
\includegraphics[clip=true,angle=0,width=0.666\columnwidth]
{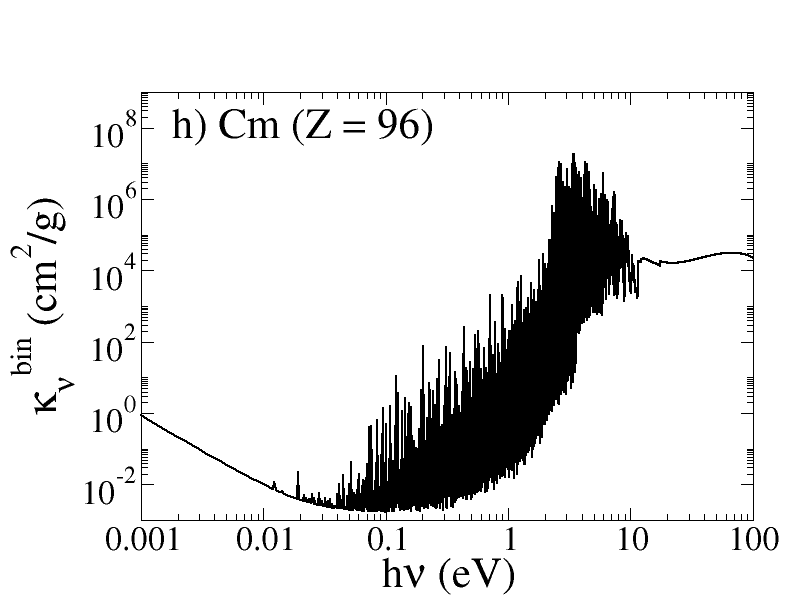}
%\hfill
\includegraphics[clip=true,angle=0,width=0.666\columnwidth]
{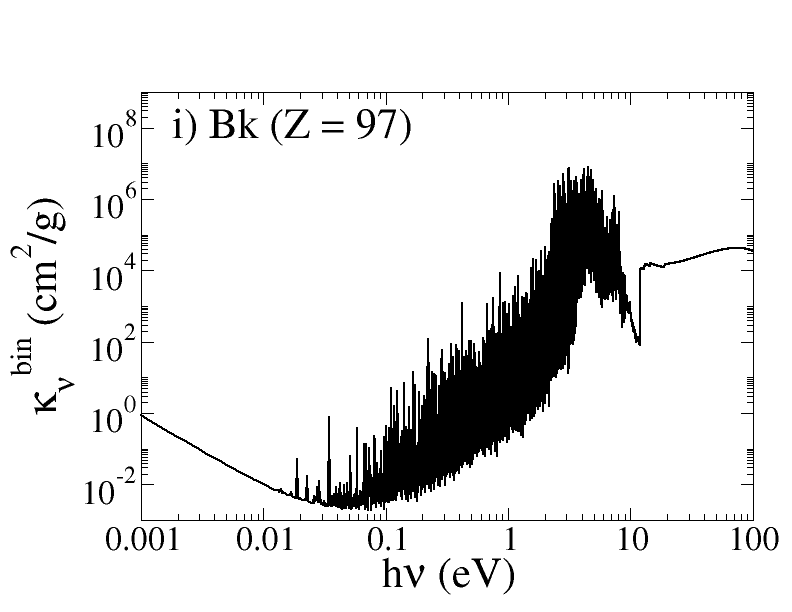}
}
\hbox{
\includegraphics[clip=true,angle=0,width=0.666\columnwidth]
{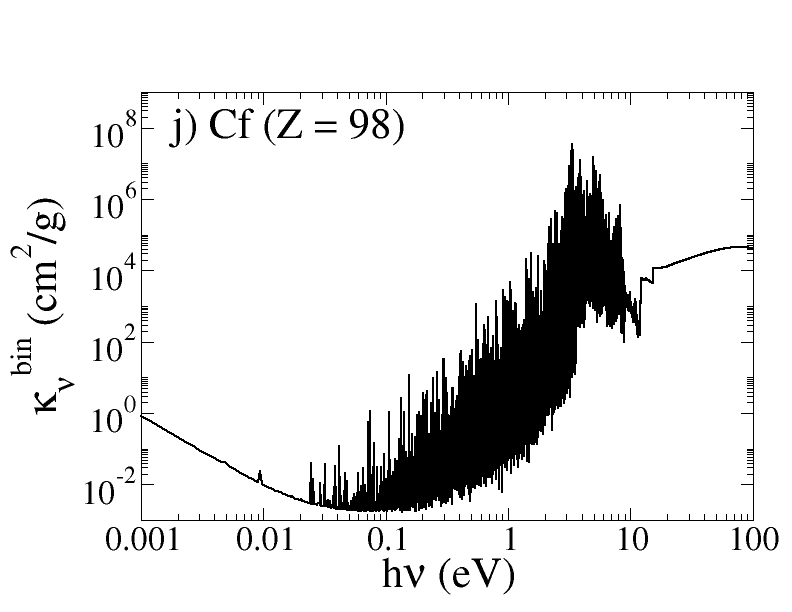}
%\hfill
\includegraphics[clip=true,angle=0,width=0.666\columnwidth]
{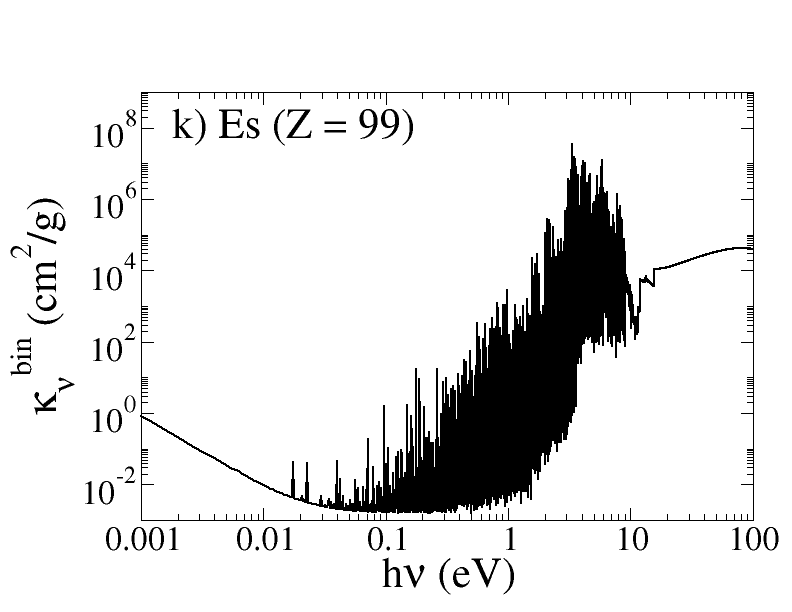}
%\hfill
\includegraphics[clip=true,angle=0,width=0.666\columnwidth]
{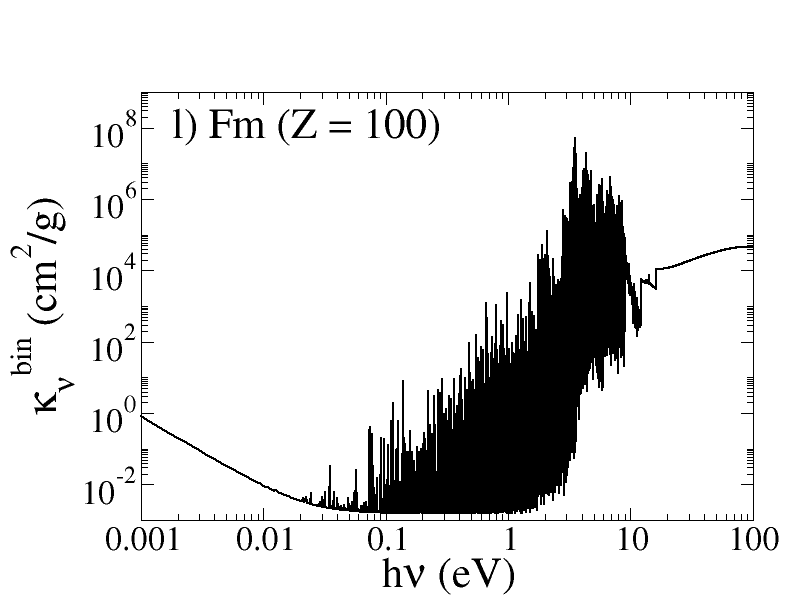}
}
\hbox{
\includegraphics[clip=true,angle=0,width=0.666\columnwidth]
{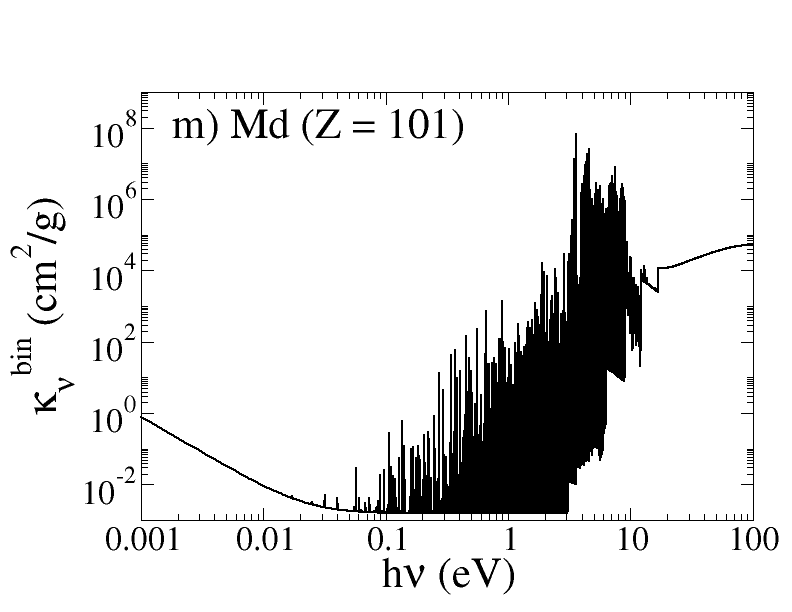}
%\hfill
\includegraphics[clip=true,angle=0,width=0.666\columnwidth]
{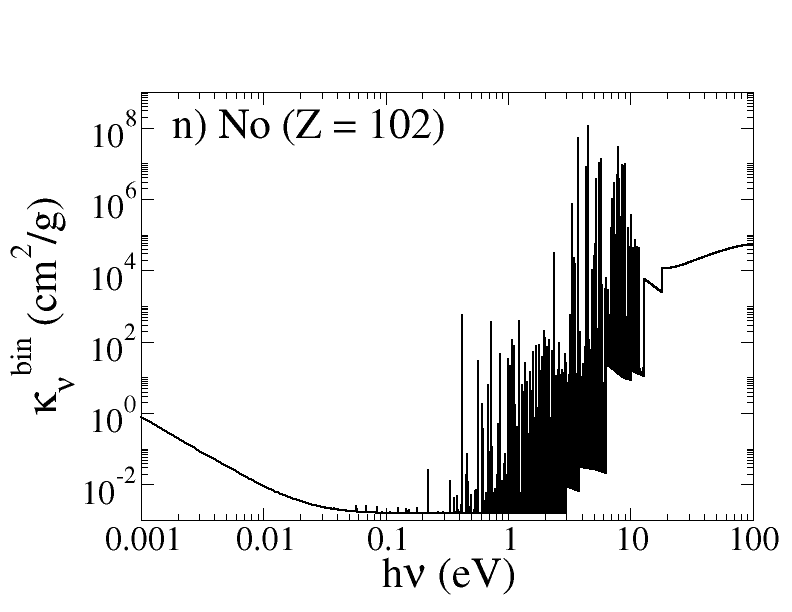}
\hfill
}
\caption{The LTE line-binned opacity for all 14 actinide elements
at $T = 0.3$~eV and $\rho = 10^{-13}$~g/cm$^3$. For these
conditions, the bound-bound contribution to the opacity is dominated by
the second, i.e. singly ionized, ion stage of each element.
Panels a--n display results for $Z = $89--102 in numerical order.}
\label{fig:opac_allZ2}
\end{figure*}

To illustrate how the opacities behave as a function of element,
we present a complete set of actinide opacities at a characteristic
ejecta density of $\rho = 10^{-13}$~g/cm$^3$ in Fig.~\ref{fig:opac_allZ2}.
In this case, a temperature of $T = 0.3$~eV was chosen in order
to highlight opacities with a b-b contribution that is
dominated by the second, i.e. singly ionized, ion stage of each element.

\begin{figure*}
\includegraphics[clip=true,angle=0,width=\columnwidth]
{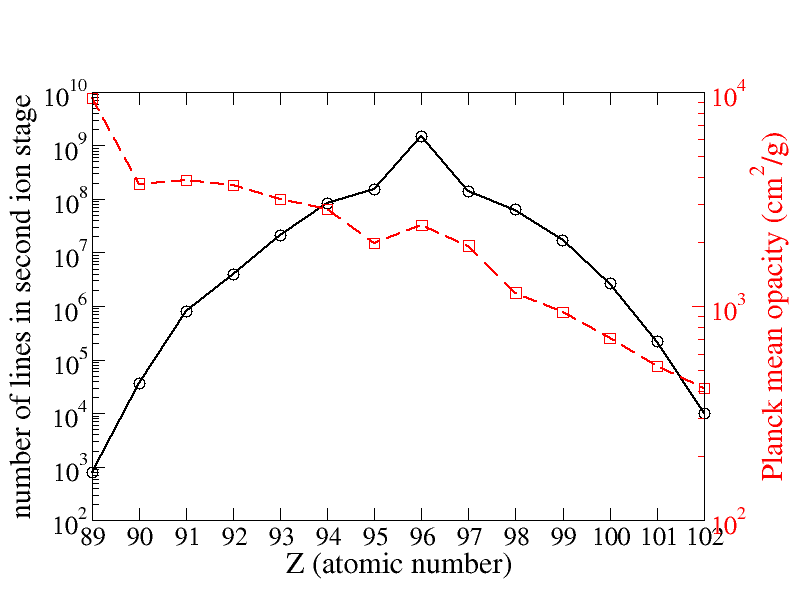}
\hfill
\includegraphics[clip=true,angle=0,width=\columnwidth]
{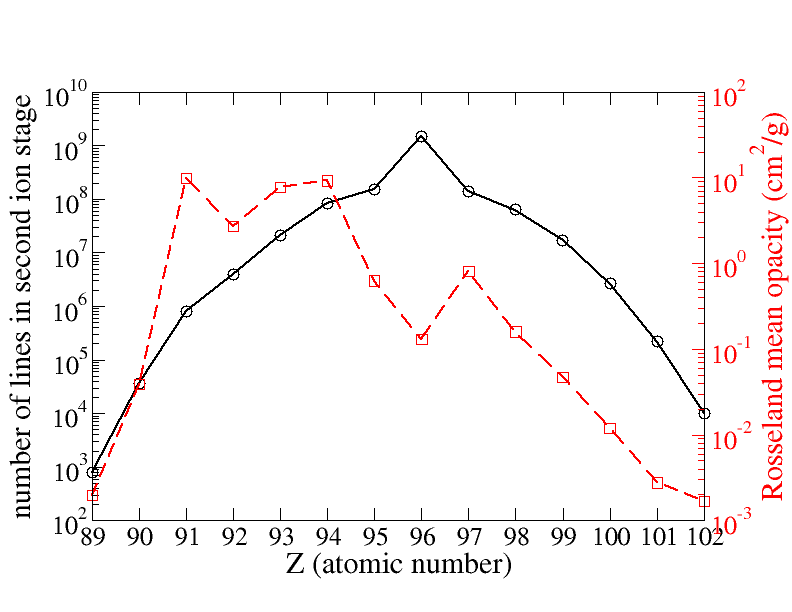}
\caption{The number of lines in the second ion stage versus atomic number, $Z$
(see, also, Table~\ref{tab:configs}).
The mean opacity associated with the line-binned opacities presented
in Fig.~\ref{fig:opac_allZ2} is also plotted versus $Z$.
Results are presented for all 14 actinide elements ($Z = 89$--102).
In both panels, the number of lines for the second ion stage is represented
by the black solid curve (with circles). This curve is associated with
the left-hand $y$ axis in each panel. The red dashed curves (with squares),
assocated with the right-hand $y$ axis in each panel, represent
the Planck mean opacity in the left panel and the Rosseland mean opacity
in the right panel. The mean opacities were calculated at
$T = 0.3$~eV and $\rho = 10^{-13}$~g/cm$^3$, corresponding to the conditions
used in Fig.~\ref{fig:opac_allZ2}.
}
\label{fig:opac_allZ2_anal}
\end{figure*}

The qualitative trends in the various panels of this figure strongly resemble
those presented in the corresponding plot (Fig.~7) of Paper~I for the singly
ionized lanthanides. For example, the bound-bound features generally increase
in strength with photon energy for each element, peaking at about 3~eV,
indicating that the emission of photons in the visible range would be
strongly suppressed for these conditions. However, as in the case
of lanthanide opacities,
there are important quantitative differences in these detailed line-binned
features when comparing one element to another,
which we illustrate in Fig.~\ref{fig:opac_allZ2_anal}.
In each of the two panels in Figure~\ref{fig:opac_allZ2_anal},
the solid black curve (with circles) represents the number of
lines in the second ion stage, which is the dominant stage for
these conditions, for each element. The number of lines for the various
ion stages can also be found in Table~\ref{tab:configs}.
Superimposed on this black curve is a red dashed curve (with squares)
that represents the mean opacity obtained from the frequency-dependent
opacities in Figure~\ref{fig:opac_allZ2}. The red dashed curve in the left
panel represents the Planck mean opacity, while
the red dashed curve in the right panel represents the Rosseland mean opacity.

As expected, the number of lines reaches a maximum near the center of the
actinide range (occuring at Cm~{\sc ii} in this example),
due to the presence of a ground-state configuration
with a $5f^7$ subshell. This half-filled subshell produces the maximum number
of fine-structure levels, compared to other ground-state configurations with
different $5f$ occupation numbers, due to the rules of quantum
mechanics and angular momentum coupling. Also as expected from the
lanthanide-opacity analysis in Paper~I, the mean-opacity curves display
some structure, rather than near-constant behavior,
illustrating that there exist significant differences in the
detailed line structures for each element.

In fact, both opacity curves indicate relatively high mean values
for the first half of the actinide range, followed by
a local minimum near the center of the actinide range, and then a strong
monotonic decrease over the latter half of the range.
These trends are very similar to those exhibited by the singly
ionized lanthanides in Fig.~9 of Paper~I.
The high values over the first part of the range indicate that the energy-level
spacings of the low-lying levels of these actinides are conducive
to the absorption of photons in relevant energy ranges for KN modeling.
On the other hand, the monotically decreasing behavior indicates that the latter
half of the actinides should not be as effective at absorbing photons.

We also note that this elemental sensitivity is consistent with previous
studies of KN emission sensitivity to the lanthanide opacities
\citep{even20,tanaka20}
in which it was shown that Nd has an outsize effect 
due to the energy positions and strengths
of its line absorption features compared to the other lanthanides.
The effect of U was also studied by \citealt{even20}, as it was the sole,
representative actinide in that work, and it was found
to have a similar outsize effect on KN emission.
This common behavior of Nd and U is expected because these two elements
are homologues, with low-lying energy levels containing $4f$ and $5f$ electrons,
respectively.

\subsection{Opacity tables}
\label{sub:opac_tab}

In order to perform radiation-transport calculations in an efficient
manner, opacity tables were generated for the 14 actinide elements discussed
above, using prescribed temperature and density grids that span
the range of conditions of interest.
The temperature grid consists of 27 values (in eV):
0.01, 0.07, 0.1, 0.14, 0.17, 0.2, 0.22, 0.24, 0.27, 0.3, 0.34, 0.4, 0.5, 0.6,
0.7, 0.8, 0.9, 1.0, 1.2, 1.5, 2.0, 2.5, 3.0, 3.5, 4.0, 4.5, and 5.0.
(As mentioned in Section~\ref{subsec:detailed_atomic}, the present models are
only valid up to $T \sim 2$~eV, due to the inclusion of just the first
four ion stages in the atomic physics calculations.)
The density grid contains 17 values ranging from 10$^{-20}$
to 10$^{-4}$~g/cm$^3$, with one value per decade.
Our photon energy grid is the same 14,900-point grid
that is used in standard Los Alamos tabular opacity efforts,
e.g.~\citealt{colgan_oplib}. The grid is actually a temperature-scaled
$u = h\nu/kT$ grid with a non-uniform spacing that is designed to provide
accurate Rosseland and Planck mean opacities. A description of this grid is
available in Table~1 of \citealt{frey13}.

\section{Investigation of Actinide Opacities in the Modeling of Light Curves
and Spectra}
\label{sec:lcs}

In this section, we explore the effect of actinide opacities
on the modeling of KN light curves and spectra.
We are particularly interested in searching for spectral signatures that could
be used to distinguish between differing actinide abundances in the dynamical
ejecta. Such distinctions may provide unique insight into the relevant nuclear
properties unreachable by current facilities \citep{mumpower16,horowitz19}.

In order to be consistent with our previous investigation of lanthanide
opacities, we first consider the KN model described in Section~4.2 of Paper~I,
i.e. a dynamical ejecta with a mass of $1.4\times10^{-2}$ M$_{\odot}$
and mean ejecta speed of 0.125$c$ (maximum speed of 0.25$c$).
Based on the spectral results from those simulations, we next expand
our study to consider a broader range of conditions using a prescribed
grid of three ejecta masses and three speeds.
For all of the light-curve and spectral simulations that follow,
we consider dynamical ejecta
with the ``main'' $r$-process elemental abundance pattern described
in \citealt{even20} (see, for example, Fig.~1 therein),
with the total actinide mass fraction held fixed at a value of 0.0884.
However, we allow the abundances for the individual actinides to vary according
to the three distributions described below.
Although we alter the abundance fractions, we assume that the nuclear heating
rate associated with the $r$-process elements is chosen to be the same for all
models, following the prescription described in \citet{wollaeger18}.
As in the case of Paper~I, the radiative transfer for these
simulations is carried out with the Monte
Carlo code {\tt SuperNu} \citep{wollaeger14},
with some improvements to the accuracy of the discrete
diffusion optimization.
The span of wavelength simulated is 1,000 to 128,000~\AA\
with 1,024 logarithmically spaced groups
(see Paper~I for details).

To estimate the range of actinide production,
we perform an $r$-process nucleosynthesis simulation with
Portable Routines for Integrated nucleoSynthesis Modeling (PRISM)
\citep{mumpower17,vassh19} using a NSM trajectory indicative of typical,
very neutron rich (low electron fraction) dynamical ejecta
\citep{piran13,korobkin12,rosswog13}.
Actinide production is sensitive to a variety of nuclear model
inputs~\cite{mumpower2016review}. To provide a range of possible outcomes,
we consider several state-of-the-art predictions in this work. For masses,
we use KTUY~\cite{koura05}, FRDM2012~\cite{moller16},
HFB-17~\cite{goriely09}, and UNEDF1~\cite{kortelainen12}. All other properties
are computed self-consistently with the masses according to the prescription
of Ref.~\cite{mumpower15}.  Neutron capture rates are computed with the
CoH$_{3}$ statistical Hauser-Feshbach code~\cite{kawano16,mumpower17}.
Beta-decay rates are based on the strength data of~\cite{moller19},
and delayed neutron emission is computed with the coupled Los Alamos
Quasi-particle Random Phase Approximation plus Hauser-Feshbach
code~\cite{mumpower16a}. Fission barriers are consistent with the mass model
used, respectively, \cite{koura14,moller15,goriely07}, except
in the case of the UNEDF1 model, as no fission barriers are available.
In this case, we use the barriers of Ref.~\cite{moller15}. Spontaneous,
$\beta$-delayed and neutron-induced fission channels are considered as in
previous work~\cite{zhu18,mumpower18,vassh20,sprouse21}. Theoretical
alpha decays of the heaviest nuclei are estimated using the Viola-Seaborg
relation and depend on the masses~\cite{viola66}. Evaluated data are taken
where available from the Atomic Mass Evaluation (2016)~\cite{wang17b} and
NuBase (2016)~\cite{audi17}.

\begin{figure}
\includegraphics[clip=true,angle=0,width=\columnwidth]
{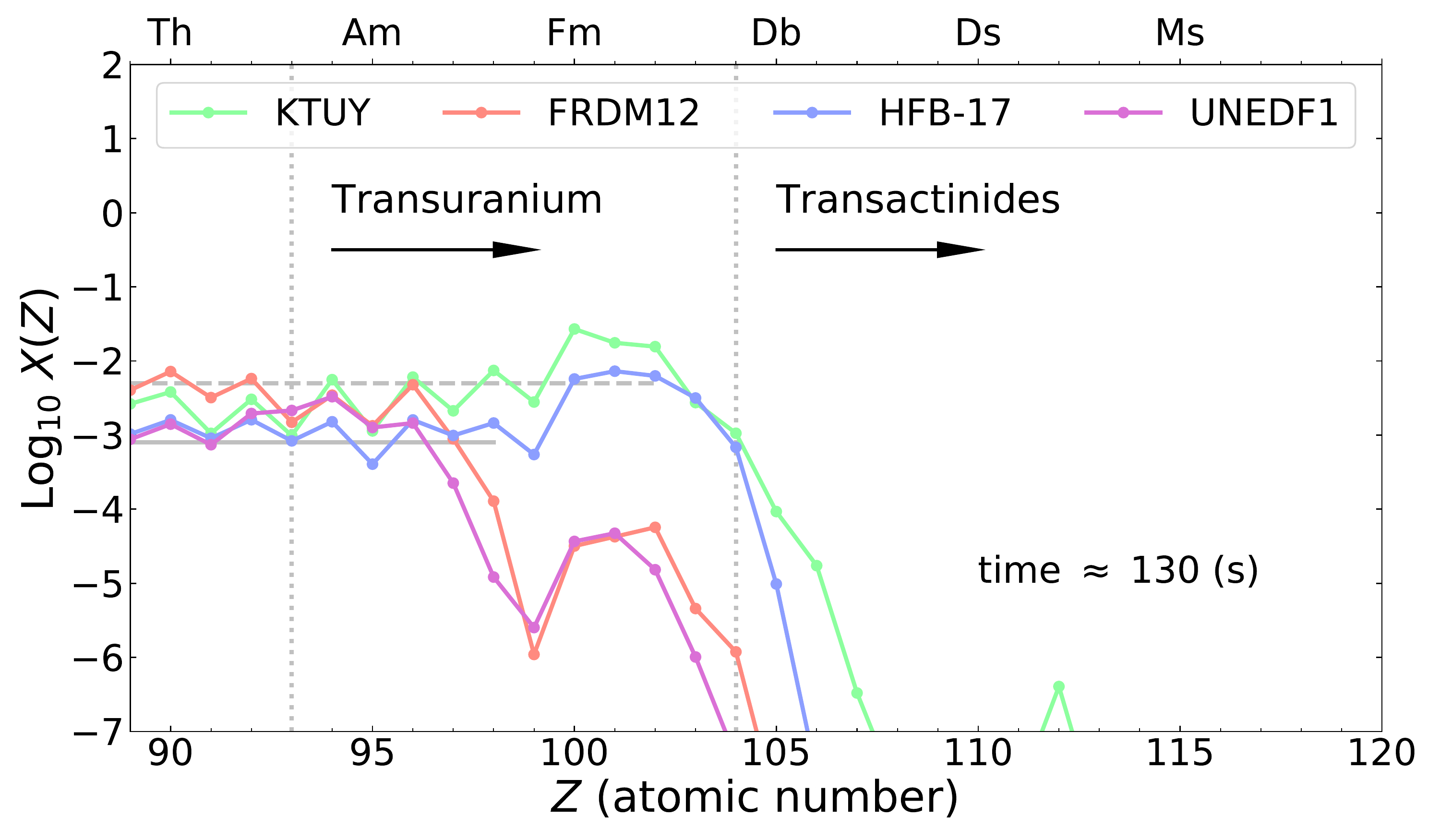}
\caption{The range of actinide and transactinide element production
in dynamical ejecta using a variety of nuclear models. The quantity
$X(Z)$ is the mass fraction of an element with atomic number $Z$.
The two horizontal gray lines illustrate the simplified distributions
that are considered in this work: the solid gray line corresponds
to a uniform distribution for actinides with $Z=89$--98
and the dashed gray line corresponds to a uniform distribution
for $Z=89$--102.
}
\label{fig:nsm1_sheprod}
\end{figure}

Fig~\ref{fig:nsm1_sheprod} shows the resultant range of actinide and
transactinide elemental mass fractions, $X(Z)$, during this simulation. 
These curves were generated with a variety of nuclear models
\citep{koura05,moller16,goriely09,kortelainen12}.
These nuclear abundances change in time due to $\alpha$-decay, $\beta$-decay
and spontaneous fission \citep{zhu18}.
Based on the $Z$ dependence of the curves in this figure,
we chose to simplify our analysis in this work
by considering the following three static-abundance distributions:
pure U (to be consistent with the analysis by \cite{even20}),
a uniform distribution of actinides
within the limited range of $Z = 89$--98 (which corresponds to
the FRDM12 and UNEDF1 models in Fig.~\ref{fig:nsm1_sheprod}),
and a uniform distribution
of actinidies encompassing the entire range of $Z = 89$--102
(similar to KTUY and HFB-17 in Fig.~\ref{fig:nsm1_sheprod}).
The latter two distributions may arise if the fission barrier heights are
locally large in this region, leading to a robust production of actinides
during a merger event \citep{holmbeck19a,holmbeck19b}.
For convenience, these two uniform distributions are illustrated
in Fig.~\ref{fig:nsm1_sheprod} by (solid and dashed) horizontal gray lines,
which differ in value by a ratio of 10/14.

\subsection{Emission study using a single ejecta mass and speed}
\label{sub:study_simple}

\begin{figure}
\includegraphics[clip=true,angle=0,width=1.0\columnwidth]
{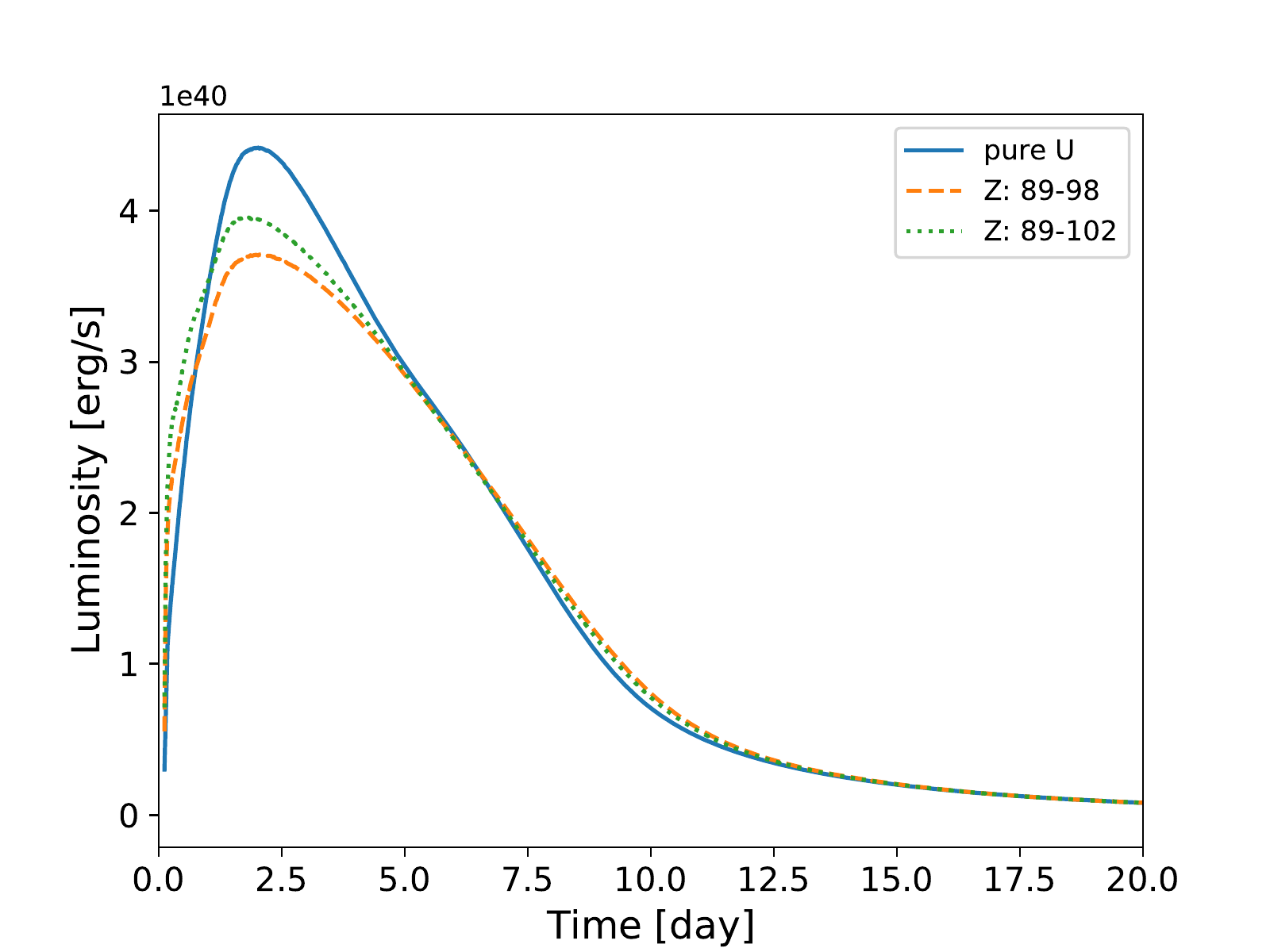}
\caption{
Bolometric luminosity for the first dynamical ejecta model,
with a mass of $1.4\times10^{-2}$ M$_{\odot}$ and mean ejecta speed of 0.125$c$.
Results are displayed for three different actinide abundance distributions:
pure U (solid blue curve),
a uniform distribution including only a
limited range of actinides from $Z = 89$--98 (dashed orange curve),
and a uniform distribution including the entire range of actinides
from $Z = 89$--102 (green dotted curve).
}
\label{fig:lc1}
\end{figure}

As mentioned above, in order to investigate the effect of the three different
actinide distributions,
we first consider the KN model that was used in previous study of 
lanthanide opacities,
i.e. the dynamical ejecta has a mass of $1.4\times10^{-2}$ M$_{\odot}$
and mean ejecta speed of 0.125$c$ (maximum speed of 0.25$c$).
We present the bolometric luminosity for this model in Fig.~\ref{fig:lc1}.
From this light-curve comparison, we observe that using U as a surrogate
for the entire
set of actinides overpredicts the peak of the luminosity by
only 10--20\% relative to the other two, more complete, actinide distributions.
The latter two curves display even less discrepancy at the peak.
Outside of the peak region, the three curves display almost no sensitivity
to the choice of actinide distribution.

\begin{figure*}
\includegraphics[clip=true,angle=0,width=1.0\columnwidth]
{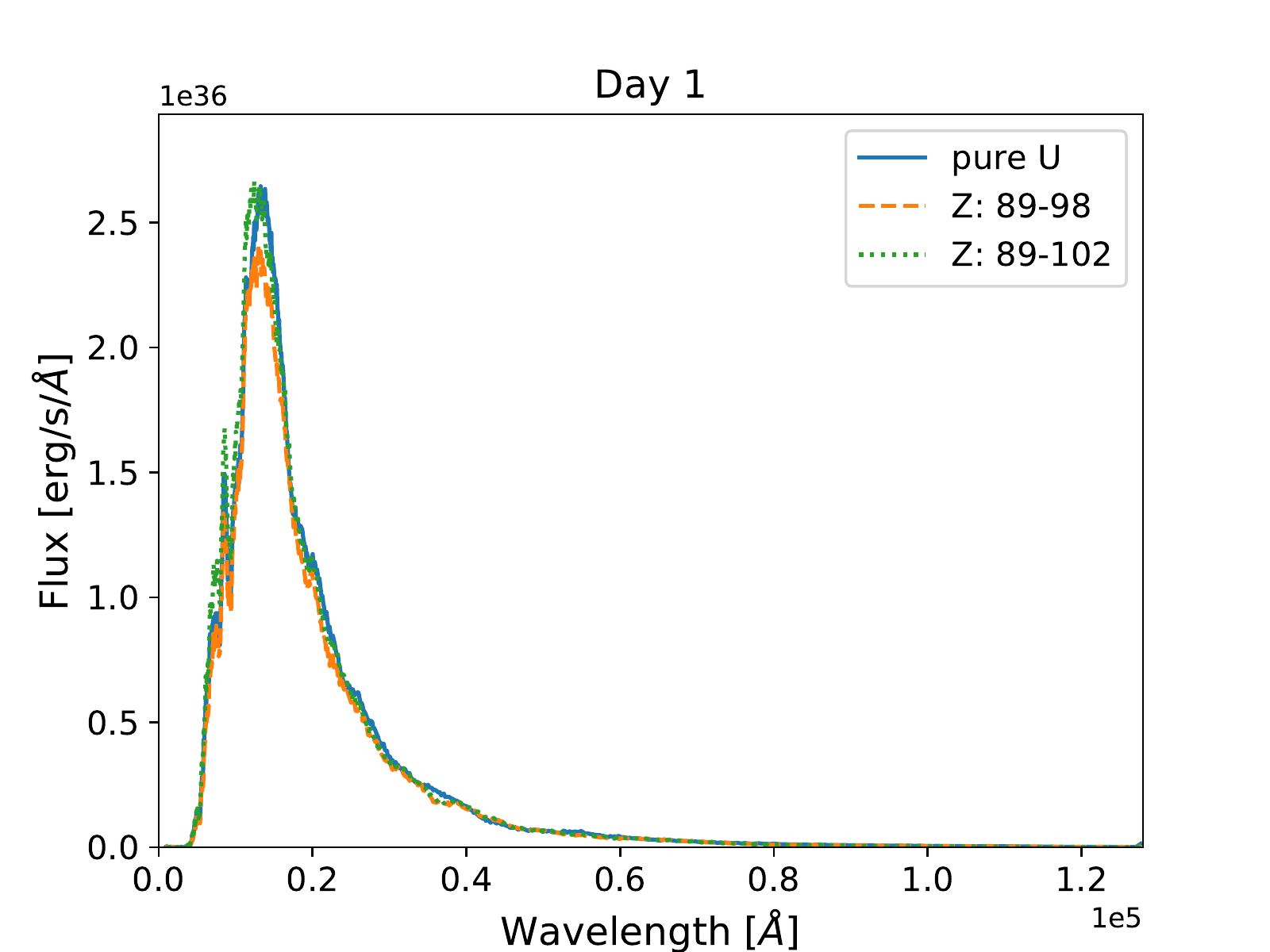}
\hfill
\includegraphics[clip=true,angle=0,width=1.0\columnwidth]
{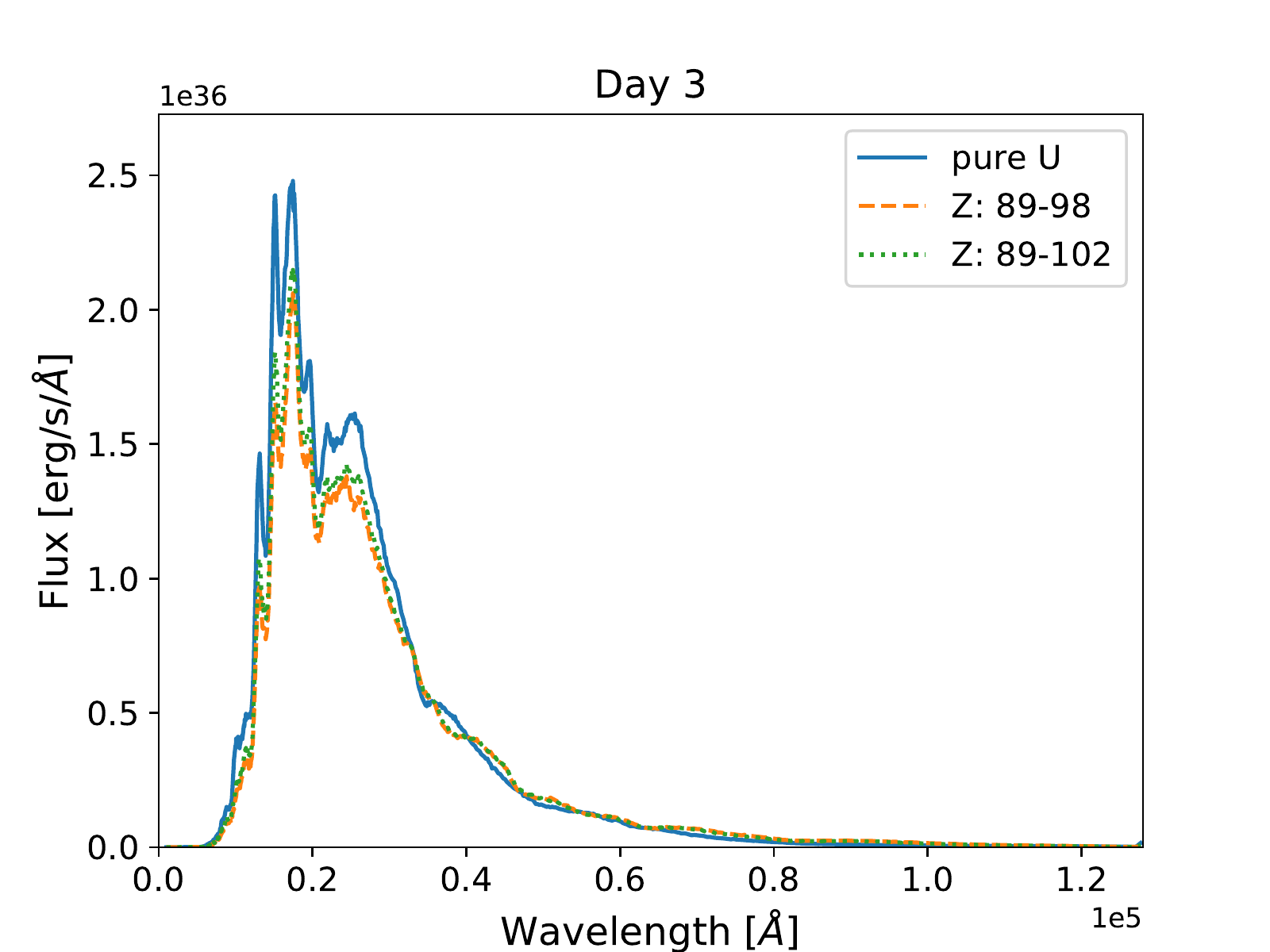}
\includegraphics[clip=true,angle=0,width=1.0\columnwidth]
{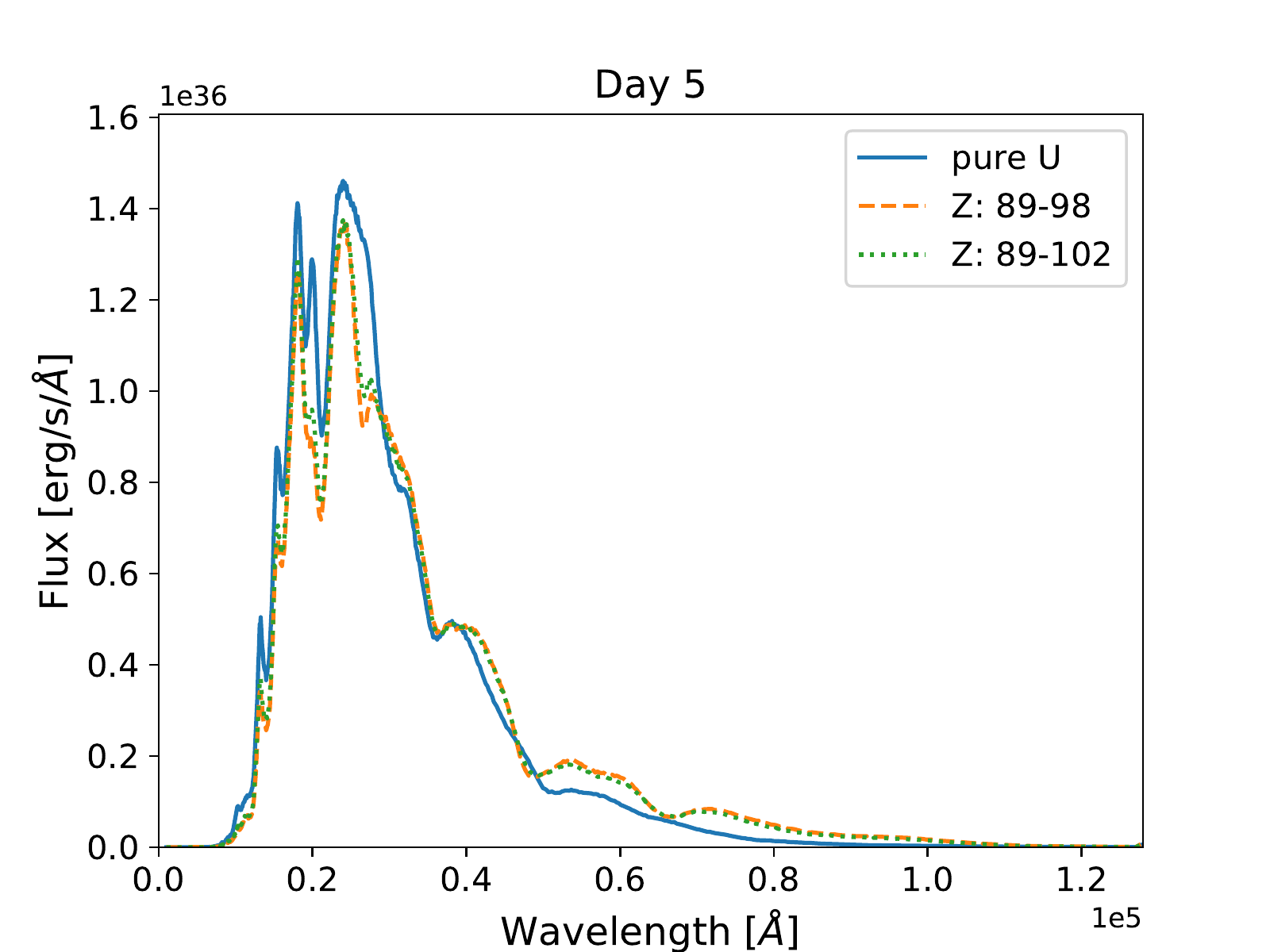}
\hfill
\includegraphics[clip=true,angle=0,width=1.0\columnwidth]
{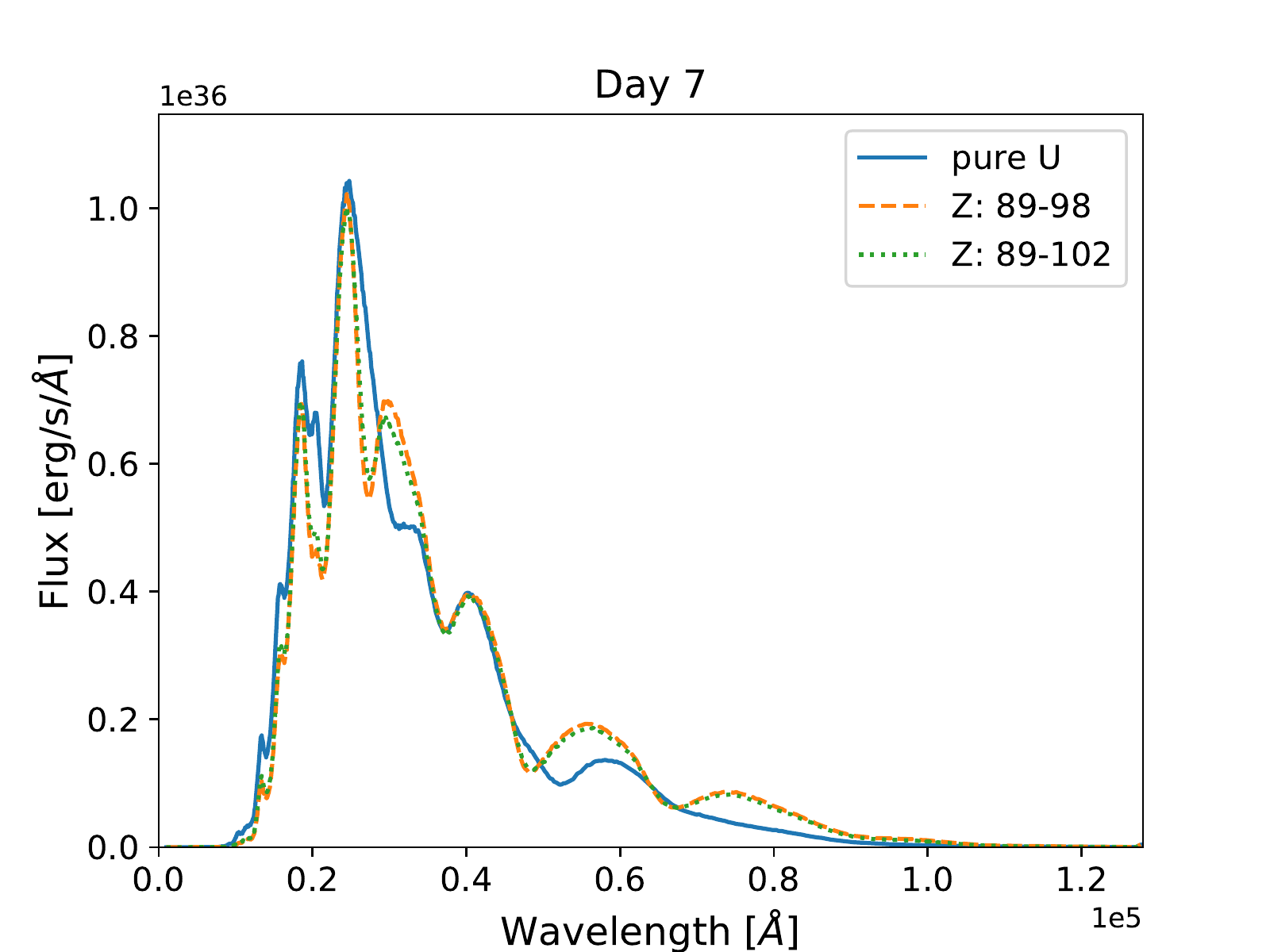}
\caption{
Spectra at 1, 3, 5 and 7~days for the first dynamical ejecta model,
with a mass of $1.4\times10^{-2}$ M$_{\odot}$ and mean ejecta speed of 0.125$c$.
Results are displayed for three different actinide abundance distributions:
pure U (solid blue curve),
a uniform distribution including only a
limited range of actinides from $Z = 89$--98 (dashed orange curve),
and a uniform distribution including the entire range of actinides
from $Z = 89$--102 (green dotted curve).
}
\label{fig:spec1}
\end{figure*}

In Fig.~\ref{fig:spec1}, we present the corresponding spectra at
1, 3 , 5 and 7~days.
Overall, we observe moderately higher (20--30\%) emission in the optical
peaks for the pure-U curve,
consistent with the luminosity comparison given in the previous
paragraph. This discrepancy provides
some indication of the level of inaccuracy that can occur in spectral
modeling when approximating the opacity of the entire set of actinides with
that of only uranium.
At early times (1 and 3~days), all three curves display
similar spectral features, but at later times, there is some differentiation
between the pure-U curve and the other two curves that are obtained
from more realistic actinide distributions.
In particular, the latter two curves display a few 
spectral features above 40,000~\AA\ that do not appear in the pure-U curve.
On a practical note, these features occur in the infrared range and 
therefore are difficult to observe with ground-based telescopes due
to absorption by various molecules in the Earth's atmosphere.

\begin{figure*}
\includegraphics[clip=true,angle=0,width=1.0\columnwidth]
{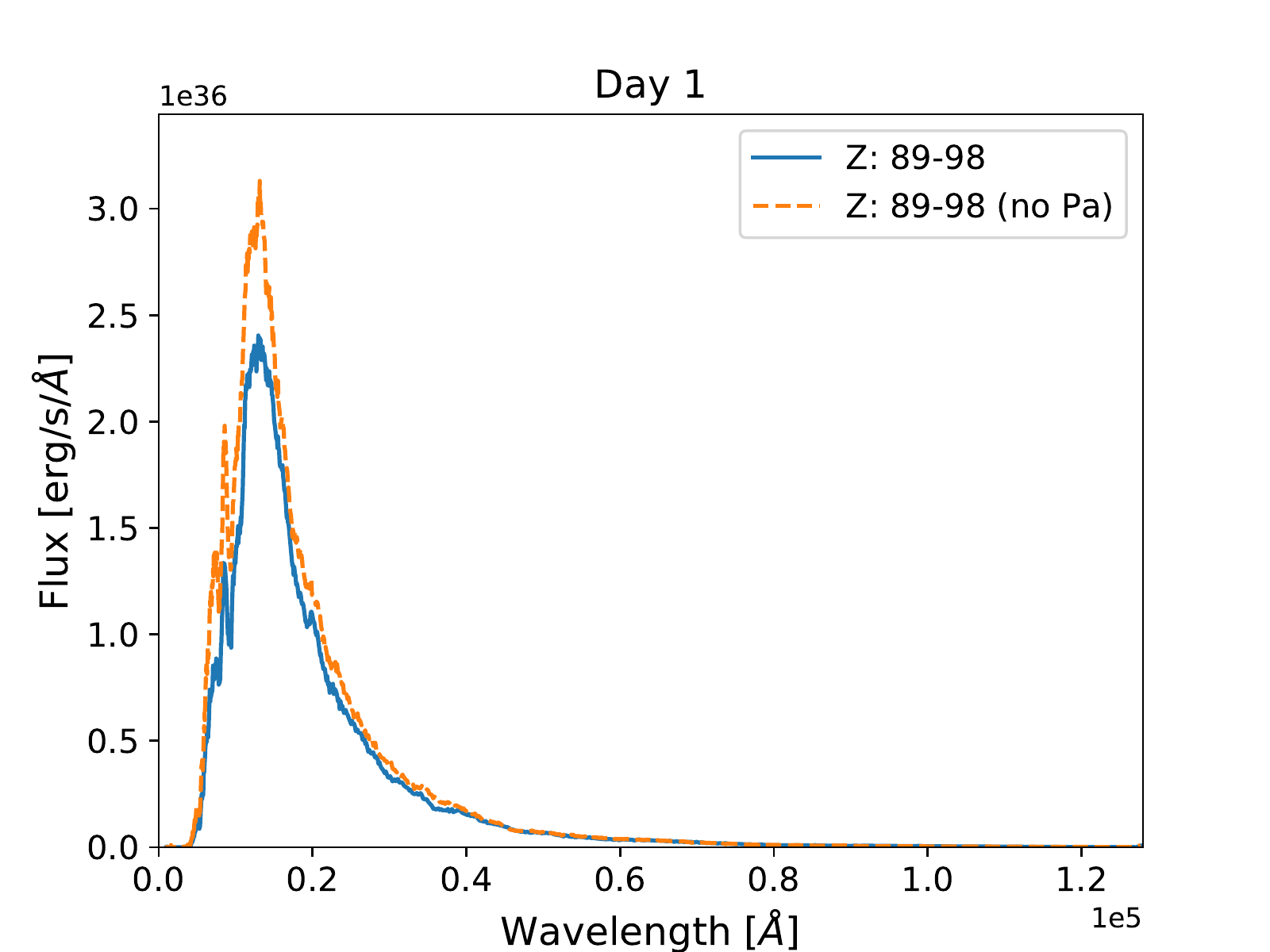}
\includegraphics[clip=true,angle=0,width=1.0\columnwidth]
{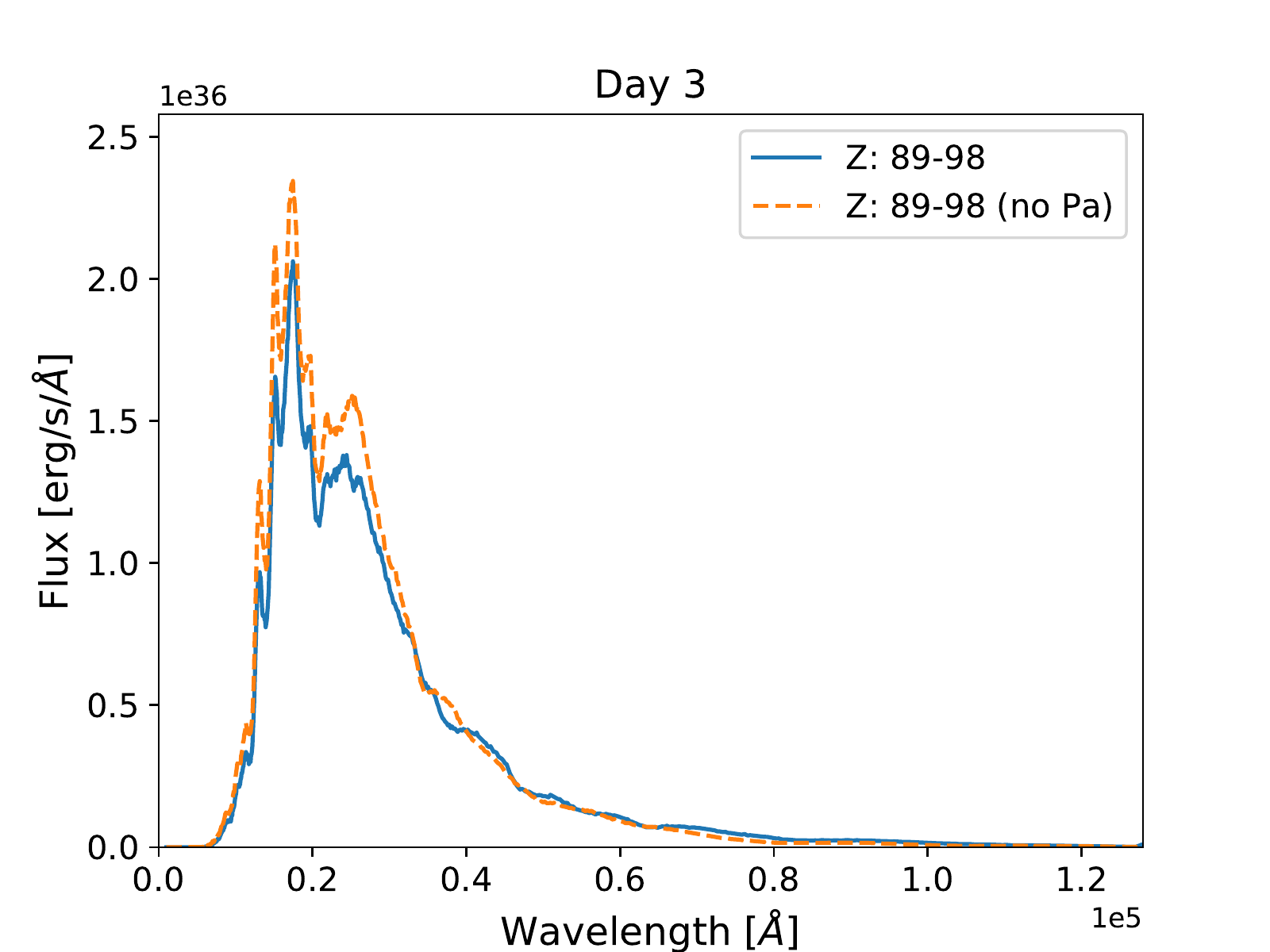}
\includegraphics[clip=true,angle=0,width=1.0\columnwidth]
{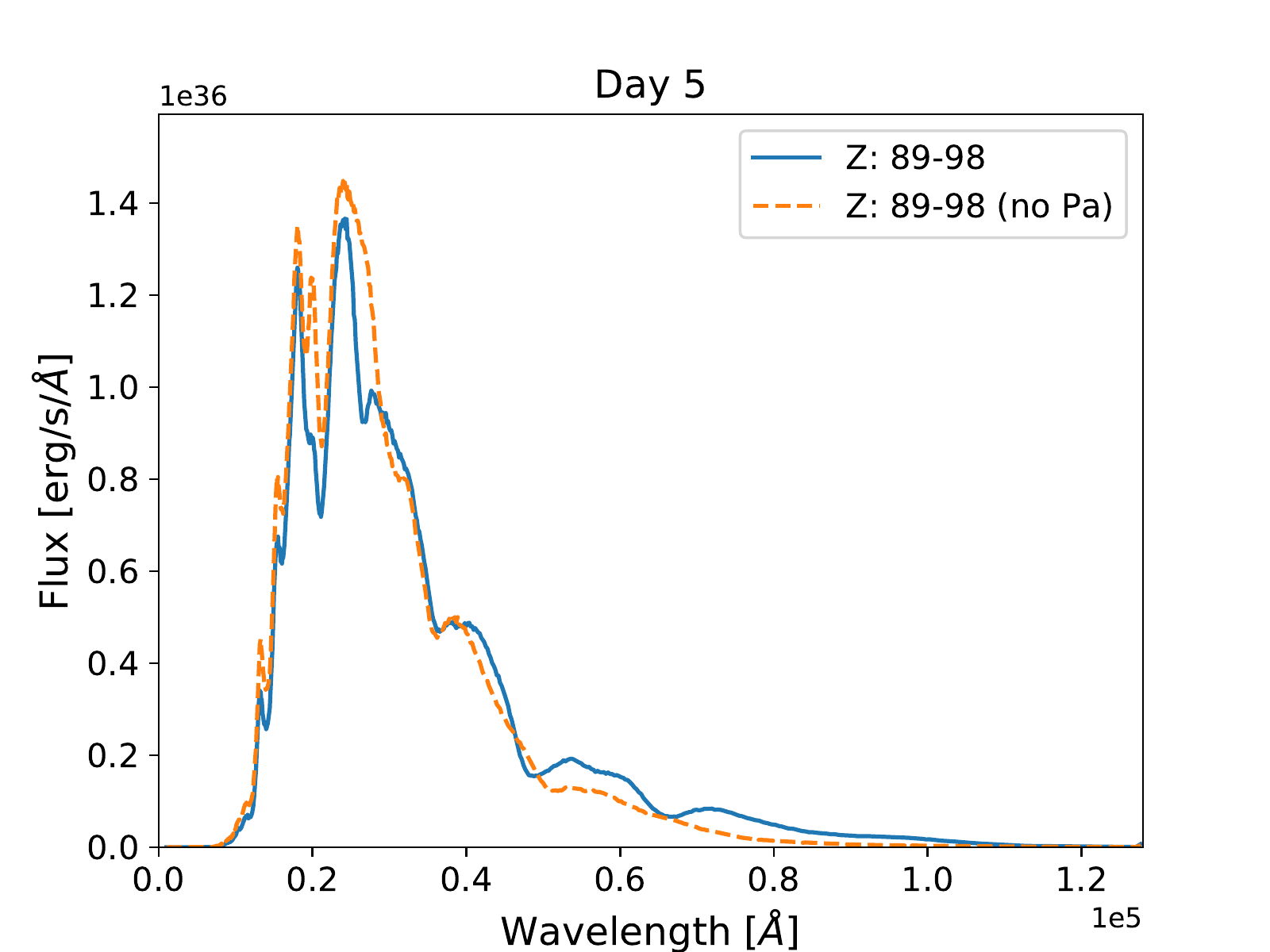}
\includegraphics[clip=true,angle=0,width=1.0\columnwidth]
{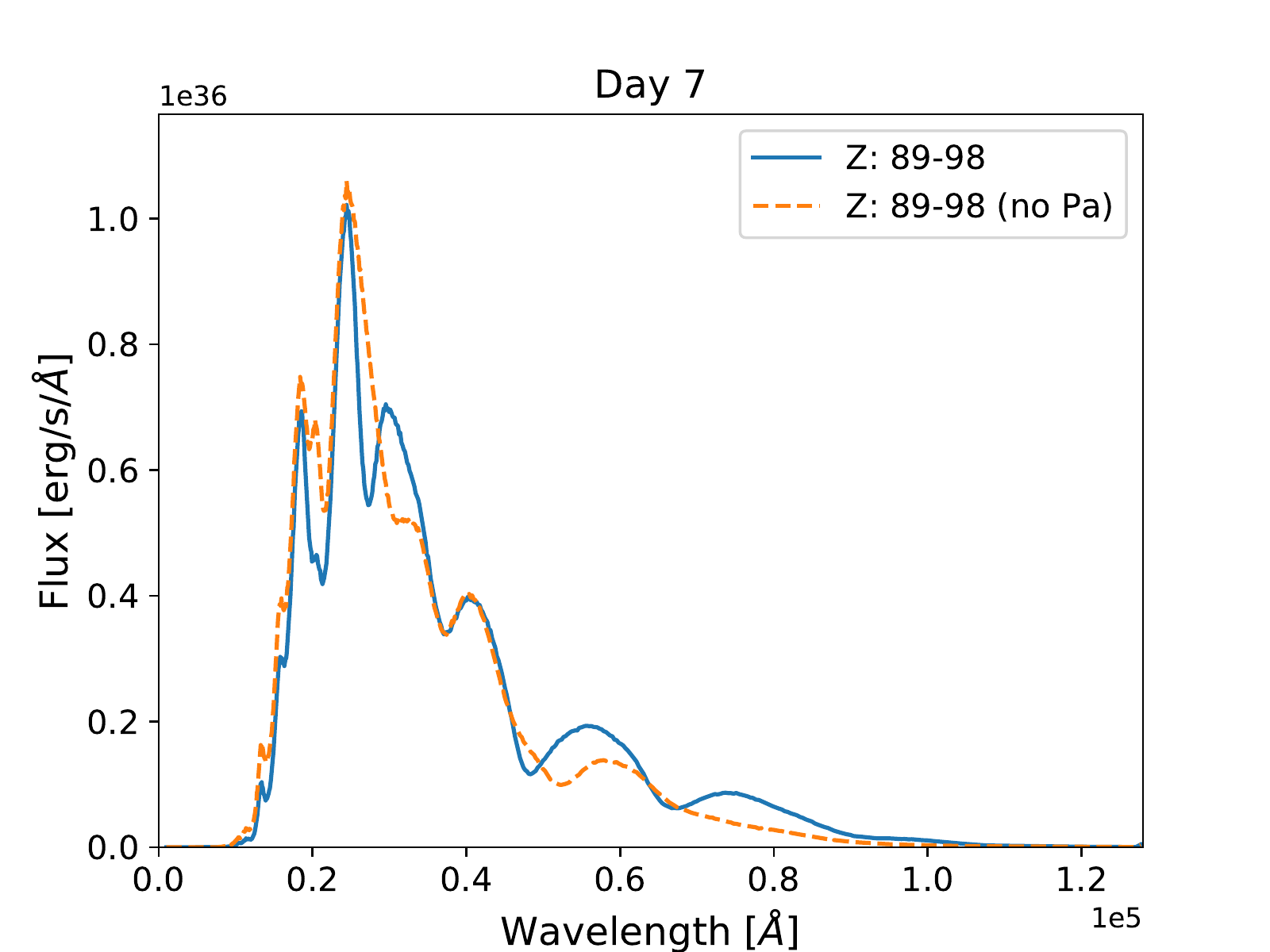}
\caption{
Spectra at 1, 3, 5 and 7~days for the first dynamical ejecta model,
with a mass of $1.4\times10^{-2}$ M$_{\odot}$ and mean ejecta speed of 0.125$c$.
Results are displayed for two different actinide abundance distributions:
a limited range of actinides from $Z = 89$--98 (solid blue curve)
and the same limited range of actinides, but with Pa $(Z = 91)$
removed from the calculation (dashed orange curve).
}
\label{fig:spec_no_Pa}
\end{figure*}

In order to better understand the source of these far-IR features,
we recalculated the spectrum presented in Fig.~\ref{fig:spec1},
using the limited-range distribution ($Z = 89$--98),
but with each of the ten actinide elements removed from the calculation,
one at a time. We found that a single actinide,
Pa $(Z=91)$, is largely responsible for these long-wavelength features,
as displayed in Fig.~\ref{fig:spec_no_Pa}.
The day-5 and day-7 panels in this figure clearly indicate the importance
of Pa for these weak features above 40,000~\AA. Furthermore, all
four panels indicate that Pa has a moderate, but noticeable, effect
on the strong infrared peaks below 40,000~\AA. We defer a more detailed study
of this effect to future work.

\subsection{Emission study using a grid of ejecta masses and speeds}
\label{sub:study_grid}

The appearance of those features above 40,000~\AA\ provide motivation
for a more in-depth search for spectral features
that might be observable and that could be used
to distinguish between the partial- and
full-actinide distributions. So we next consider a grid of three
ejecta masses (0.003, 0.01, 0.03 M$_{\odot}$) and three ejecta
speeds (0.05$c$, 0.15$c$, 0.3$c$), which,
when combined with the three actinide distributions,
yields a total of 27 models to investigate. This 3x3 grid represents a subset
of the simulation cube considered in our earlier work \citep{wollaeger21},
with the minimum and maximum masses omitted in order to focus
on the most likely KN events.

\begin{figure*}
\includegraphics[clip=true,angle=0,width=0.666\columnwidth]
{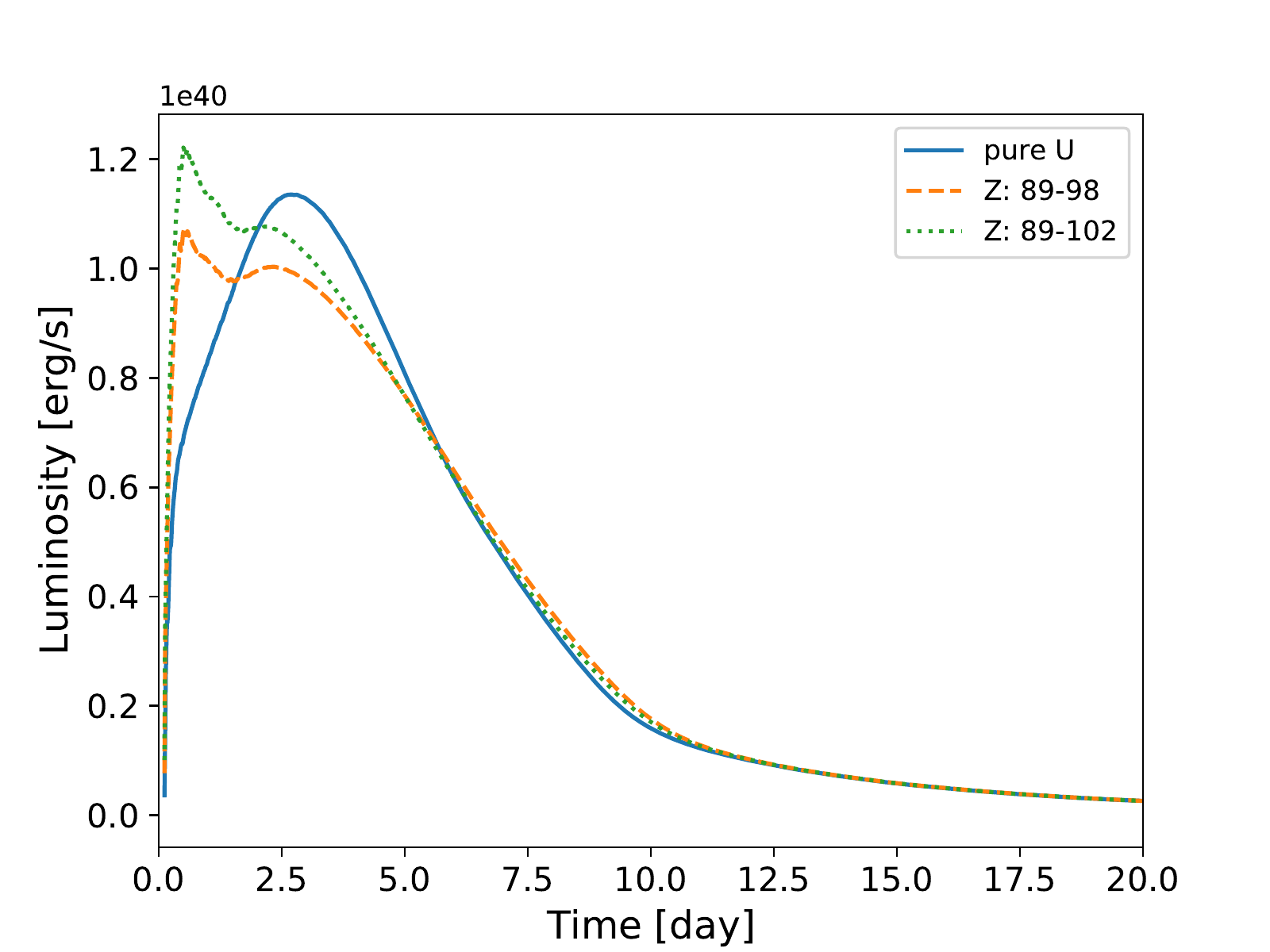}
\includegraphics[clip=true,angle=0,width=0.666\columnwidth]
{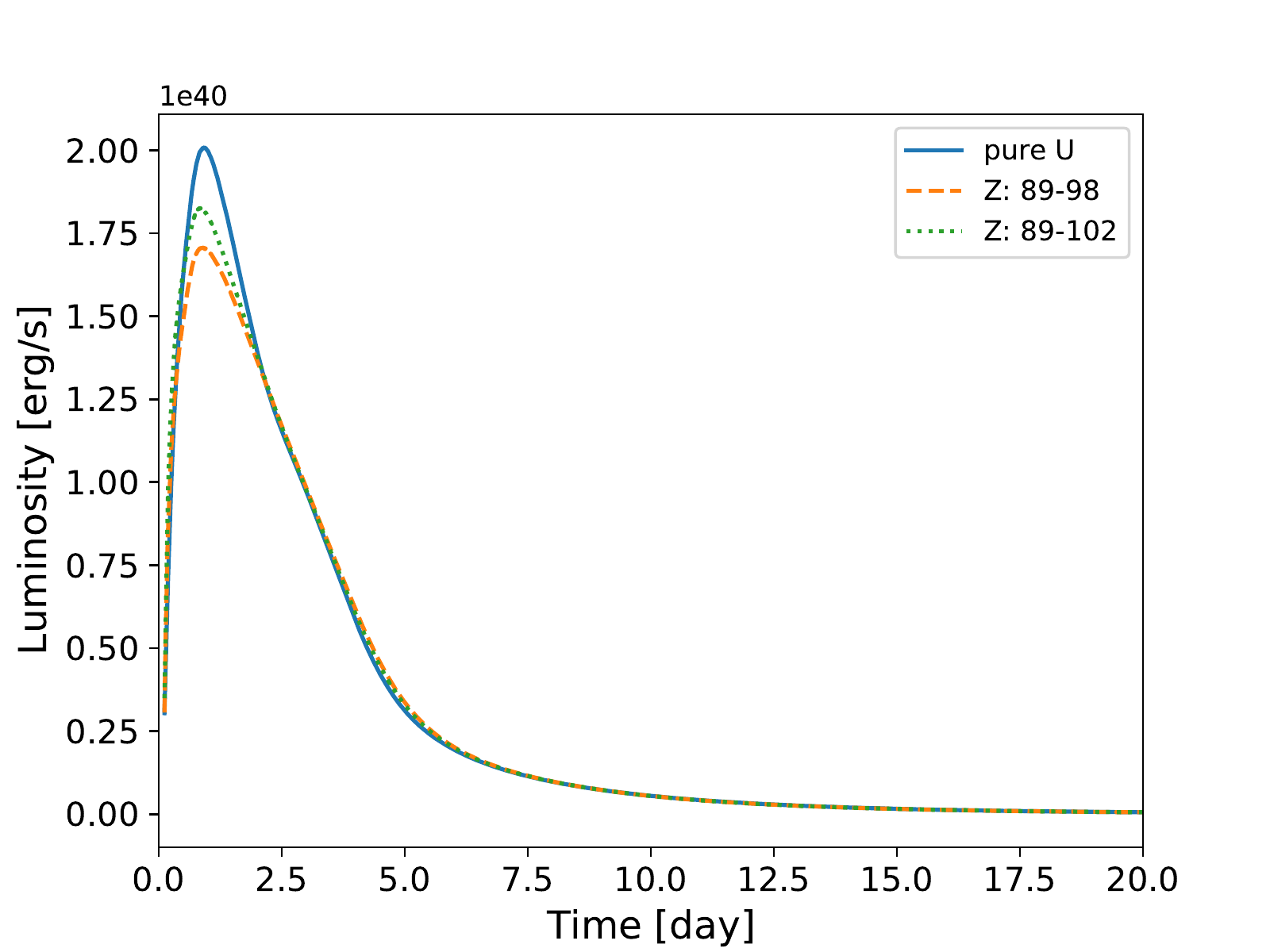}
\includegraphics[clip=true,angle=0,width=0.666\columnwidth]
{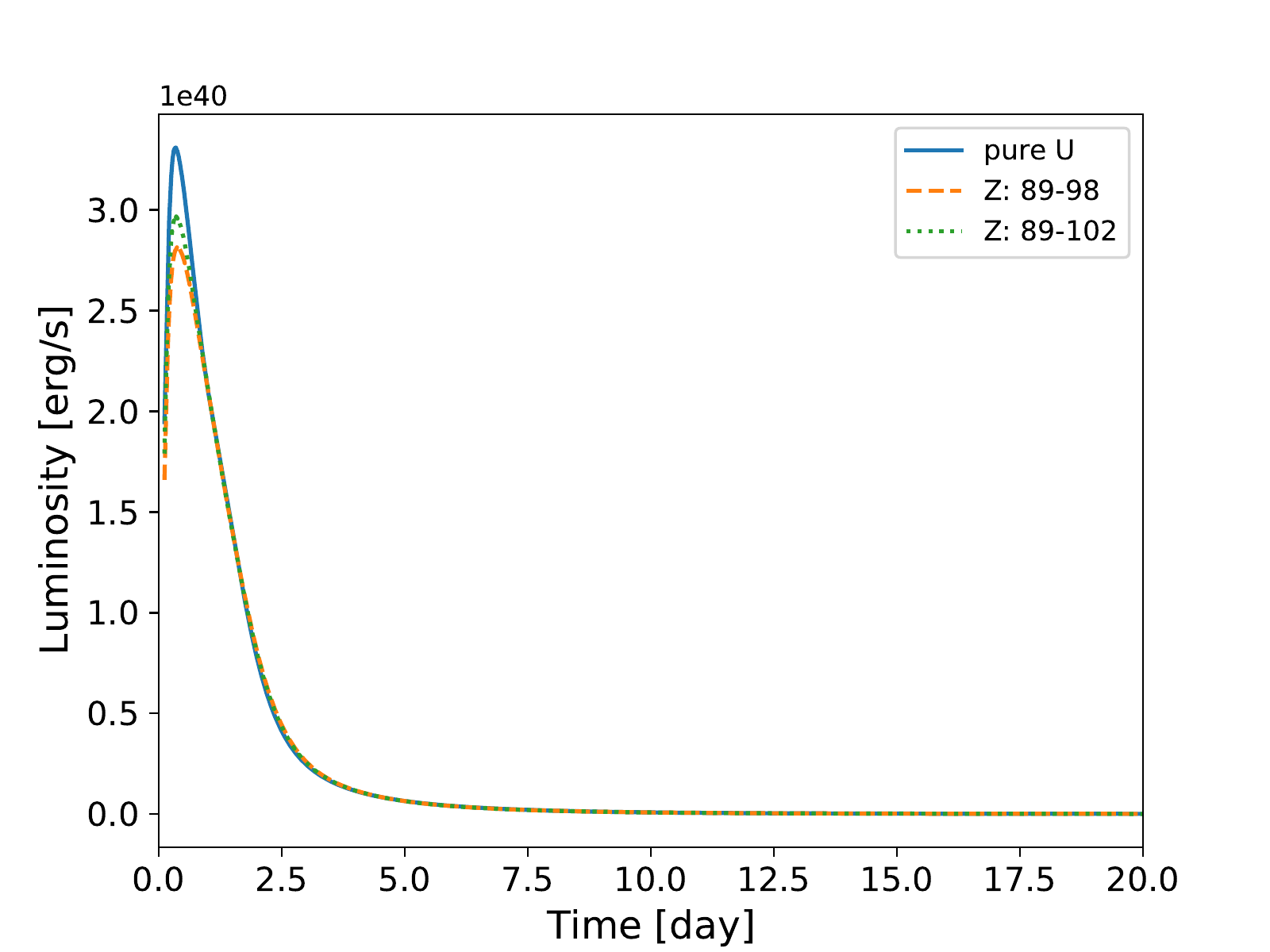}
\includegraphics[clip=true,angle=0,width=0.666\columnwidth]
{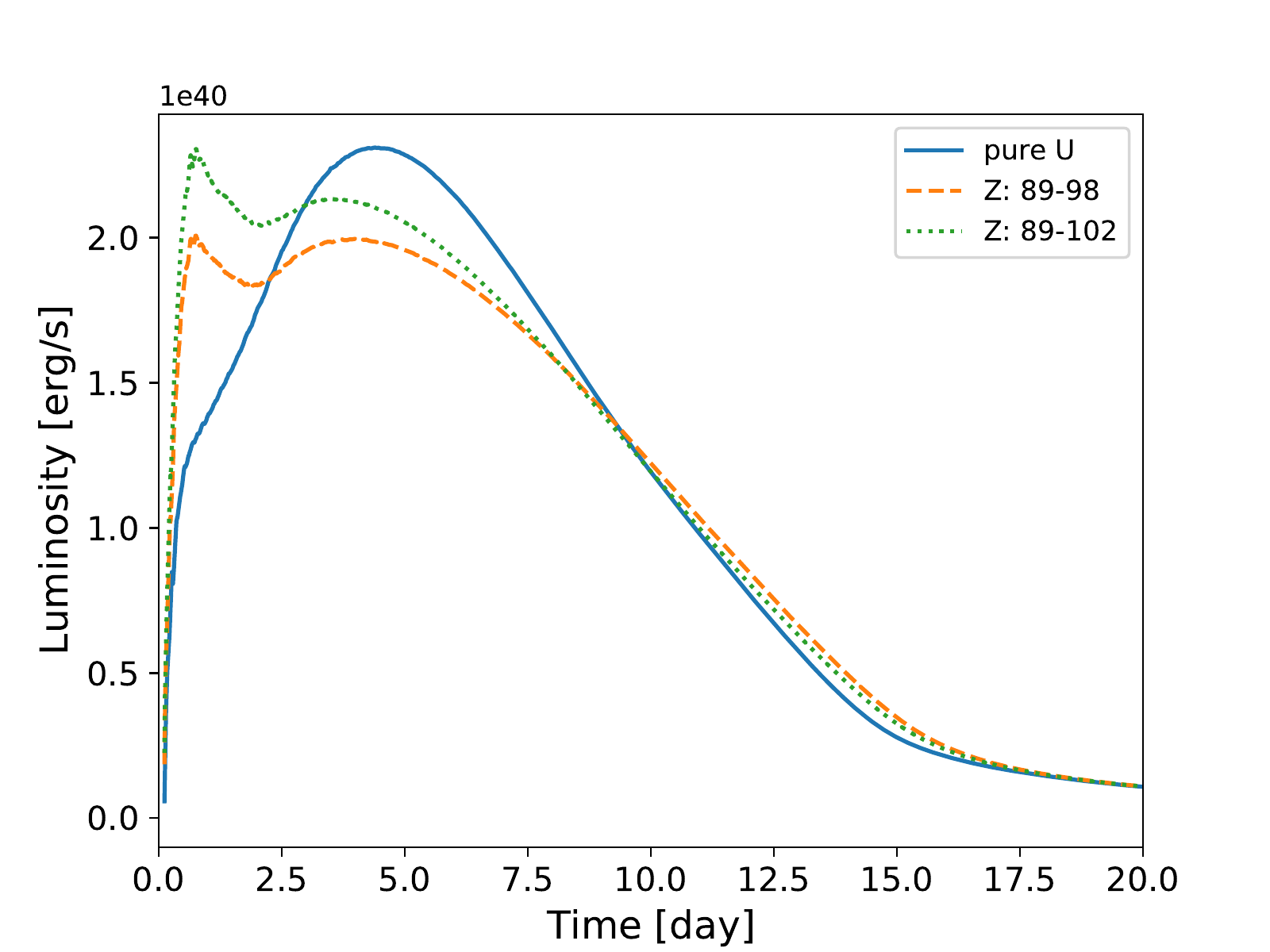}
\includegraphics[clip=true,angle=0,width=0.666\columnwidth]
{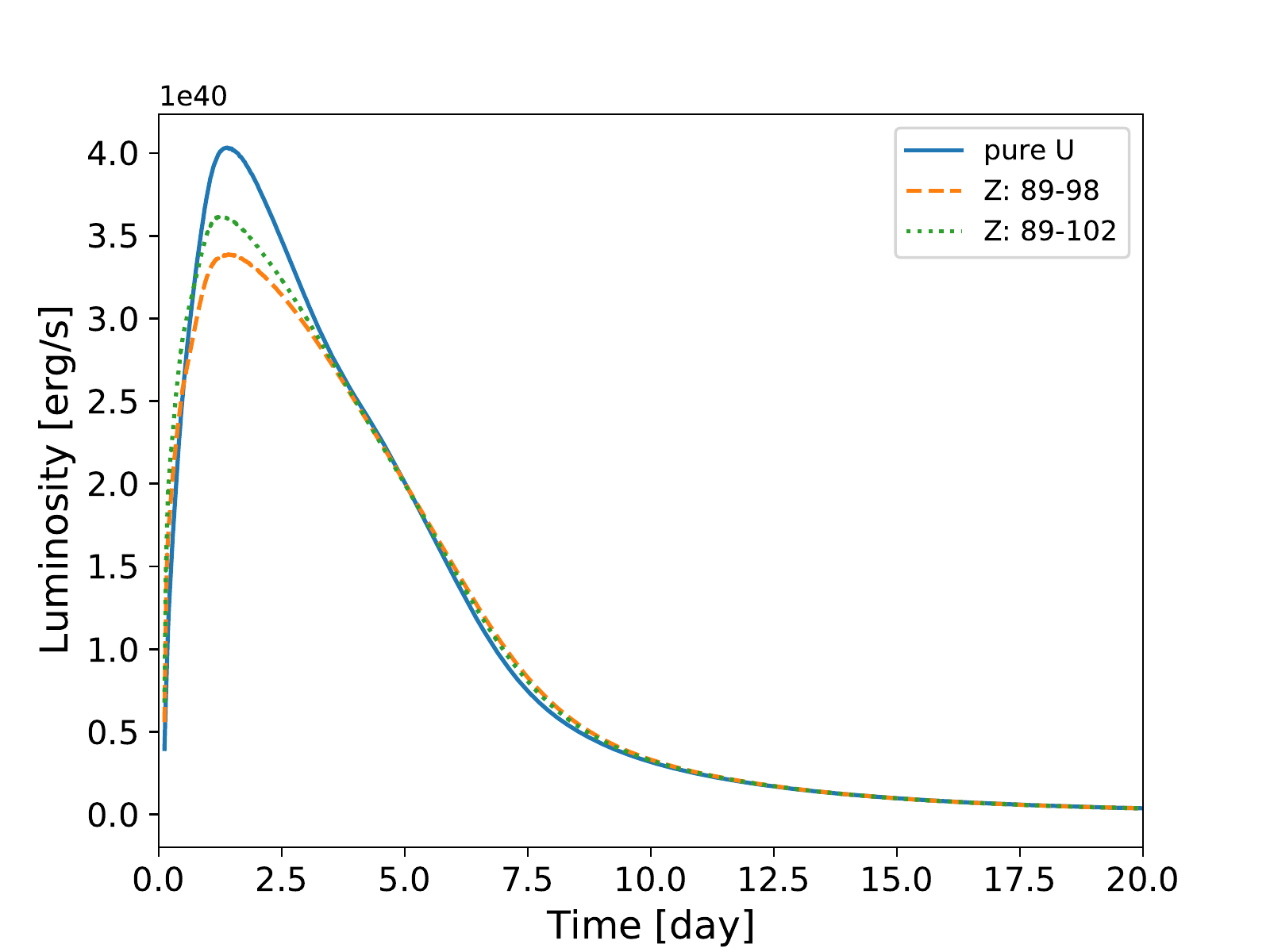}
\includegraphics[clip=true,angle=0,width=0.666\columnwidth]
{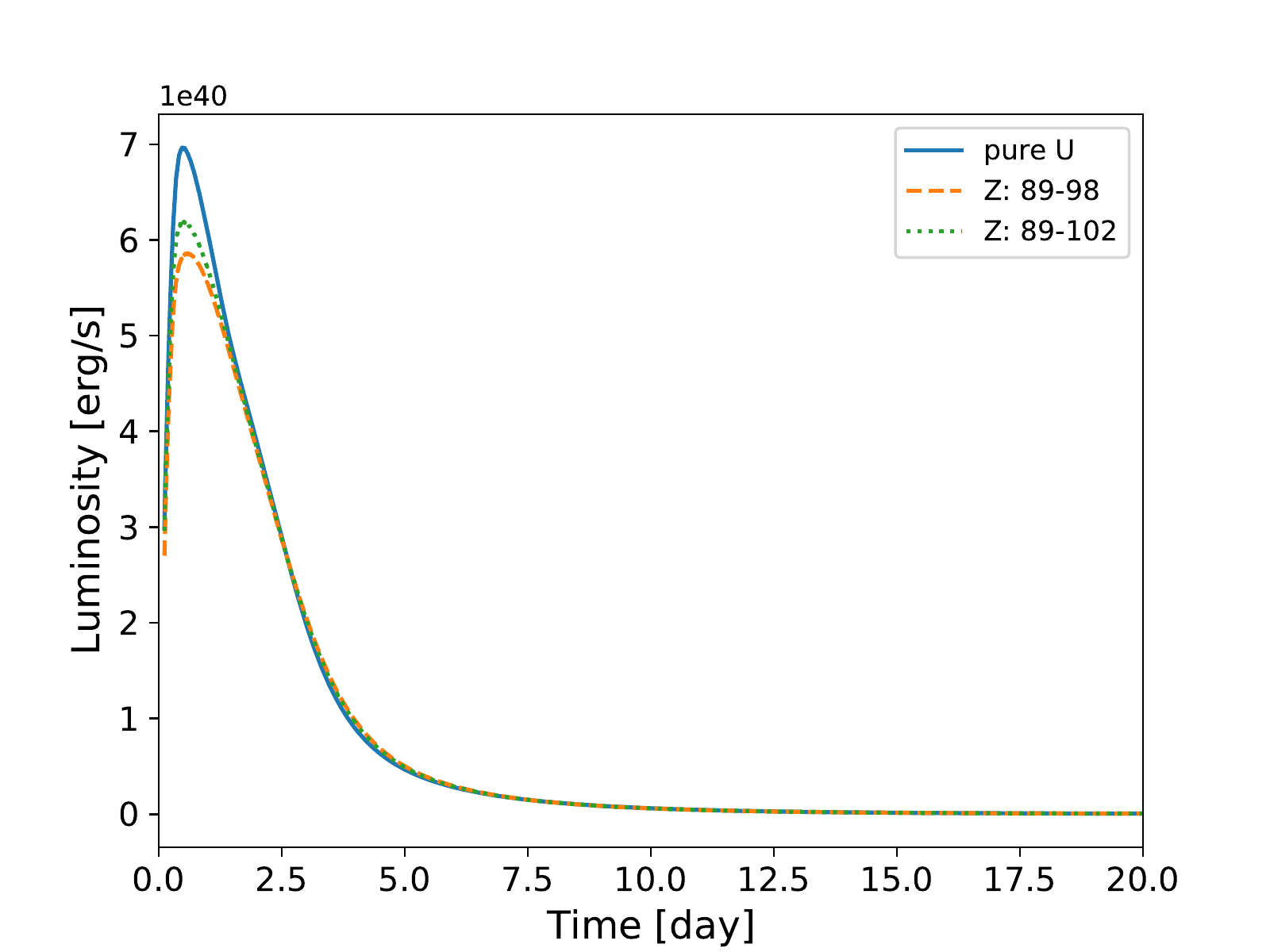}
\includegraphics[clip=true,angle=0,width=0.666\columnwidth]
{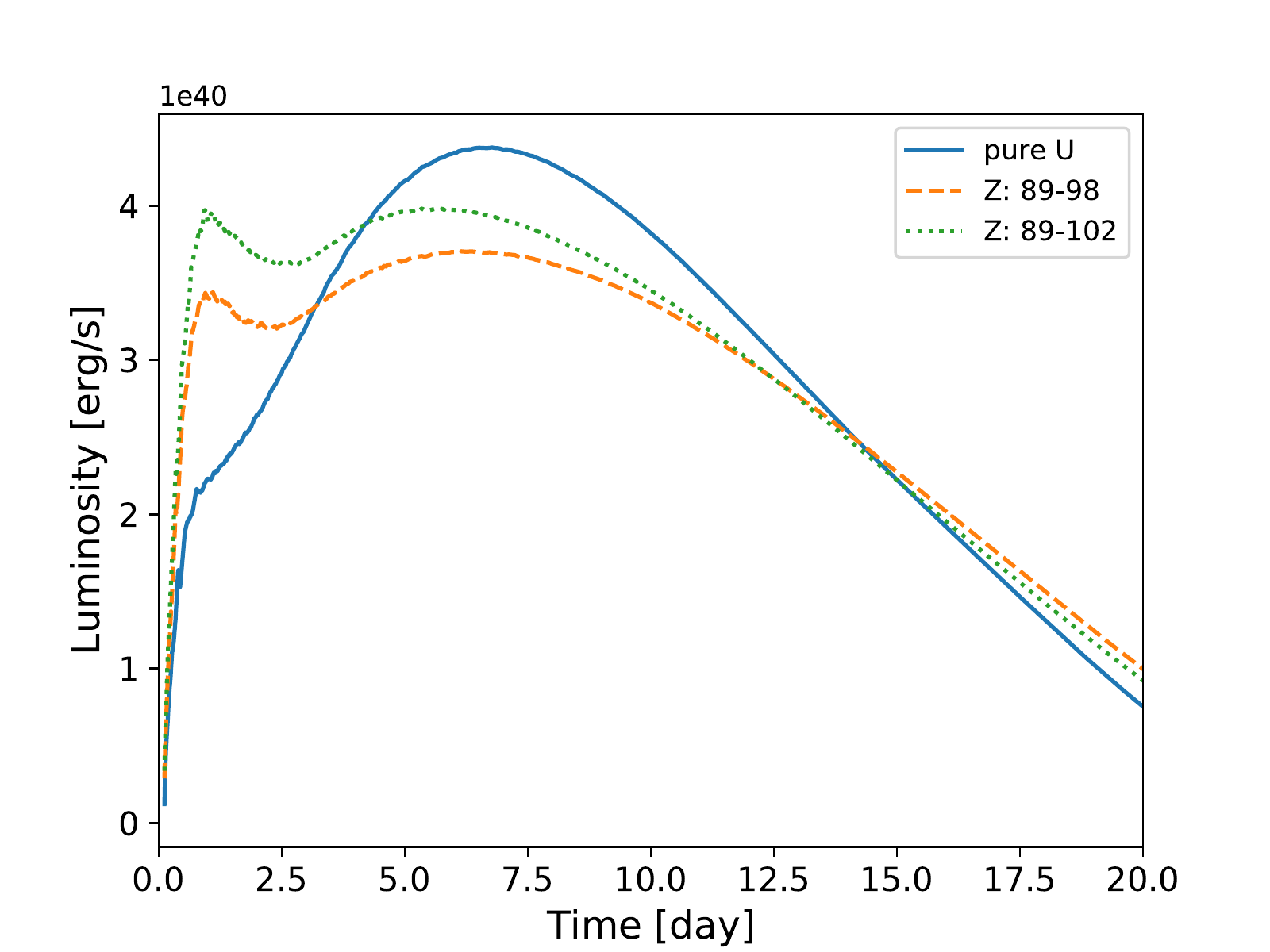}
\includegraphics[clip=true,angle=0,width=0.666\columnwidth]
{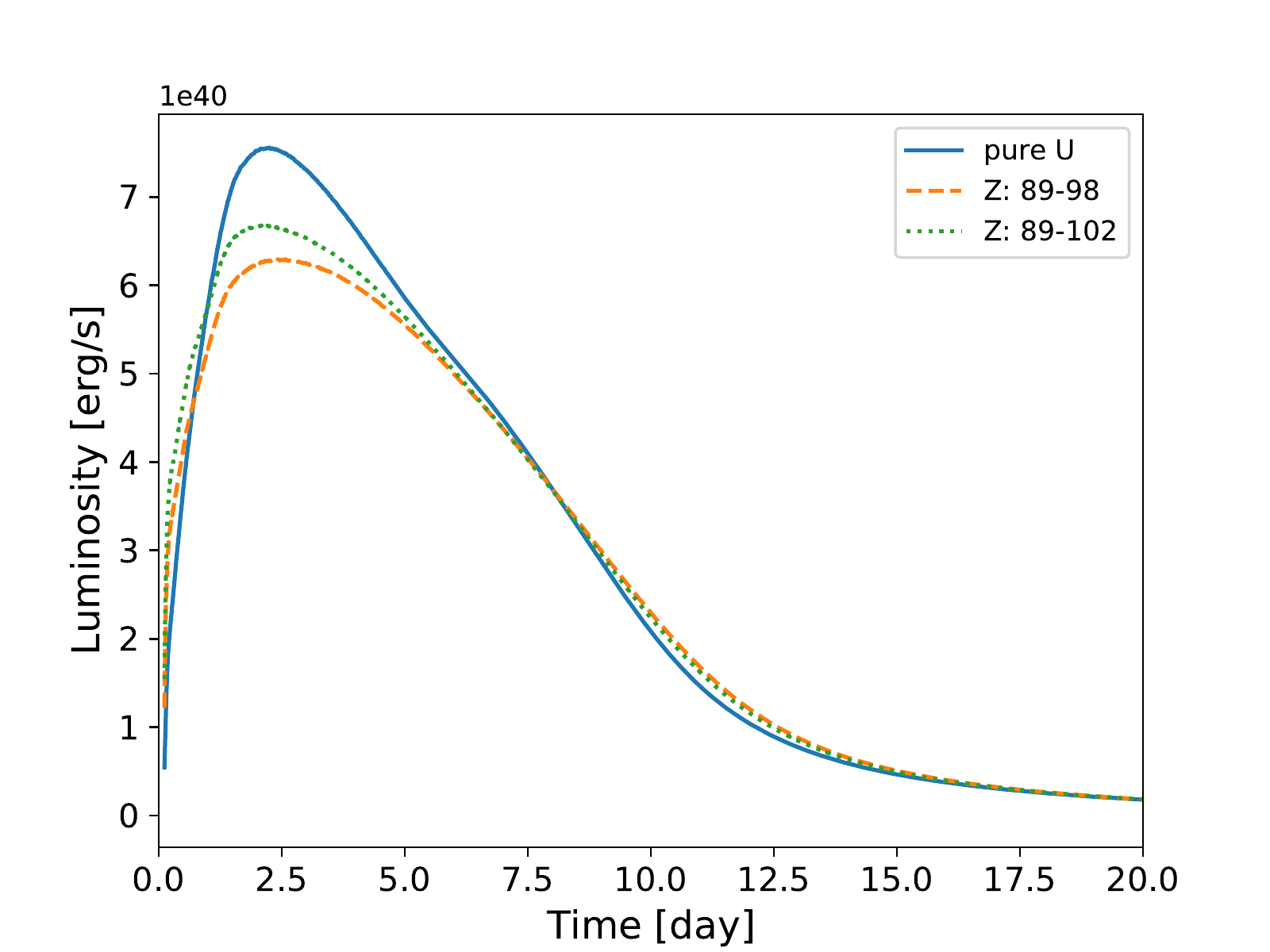}
\includegraphics[clip=true,angle=0,width=0.666\columnwidth]
{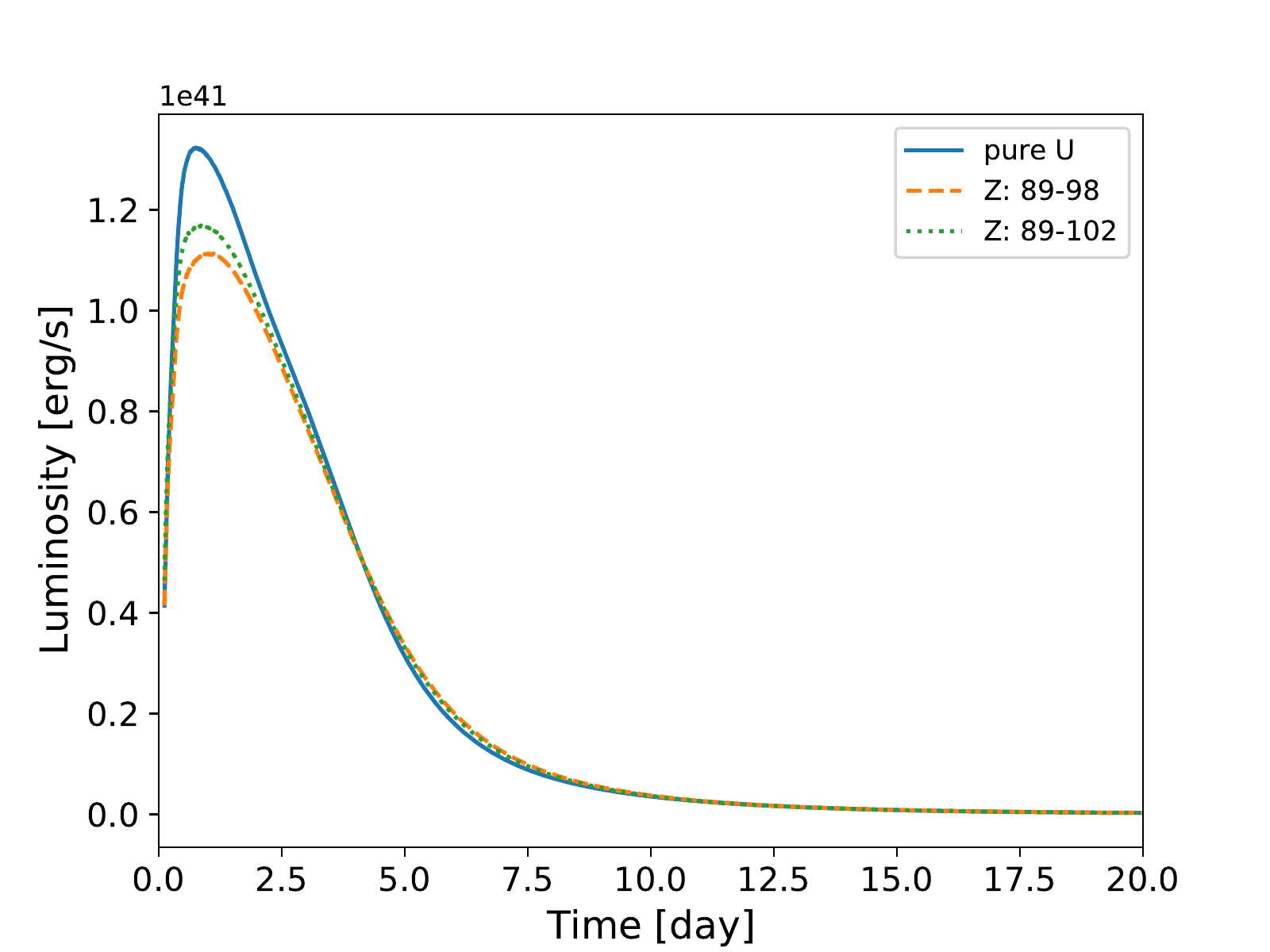}
\caption{
Bolometric luminosity for the 3x3 grid of masses and velocities
considered in this work. The mass is constant in each row,
while the velocity is constant in each column. Results for the lowest
mass-velocity pair (0.003 M$_{\odot}$, 0.05$c$) are displayed
in the upper left-hand corner
and for the highest mass-velocity pair (0.03 M$_{\odot}$, 0.3$c$),
in the lower right-hand corner.
Results are displayed for three different actinide abundance distributions:
pure U (solid blue curve),
a uniform distribution including only a
limited range of actinides from $Z = 89$--98 (dashed orange curve),
and a uniform distribution including the entire range of actinides
from $Z = 89$--102 (green dotted curve).
}
\label{fig:lums_3_3}
%\end{sidewaysfigure}
\end{figure*}

In Fig.~\ref{fig:lums_3_3} we present the light curves for this
3x3 grid, with the ejecta mass held constant in each row and the density
held constant in each column.
As expected, the lowest velocity simulations (first column) produce
broad light curves and the highest velocity simulations (third column)
produce much narrower light curves.
For any given pair of mass-velocity values,
it is straightforward to see that the partial- and
full-actinide distributions track each other very closely, with the
latter producing slightly higher peak luminosity, while the pure-U
curve produces the highest peak values and displays a qualitatively
different behavior for the three cases with the lowest velocity.

\begin{figure*}
\includegraphics[clip=true,angle=0,width=1.0\columnwidth]
{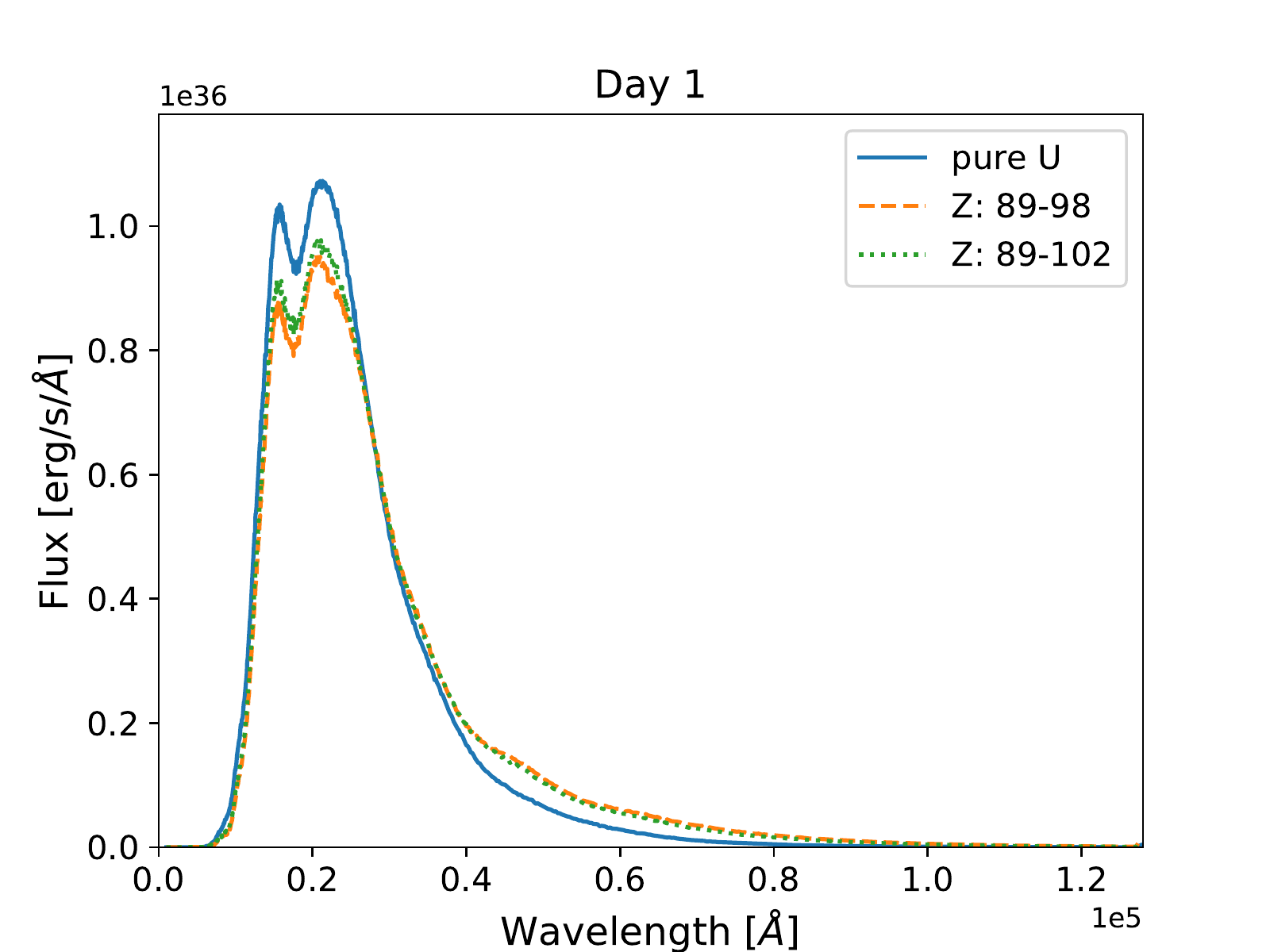}
\includegraphics[clip=true,angle=0,width=1.0\columnwidth]
{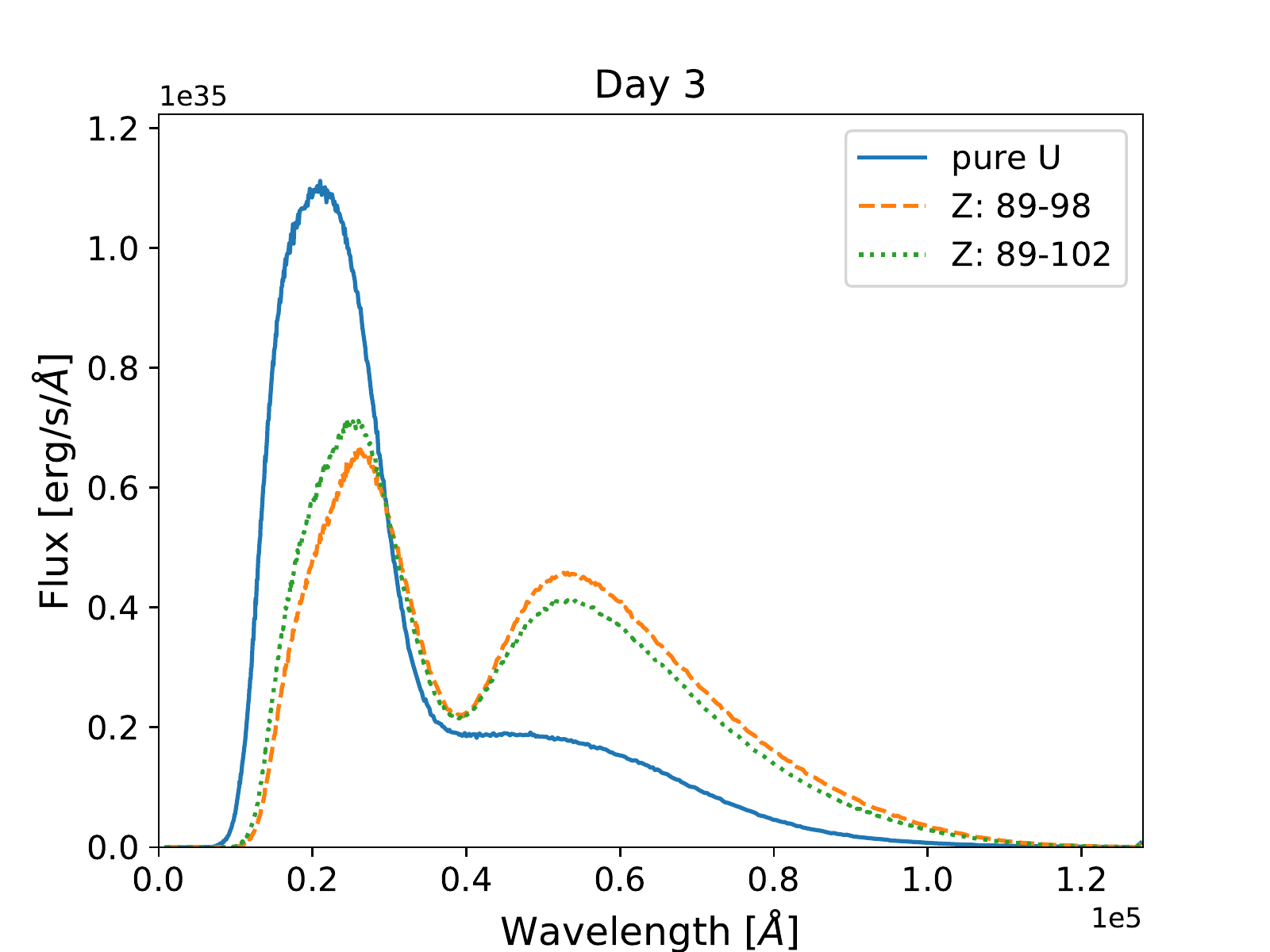}
\includegraphics[clip=true,angle=0,width=1.0\columnwidth]
{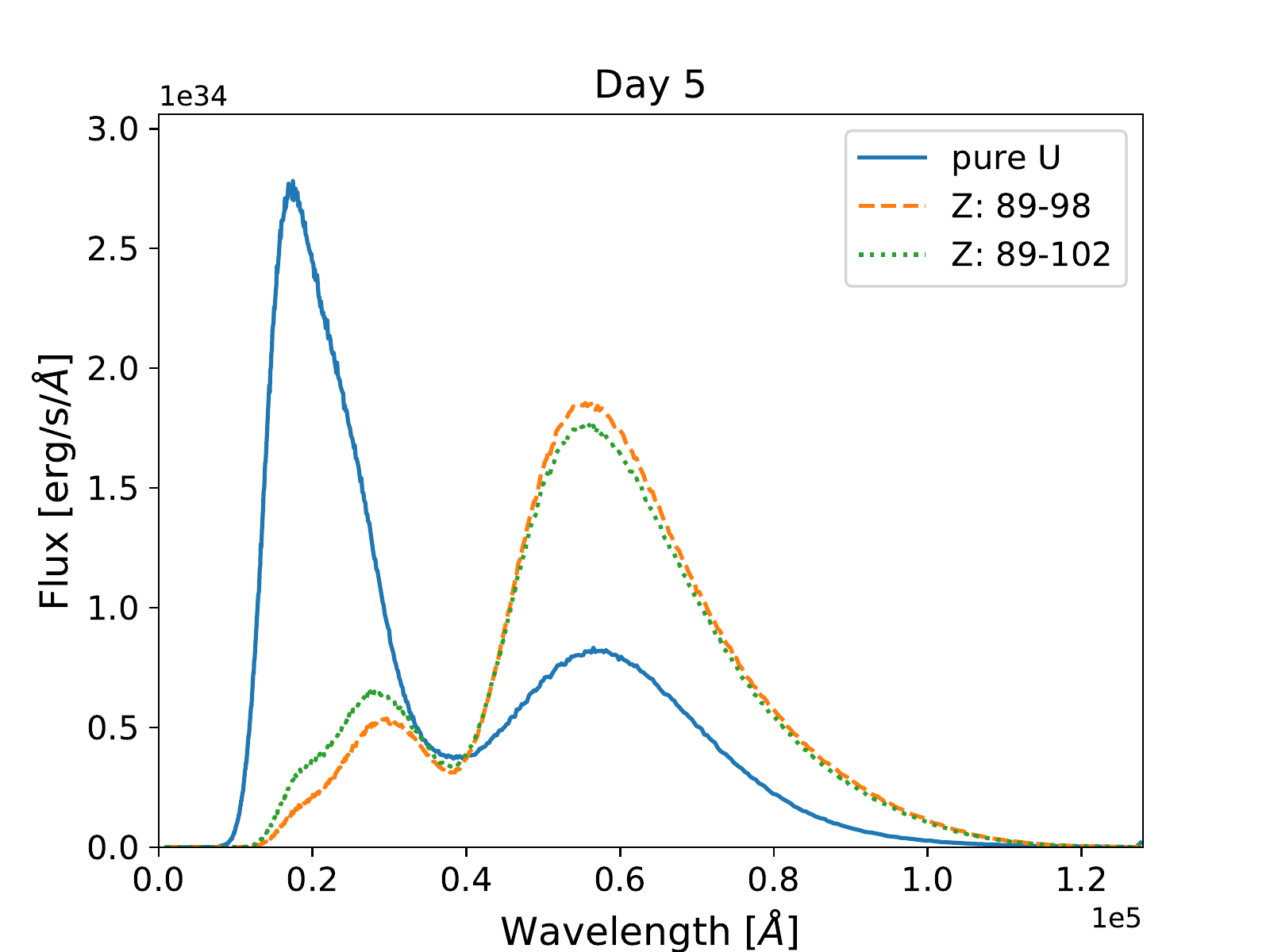}
\includegraphics[clip=true,angle=0,width=1.0\columnwidth]
{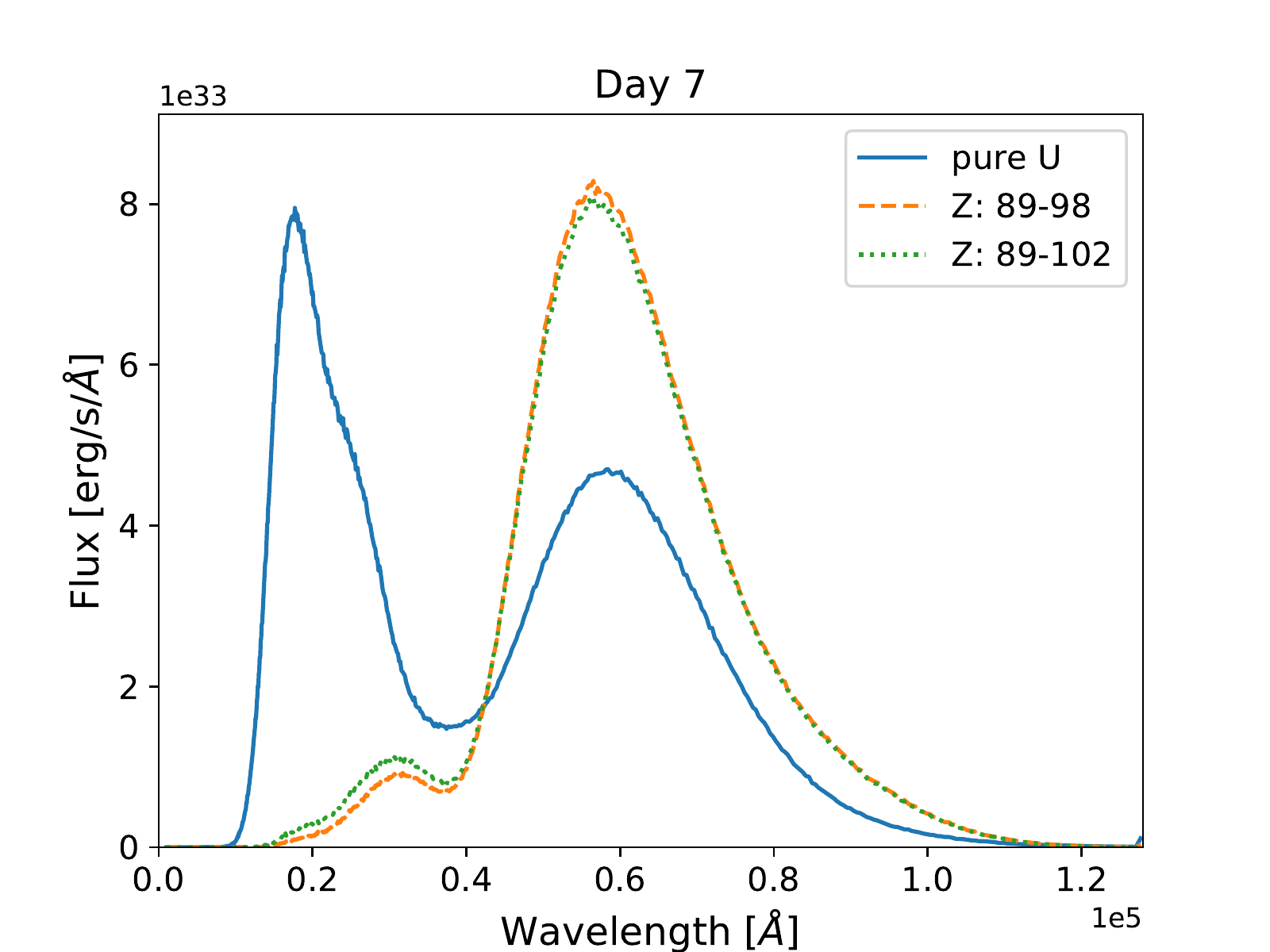}
\caption{
Spectra at 1, 3, 5 and 7 days for an ejecta mass of 0.003 M$_{\odot}$ and 
mean ejecta speed of 0.3$c$.
Results are displayed for three different actinide abundance distributions:
pure U (solid blue curve),
a uniform distribution including only a
limited range of actinides from $Z = 89$--98 (dashed orange curve),
and a uniform distribution including the entire range of actinides
from $Z = 89$--102 (green dotted curve).
}
\label{fig:spec_m003v30}
\end{figure*}

The corresponding spectra for the 3x3 grid can be similarly organized,
i.e. keeping the velocity fixed produces similar spectra. As a specific example,
we present in Fig.~\ref{fig:spec_m003v30} the spectra at 1, 3, 5 and 7~days
for the case of 0.003 M$_{\odot}$ and 0.30$c$.
From this figure, we again see that the partial- and full-actinide
results are very similar, while the pure-U curve displays different
behavior over most of the simulation time.

For the sake of brevity, we do not display the additional four-panel
spectral figures for  the remaining eight mass-velocity pairs.
Instead, in order to analyze these results in a more convenient
and efficient manner, we consider the L1 spectral norm, given by
\begin{equation}
{\rm L1} = \int_{\lambda_{\rm min}}^{\lambda_{\rm max}}\,d\lambda\,
| F_x(\lambda) - F_{\rm baseline}(\lambda) |
/
\int_{\lambda_{\rm min}}^{\lambda_{\rm max}}\,d\lambda\,
F_{\rm baseline}(\lambda) \,,
\label{l1_norm}
\end{equation}
where $F_x(\lambda)$ is the flux computed with either the partial-
or complete-actinide distribution, and the baseline reference,
$F_{\rm baseline}(\lambda)$,
is chosen to be the flux computed with the pure-U distribution.
For each of the two actinide distributions ($89 \le Z \le 98$ or
$89 \le Z \le 102$), L1-norm curves are computed for the nine possible pairs
of mass-velocity values described above, as a function of time.

\begin{figure*}
%\hfill
\centering
\includegraphics[clip=true,angle=0,width=1.0\columnwidth]
{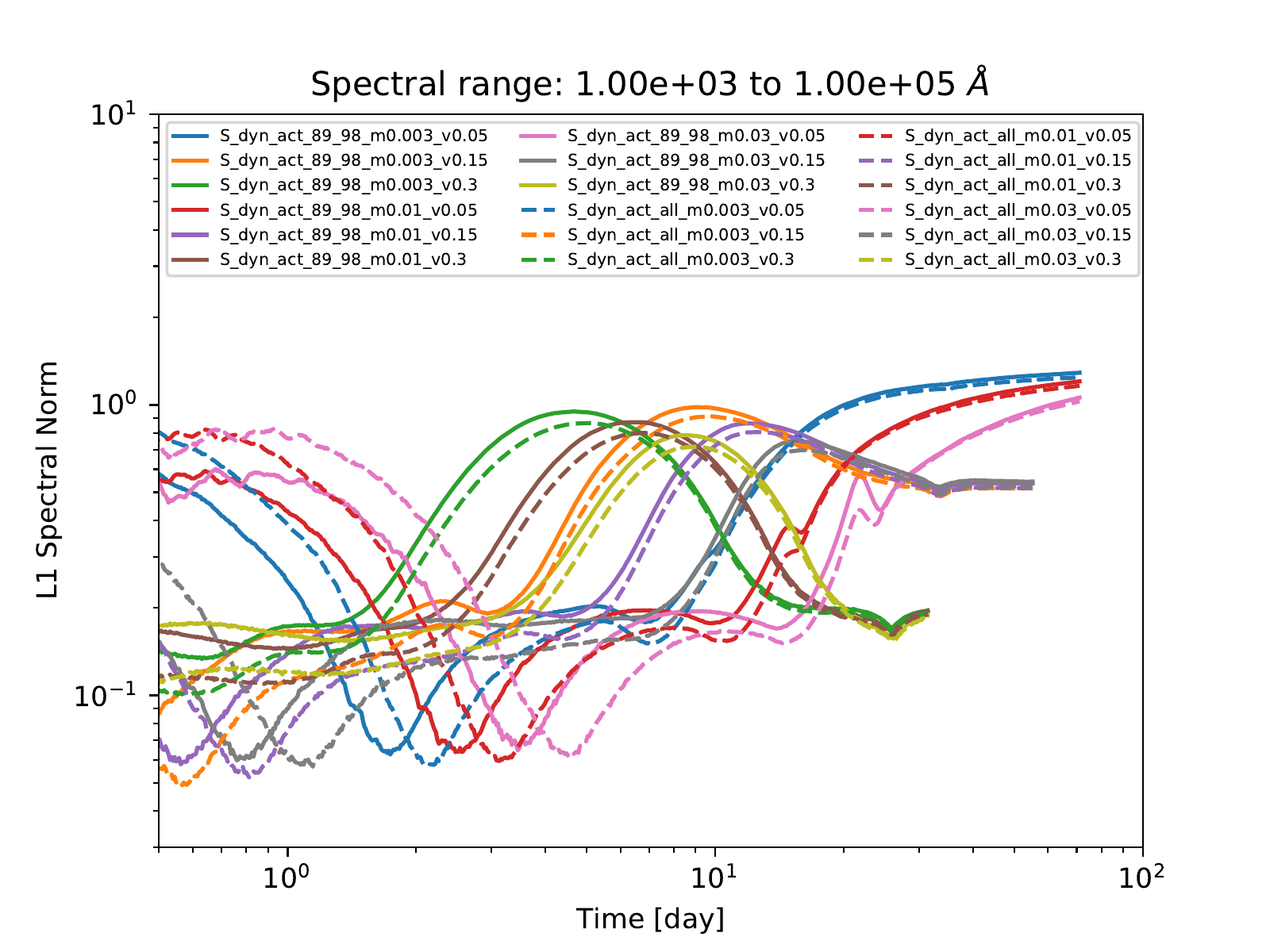}
%\hfill
\\
\includegraphics[clip=true,angle=0,width=1.0\columnwidth]
{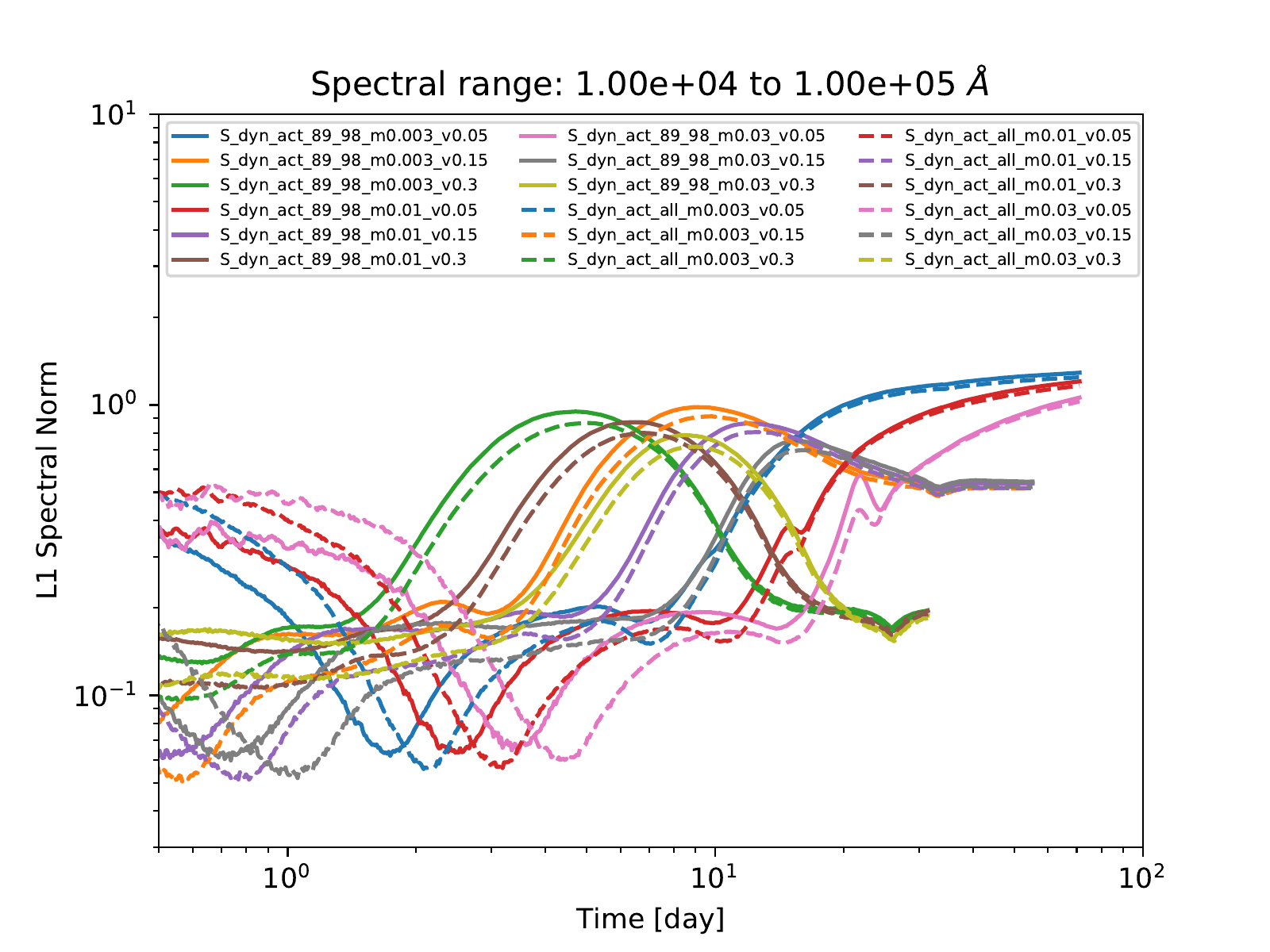}
\hfill
\includegraphics[clip=true,angle=0,width=1.0\columnwidth]
{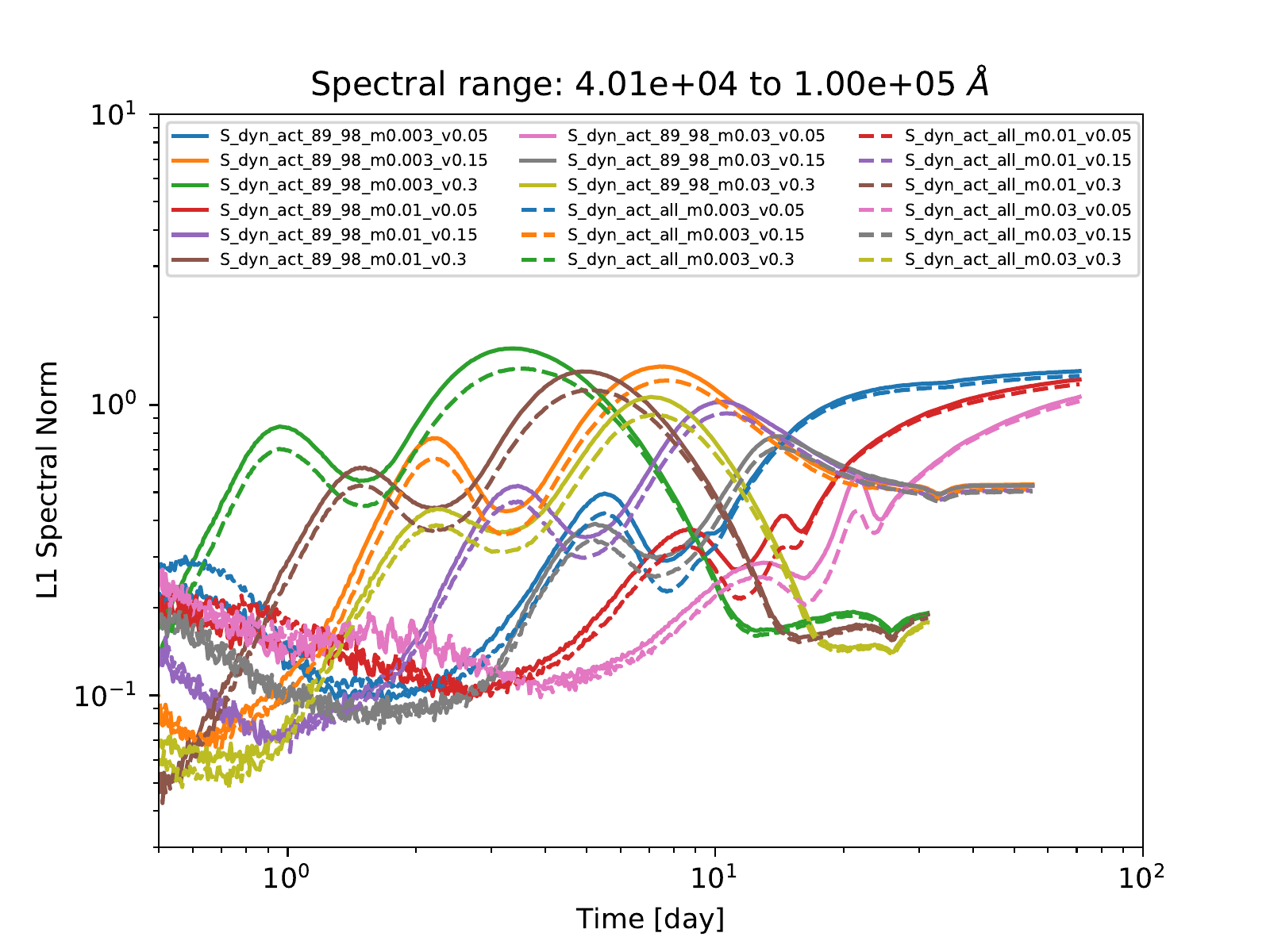}
\caption{
L1 spectral norm as a function of time, computed over three wavelength ranges:
the full wavelength range from 10$^3$--10$^5$~\AA\ (upper panel),
a more limited wavelength range from 10$^4$--10$^5$~\AA\ (bottom left-hand
panel) and
an even more limited wavelength range from 4$\times$10$^4$--10$^5$~\AA\
(bottom right-hand panel).
The L1 norm is calculated according to equation~(\ref{l1_norm}),
with the pure-U spectrum chosen to be the baseline.
In each panel, the nine solid curves represent
the L1 norm for the partial-actinide distribution and the nine dashed curves
represent the L1 norm for the complete-actinide distribution.
The nine curves correspond to different pairs of ejecta mass and speed
(see text for details).
}
\label{fig:l1_intgrl}
\end{figure*}

We present comparisons of the L1 spectral norm in Fig.~\ref{fig:l1_intgrl}.
Results are presented for the L1 norm computed over three wavelength ranges:
the full wavelength range
from 10$^3$--10$^5$~\AA\ (essentially the total spectrum),
a more limited wavelength range from 10$^4$--10$^5$~\AA\
(from near-IR to longer wavelengths) and
an even more limited wavelength range from 4$\times$10$^4$--10$^5$~\AA\
(from mid-IR to longer wavelengths).
In each panel, the solid curves represent
the L1 norm computed for the partial-actinide distribution and the dashed curves
represent the L1 norm computed for the complete-actinide distribution.
Thus, each panel contains nine solid curves and nine dashed curves,
corresponding to the nine possible pairs of mass and speed values
described above.

Before offering some analysis of these results, we first
provide a brief explanation of the rich amount of information that can be
displayed in these time-dependent curves.
There are three particular items of interest for this study:
(1) Choosing the pure-U result as the baseline flux is useful in an absolute
sense because (for a fixed mass-velocity pair) a given L1 curve displays
the inaccuracy that occurs due to choosing only uranium as a surrogate for a
more complete set of actinides. (2) On the other hand,
for relative L1-norm comparisons, the choice of baseline flux is not 
too important. More specifically, for a fixed mass-velocity pair, significant
differences that occur when comparing the two L1 curves resulting
from each actinide
distribution indicate the existence of spectral features that could
be used to distinguish between the two abundance patterns.
(3) Alternatively, for a fixed actinide distribution,
differences that occur when comparing curves produced with
different mass-velocity pairs indicate possible spectral features
of general interest due to the presence of actinide elements.

Returning to Fig.~\ref{fig:l1_intgrl}, and keeping in mind the above
three items of interest, we note the following patterns:
(1) the maximum value for each solid or dashed curve occurs between
about 1--20 days,
depending on the particular mass-velocity pair, and the maximum value
is approximately one, which indicates a difference of a factor of
approximately two in the integrated flux.
(2)
For a fixed mass-velocity pair, the two abundance curves agree reasonably well,
indicating that there is not much spectral sensitivity to
the two actinide distributions considered here.
(3)
For either actinide distribution, the various mass-velocity curves display
very similar behavior when considering fixed mass or fixed velocity,
provided that the curves are appropriately shifted forward or backward in time.
The lack of abrupt changes in the maximum
L1-norm values (provided that the time-shift effect is taken into account;
see more details below)
indicates that different spectral features are not appearing and disappearing
as a function of different mass-velocity conditions.

Returning to the time-shift behavior, we note that this 
pattern is expected from basic physics considerations of a homologous
expansion, i.e.
increasing the ejecta velocity makes line expression occur sooner in time,
while increasing the ejecta mass (or density) produces a slower 
evolution of the expansion
and the spectral features evolve on a longer time scale.
More specifically, the trends in the L1 norm are expected when
considering the evolution of the photosphere.
The optical depth of the ejecta scales as $L \times \rho \times \kappa$,
where $L$ is the sizescale of the ejecta, $\rho$ is the density and $\kappa$
is the opacity. Assuming that the expansion dominates the sizescale, the optical
depth scales as $M_{\rm ejecta} \, \kappa/v^2_{\rm ejecta}$. The variations
in the models in Fig.~\ref{fig:l1_intgrl} roughly follow these trends.
Arnett derived a more
detailed picture for Type~Ia supernovae \citep{arnett79,arnett82},
where the shape function of the emission could be reduced by a similar parameter
$y=(M_{\rm ejecta} \, \kappa/v_{\rm ejecta})^{1/2}$. Although we do not provide
a full comparison with the Arnett derivation here, we note that the trends in
our KN simulations are close to those predicted with that analytic model.

\begin{figure}
\includegraphics[clip=true,angle=0,width=1.0\columnwidth]
{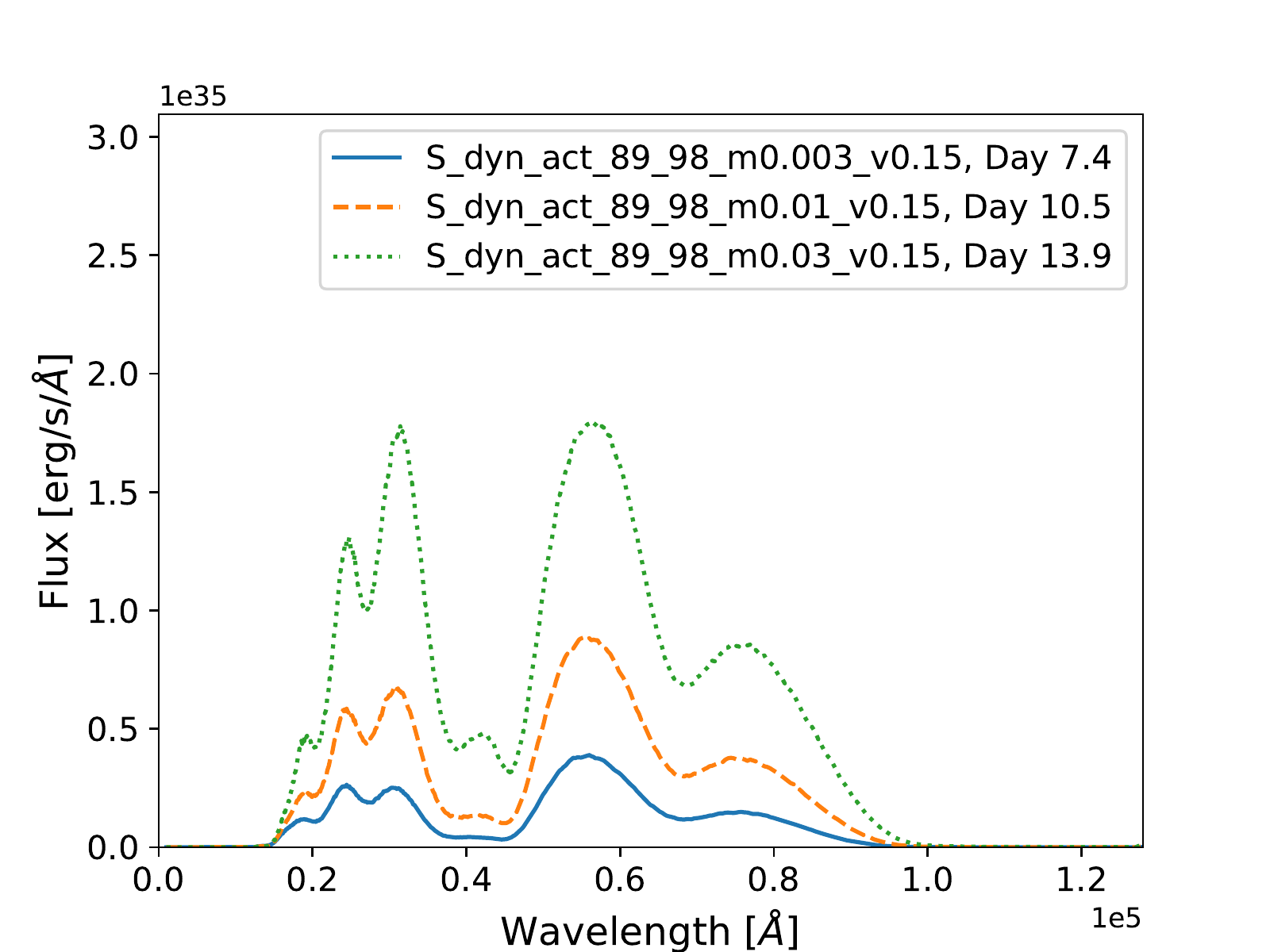}
\caption{
Spectra for three different mass-velocity pairs: the mean ejecta
speed is fixed at 0.15$c$ and the ejecta masses are 
0.003, 0.01, 0.03 M$_{\odot}$, represented by solid blue,
dashed orange, and dotted blue curves, respectively.
The spectra are calculated at different times, corresponding to the
peak L1-norm values displayed in the lower right-hand panel
of Fig.~\ref{fig:l1_intgrl},
i.e. at 6.9, 9.6 and 13.4 days, respectively.
These spectra are all generated with the limited range of actinide
abundances ($Z = 89$--98).
}
\label{fig:spec_mass_l1}
\end{figure}

In order to illustrate these trends,
we present in Fig.~\ref{fig:spec_mass_l1} a comparison of
three spectra for which the mean ejecta speed is fixed at the
middle grid value of 0.15$c$ and the ejecta mass takes on each
of the three grid values (0.003, 0.01, 0.03 M$_{\odot}$).
These spectra are generated with the partial-actinide distribution,
but the results are expected to be the same for the complete-actinide
distribution. The three spectra are plotted at the times that 
correspond to their maximum
L1-norm values in the lower right-hand panel of
Fig.~\ref{fig:l1_intgrl}, i.e. day 6.9, 9.6 and 13.4, respectively.
Note that the various spectral features in each curve align almost perfectly
in their wavelength positions when the time shift is applied.
When applying the same type of time-shift analysis to the opposite situation,
in which the ejecta mass is held fixed while the mean speed is allowed
to change,
we find that the spectral features also become better aligned, although
the agreement in relative feature strength
is not as good as that displayed in Fig.~\ref{fig:l1_intgrl}.
We attribute this poorer agreement to the fact that Doppler broadening,
which is important in determining the shapes of the spectral features,
depends strongly on the ejecta velocity.
Thus, taking into account the time-shift analysis and
the above three items of interest,
a main conclusion of this work is that ejecta mass and velocity play a
more significant role in modeling KN emission than the choice of
actinide abundances. Furthermore, specific mass-velocity conditions
shift the line emission earlier or later in time, but do not result
in the production of spectral features that arise from different
atomic transitions.

\section{Summary}

We have made a first attempt to calculate
a complete set of actinide opacities for use in KN modeling.
Basic trends observed in the actinide opacities follow similar patterns
that were observed previously for lanthanide opacities, e.g. a dominant
contribution from line absorption over a significant fraction
of the wavelength range that is relevant for KN emission.
The bound-bound contribution to these opacities was computed with a line-binned
approach, which allows for the generation of tabulated opacities, independent
of the particular type of hydrodynamic expansion that is chosen.
These tabular opacities will be made available on NIST-LANL opacity
website \citep{nist_opac}, concurrent with the publication of this article,
to aid in KN modeling efforts.

These actinide opacities were used in simulations of KN light curves
and spectra in order to explore the sensitivity
to different actinide abundance distributions
predicted by nuclear theory. The simulations
were carried out for a range of ejecta masses and speeds,
while the nuclear heating rate was held fixed.
We found a relative lack of importance of the heavier actinides on KN emission,
making it difficult to distinguish between the two abundance distributions
considered in this study. This behavior
is consistent with the corresponding lack of importance of the heavier
lanthanide elements observed in previous studies. This behavior
is a consequence of the energy-level
spacing of low-lying levels and the resulting positions of absorption line
features. For the lighter actinides, our modeling indicates that protactinium is
responsible for the production of faint, mid-IR features above 40,000~\AA\ 
at 5--7~days post merger,
and also has a very moderate effect on the strong (near-IR) features that
occur around the peak of the emission at 1--7~days post merger.
Compared to variation in actinide abundances, we found that the choice of
ejecta mass and velocity has a more significant effect on the behavior
of KN emission.
Future work is planned to study various sensitivities to actinide
elements in more detail.

\section*{Acknowledgements}

Helpful discussions with O.~Korobkin, A.L.~Hungerford and E.A.~Chase
are gratefully acknowledged.
This work was supported by the US Department of Energy through the Los Alamos
National Laboratory. Los Alamos National Laboratory is operated
by Triad National Security, LLC, for the National Nuclear Security
Administration of US Department of Energy (Contract No.~89233218CNA000001).
Research presented in this article was supported by the Laboratory
Directed Research and Development program of Los Alamos National Laboratory
under project number 20190021DR.

\section*{Data Availability}
The opacities described in this work will be made available at the NIST-LANL
opacity website: https://nlte.nist.gov/OPAC.

\newpage
\clearpage

%%%%%%%%%%%%%%%%%%%%%%%%%%%%%%%%%%%%%%%%%%%%%%%%%%



\bibliographystyle{mnras}
\bibliography{master}

%%%%%%%%%%%%%%%%%%%%%%%%%%%%%%%%%%%%%%%%%%%%%%%%%%

%%%%%%%%%%%%%%%%% APPENDICES %%%%%%%%%%%%%%%%%%%%%

%\clearpage 

\appendix

%\begin{appendices}

\section{List of configurations used in this work}
\label{app:config_list}

Table~\ref{tab:configs} in this appendix contains a list of configurations
that were used in calculating the energy levels and oscillator strengths for the
14 actinide elements considered
in this work. Based on the relevant conditions of kilonova ejecta, only
the first four ion stages were calculated for each element.
The list of configurations was chosen to obtain a good representation
of the lowest lying energy levels that are necessary to: (a) obtain converged
atomic level populations via Saha-Boltzmann statistics and (b) calculate
converged opacities with respect to the number of bound-bound transitions
in the photon energy range of interest.
The choice of configurations was based on the energy-level entries in the NIST
database \citep{nist}, when available, as well as ab initio atomic structure
calculations.
%\begin{deluxetable*}{lcrr}
\begin{table*}
\centering
%\tablecaption{\rm A list of configurations, number of fine-structure levels,
\caption{\rm A list of configurations, number of fine-structure levels,
and number of (electric dipole) absorption lines for the various ion stages
considered in this work. A completely filled Rn core is assumed
for the 14 actinide elements.
For the first two ion stages of nobelium, No {\sc i} and {\sc ii}, the
orbital angular momentum symbol $\ell$ represents the range of values
$\ell = s, p, d, f$ and $g$. }
\vspace*{0.5\baselineskip}
\begin{tabular}{lcrr}
\hline
%\tablehead{
Ion stage   &  Configurations &  \# of levels & \# of lines \\
%}
\hline
%\startdata
Ac {\sc i} &
$6d^1 7s^2$,
$6d^2 7s^1$,
$7s^2 7p^1$,
$6d^1 7s^1 7p^1$,
$6d^2 7p^1$, & 366 & 16,164 \\
&
$6d^3$,
$5f^1 7s^2 $,
$5f^1 6d^1 7s^1$,
$5f^1 7s^1 7p^1$,
$5f^1 6d^2$,
$5f^1 6d^1 7p^1$ \\

Ac {\sc ii} &
$7s^2$,
$6d^1 7s1$,
$6d^2$,
$7s^1 7p^1$,
$6d^1 7p^1$, & 81 & 797 \\
&
$5f^1 7s^1$,
$5f^1 6d^1$,
$7s^1 8s^1$,
$5f^1 7p^1$,
$5f^2$ \\

Ac {\sc iii} &
$7s^1$,
$6d^1$,
$5f^1$,
$7p^1$ & 7 & 8 \\

Ac {\sc iv} &
$6p^6$,
$6p^5 7s^1$,
$6p^5 6d^1$,
$6p^5 5f^1$,
$6p^5 7p^1$ & 39 & 211 \\
\hline

Th {\sc i} &
$6d^2 7s^2$,
$6d^3 7s^1$,
$5f^1 6d^1 7s^2$,
$6d^1 7s^2 7p^1$,
$6d^2 7s^1 7p^1$,
$5f^1 6d^2 7s^1$, & 2851 & 802,472 \\
&
$5f^1 7s^2 7p^1$,
$6d^4$,
$5f^1 6d^1 7s^1 7p^1$,
$5f^2 7s^2$,
$5f^1 7s^2 7p^1$,
$5f^1 6d^3$, & & \\
&
$6d^3 7p^1$,
$5f^1 6d^2 7p^1$,
$5f^2 7s^1 7p^1$,
$5f^2 6d^2$,
$5f^2 6d^1 7s^1$,
$5f^2 6d^1 7p^1$ \\

Th {\sc ii} &
$6d^1 7s^2$,
$6d^2 7s^1$,
$5f^1 7s^2$,
$5f^1 6d^1 7s^1$,
$6d^3$,
$5f^1 6d^2 $,
$6d^1 7s^1 7p^1$, & 566 & 35,724 \\
&
$5f^2 7s^1$,
$5f^1 7s^1 7p^1$,
$5f^1 6d^1 7p^1$,
$7s^2 7p^1$,
$5f^2 6d^1$,
$6d^2 7p^1$,
$5f^2 7p^1$ \\

Th {\sc iii} &
$5f^1 6d^1$,
$6d^2 $,
$5f^1 7s^1$,
$6d^1 7s^1$,
$5f^2$,
$5f^1 7p^1$,
$6d^1 7p^1$,
$7s^1 7p^1$ & 78 & 759 \\

Th {\sc iv} &
$5f^1$,
$7s^1$,
$6d^1$,
$7p^1$ & 7 & 8 \\
\hline

Pa {\sc i} &
$5f^3 7s^2$,
$5f^2 6d^1 7s^2$,
$5f^2 6d^2 7s^1$,
$5f^3 6d^1 7s^1$,
$5f^3 6d^2$,
$5f^2 6d^1 7s^1 7p^1$, & 10,985 & 10,501,071 \\
&
$5f^3 6d^1 7p^1$,
$5f^3 7s^1 7p^1$,
$5f^1 6d^2 7s^2$,
$5f^1 6d^2 7s^1 7p^1$,
$5f^2 6d^2 7p^1$ \\

Pa {\sc ii} &
$5f^2 7s^2$,
$5f^3 7s^1$,
$5f^3 6d^1$,
$5f^3 7p^1$,
$5f^2 6d^2$,
$5f^2 6d^1 7s^1$,
$5f^2 6d^1 7p^1$, & 2,940 & 816,301 \\
&
$5f^2 7s^1 7p^1$,
$5f^1 6d^2 7s^1$,
$5f^1 6d^2 7p^1$,
$6d^3 7s^1$,
$6d^3 7p^1$ \\

Pa {\sc iii} &
$5f^3$,
$5f^2 7s^1$,
$5f^2 6d^1$, & 361 & 12,394 \\
&
$5f^2 7p^1$,
$5f^1 6d^2$,
$5f^1 6d^1 7s^1$ \\

Pa {\sc iv} &
$5f^2$,
$5f^1 7s^1$,
$5f^1 6d^1$,
$5f^1 7p^1$,
$6d^2$,
$6d^1 7s^1$ & 62 & 452 \\
\hline

U {\sc i} &
$5f^4 7s^2$,
$5f^3 6d^1 7s^2$,
$5f^4 6d^1 7s^1$, & 16,882 & 20,948,831 \\
&
$5f^4 6d^2$,
$5f^3 6d^1 7s^1 7p^1$,
$5f^4 6d^1 7p^1$ \\

U {\sc ii} &
$5f^3 7s^2$,
$5f^4 7s^1$,
$5f^4 6d^1$,
$5f^4 7p^1$, & 6,929 & 4,016,742 \\
&
$5f^3 6d^2$,
$5f^3 6d^1 7s^1$,
$5f^3 6d^1 7p^1$,
$5f^3 7s^1 7p^1$ \\

U {\sc iii} &
$5f^4$,
$5f^3 7s^1$,
$5f^3 6d^1$,
$5f^3 7p^1$, & 1,650 & 233,822 \\
&
$5f^2 6d^2$,
$5f^2 6d^1 7s^1$,
$5f^1 6d^2 7s^1$ \\

U {\sc iv} &
$5f^3$,
$5f^2 7s^1$,
$5f^2 6d^1$,
$5f^2 7p^1$ & 241 & 5,784 \\
\hline

Np {\sc i} &
$5f^5 7s^2$,
$5f^4 6d^1 7s^2$,
$5f^5 6d^1 7s^1$,
$5f^5 6d^2$, & 37,504 & 102,137,419\\
&
$5f^4 6d^1 7s^1 7p^1$,
$5f^5 6d^1 7p^1$,
$5f^5 7s^1 7p^1$ \\

Np {\sc ii} &
$5f^5 7s^1$,
$5f^5 6d^1$,
$5f^5 7p^1$,
$5f^4 6d^2$, & 16,595 & 21,306,572 \\
&
$5f^4 6d^1 7s^1$,
$5f^4 6d^1 7p^1$,
$5f^4 7s^1 7p^1$ \\

Np {\sc iii} &
$5f^5$,
$5f^4 7s^1$,
$5f^4 6d^1$,
$5f^4 7p^1$, & 5,274 & 2,262,369 \\
&
$5f^3 6d^2$,
$5f^3 6d^1 7s^1$,
$5f^2 6d^2 7s^1$ \\

Np {\sc iv} &
$5f^4$,
$5f^3 7s^1$,
$5f^3 6d^1$,
$5f^3 7p^1$ & 817 & 57,765 \\
\hline

Pu {\sc i} &
$5f^6 7s^2$,
$5f^5 6d^1 7s^2$,
$5f^6 6d^1 7s^1$,
$5f^6 6d^2$, & 65,015 & 276,327,167 \\
&
$5f^5 6d^1 7s^1 7p^1$,
$5f^6 6d^1 7p^1$,
$5f^6 7s^1 7p^1$,
$5f^6 7p^2$ \\

Pu {\sc ii} &
$5f^6 7s^1$,
$5f^6 6d^1$, 
$5f^6 7p^1$, 
$5f^5 6d^2$, & 32,828 & 82,524,211 \\
&
$5f^5 6d^1 7s^1$,
$5f^5 6d^1 7p^1$,
$5f^5 7s^1 7p^1$,
$5f^5 7p^2$ \\

Pu {\sc iii} &
$5f^6$,
$5f^5 7s^1$,
$5f^5 6d^1$,
$5f^5 7p^1$,  & 13,277 & 13,511,494 \\
&
$5f^4 6d^2$,
$5f^4 6d^1 7s^1$,
$5f^3 6d^2 7s^1$,
$5f^4 7s^2$ \\

Pu {\sc iv} &
$5f^5$,
$5f^4 7s^1$,
$5f^4 6d^1$,
$5f^4 7p^1$ & 1,994 & 320,633 \\
\hline

Am {\sc i} &
$5f^7 7s^2$,
$5f^6 6d^1 7s^2$,
$5f^7 6d^1 7s^1$,
$5f^7 6d^2$,
$5f^6 6d^2 7s^1$,
$5f^6 6d^1 7s^1 7p^1$,
$5f^7 6d^1 7p^1$, & 164,455 & 1,939,681,964 \\
&
$5f^7 7s^1 7p^1$,
$5f^7 7s^1 7d^1$,
$5f^7 7s^1 8s^1$,
$5f^7 7s^1 8p^1$,
$5f^7 6d^1 7d^1$,
$5f^7 6d^1 8s^1$,
$5f^7 6d^1 8p^1$ \\

Am {\sc ii} &
$5f^7 7s^1$,
$5f^7 6d^1$,
$5f^7 7p^1$,
$5f^6 6d^2$,
$5f^6 6d^1 7s^1$, & 46,213 & 152,795,857 \\
&
$5f^6 6d^1 7p^1$,
$5f^6 7s^1 7p^1$,
$5f^7 7d^1$,
$5f^7 8s^1$ \\

Am {\sc iii} &
$5f^7$,
$5f^6 7s^1$,
$5f^6 6d^1$, & 17,058 & 13,844,004 \\
&
$5f^6 7p^1$,
$5f^5 6d^2$,
$5f^5 6d^1 7s^1$ \\

Am {\sc iv} &
$5f^6$,
$5f^5 7s^1$,
$5f^5 6d^1$,
$5f^5 7p^1$ & 3,737 & 1,045,697 \\
\hline

Cm {\sc i} &
$5f^8 7s^2$,
$5f^7 6d^1 7s^2$,
$5f^7 6d^2 7s^1$,
$5f^7 6d^2 7p^1$,
$5f^7 7s^2 7p^1$, & 231,490 & 3,707,210,225 \\
&
$5f^7 7s^1 7p^2$,
$5f^7 6d^3$,
$5f^8 6d^1 7s^1$,
$5f^8 6d^2$,
$5f^7 6d^1 7s^1 7p^1$, & & \\
&
$5f^8 6d^1 7p^1$,
$5f^8 7s^1 7p^1$,
$5f^9 7s^1$,
$5f^9 6d^1$,
$5f^9 7p^1$ \\

Cm {\sc ii} &
$5f^8 7s^1$,
$5f^8 6d^1$,
$5f^8 7p^1$,
$5f^7 6d^2$,
$5f^7 7s^2$,
$5f^7 6d^1 7s^1$,
$5f^7 6d^1 7p^1$, & 142,841 & 1,490,279,275 \\
&
$5f^7 7s^1 7p^1$,
$5f^6 6d^1 7s^2$,
$5f^6 6d^2 7s^1$,
$5f^7 7s^1 7d^1$,
$5f^7 7s^1 8s^1$,
$5f^7 7s^1 8p^1$, & & \\
&
$5f^7 6d^1 7d^1$,
$5f^7 6d^1 8s^1$,
$5f^7 6d^1 8p^1$,
$5f^9$,
$5f^8 7d^1$,
$5f^8 8s^1$,
$5f^8 8p^1$ \\

Cm {\sc iii} &
$5f^8$,
$5f^7 7s^1$,
$5f^7 6d^1$, & 23,268 & 21,734,676 \\
&
$5f^7 7p^1$,
$5f^6 6d^2$,
$5f^6 6d^1 7s^1$ \\

Cm {\sc iv} &
$5f^7$,
$5f^6 7s^1$,
$5f^6 6d^1$,
$5f^6 7p^1$ & 5,323 & 2,073,702 \\
\hline
\end{tabular}
\label{tab:configs}
\end{table*}
% TABLE SPLIT HERE
\addtocounter{table}{-1}
\begin{table*}
\centering
\caption{\rm Continued\ldots}
\vspace*{0.5\baselineskip}
\begin{tabular}{lcrr}
\hline
%\tablehead{
Ion stage   &  Configurations &  \# of levels & \# of lines \\
%}
\hline
%\startdata
Bk {\sc i} &
$5f^9 7s^2$,
$5f^8 6d^1 7s^2$,
$5f^8 6d^2 7s^1$,
$5f^8 7s^2 7p^1$,
$5f^9 6d^1 7s^1$,
$5f^9 6d^2$,
$5f^8 6d^1 7s^1 7p^1$, & 130,293 & 1,240,175,537 \\
&
$5f^9 6d^1 7p^1$,
$5f^9 7s^1 7p^1$,
$5f^9 7p^2$,
$5f^9 6d^1 7d^1$,
$5f^9 6d^1 8s^1$,
$5f^9 6d^1 8p^1$,
$5f^9 7s^1 7d^1$, & & \\
&
$5f^9 7s^1 8s^1$,
$5f^9 7s^1 8p^1$,
$5f^{10} 6d^1$,
$5f^{10} 7s^1$,
$5f^{10} 7p^1$,
$5f^{10} 7d^1$,
$5f^{10} 8s^1$,
$5f^{10} 8p^1$ \\

Bk {\sc ii} &
$5f^9 7s^1$,
$5f^9 6d^1$,
$5f^9 7p^1$,
$5f^8 7s^2$,
$5f^8 6d^2$,
$5f^8 6d^1 7s^1$, & 44,051 & 142,449,866 \\
&
$5f^8 6d^1 7p^1$,
$5f^8 7s^1 7p^1$,
$5f^9 7d^1$,
$5f^9 8s^1$,
$5f^9 8p^1$,
$5f^{10}$ \\

Bk {\sc iii} &
$5f^9$,
$5f^8 7s^1$,
$5f^8 6d^1$,
$5f^8 7p^1$, & 24,992 & 21,470,795 \\
&
$5f^7 6d^2$,
$5f^7 6d^1 7s^1$,
$5f^7 7s^2$ \\

Bk {\sc iv} &
$5f^8$,
$5f^7 7s^1$,
$5f^7 6d^1$,
$5f^7 7p^1$ & 5,983 & 2,545,975 \\
\hline

Cf {\sc i} &
$5f^{10} 7s^2$,
$5f^9  6d^1 7s2$,
$5f^9  6d^1 7s1 7p1$,
$5f^{10} 6d^1 7s1$,
$5f^{10} 7s^1 7p1$,
$5f^{10} 7s^1 7d1$, & 123,715 & 876,487,942 \\
&
$5f^{10} 7s^1 8s^1$,
$5f^{10} 7s^1 8p^1$,
$5f^9  7s^2 7p^1$,
$5f^9  6d^2 7s^1$,
$5f^{10} 7p^2$,
$5f^{10} 6d^2$, & & \\
&
$5f^{10} 6d^1 7p^1$,
$5f^{10} 6d^1 7d^1$,
$5f^{10} 6d^1 8s^1$,
$5f^{10} 6d^1 8p^1$,
$5f^9  6d^2 7p^1$ \\

Cf {\sc ii} &
$5f^{10} 7s^1$,
$5f^{10} 6d^1$,
$5f^{10} 7p^1$,
$5f^9  6d^2$,
$5f^9  7s^2$,
$5f^9  6d^1 7s^1$, & 28,805 & 63,361,823 \\
&
$5f^9  6d^1 7p^1$,
$5f^9  7s^1 7p^1$,
$5f^{10} 7d^1$,
$5f^{10} 8s^1$,
$5f^{10} 8p^1$,
$5f^{11} $ \\

Cf {\sc iii} &
$5f^{10} $,
$5f^9 7s^1$,
$5f^9 6d^1$,
$5f^9 7p^1$, & 21,129 & 12,894,033 \\
&
$5f^8 6d^2$,
$5f^8 6d^1 7s^1$,
$5f^8 7s^2$ \\

Cf {\sc iv} &
$5f^9$,
$5f^8 7s^1$,
$5f^8 6d^1$,
$5f^8 7p^1$ & 5,194 & 1,943,961 \\
\hline

Es {\sc i} &
$5f^{11} 7s^2$,
$5f^{10} 6d^1 7s^2$,
$5f^{10} 6d^1 7s^1 7p^1$,
$5f^{11} 6d^1 7s^1$,
$5f^{11} 7s^1 7p^1$,
$5f^{11} 7s^1 7d^1$,
$5f^{11} 7s^1 8s^1$,
$5f^{11} 7s^1 8p^1$, & 59,898 & 213,624,256 \\
&
$5f^{10} 7s^2 7p^1$,
$5f^{10} 6d^2 7s^1$,
$5f^{11} 7p^2$,
$5f^{11} 6d^1 7p^1$,
$5f^{11} 6d^1 7d^1$,
$5f^{11} 6d^1 8s^1$,
$5f^{11} 6d^1 8p^1$, & & \\
&
$5f^{10} 6d^2 7p^1$,
$5f^{11} 6d^2$,
$5f^{12} 7s^1$,
$5f^{12} 6d^1$,
$5f^{12} 7p^1$,
$5f^{12} 7d^1$,
$5f^{12} 8s^1$,
$5f^{12} 8p^1$ \\

Es {\sc ii} &
$5f^{11} 7s^1$,
$5f^{11} 6d^1$,
$5f^{11} 7p^1$,
$5f^{10} 6d^2$,
$5f^{10} 7s^2$,
$5f^{10} 6d^1 7s^1$, & 14,693 & 17,006,055 \\
&
$5f^{10} 6d^1 7p^1$,
$5f^{10} 7s^1 7p^1$,
$5f^{11} 7d^1$,
$5f^{11} 8s^1$,
$5f^{11} 8p^1$,
$5f^{12}$ \\

Es {\sc iii} &
$5f^{11} $,
$5f^{10} 7s^1$,
$5f^{10} 6d^1$,
$5f^{10} 7p^1$ & 1,837 & 259,812 \\

Es {\sc iv} &
$5f^{10}$,
$5f^9 7s^1$,
$5f^9 6d^1$,
$5f^9 7p^1$ & 3,549 & 915,339 \\
\hline

Fm {\sc i} &
$5f^{12} 7s^2$,
$5f^{11} 6d^1 7s^2$,
$5f^{11} 6d^1 7s^1 7p^1$,
$5f^{12} 6d^1 7s^1$,
$5f^{12} 7s^1 7p^1$,
$5f^{12} 7s^1 7d^1$,
$5f^{12} 7s^1 8s^1$,
$5f^{12} 7s^1 8p^1$, & 21,847 & 28,998,514 \\
&
$5f^{11} 7s^2 7p^1$,
$5f^{11} 6d^2 7s^1$,
$5f^{12} 7p^2$,
$5f^{12} 6d^2$,
$5f^{12} 6d^1 7p^1$,
$5f^{12} 6d^1 7d^1$,
$5f^{12} 6d^1 8s^1$, & & \\
&
$5f^{12} 6d^1 8p^1$,
$5f^{11} 6d^2 7p^1$,
$5f^{13} 7s^1$,
$5f^{13} 6d^1$,
$5f^{13} 7p^1$,
$5f^{13} 7d^1$,
$5f^{13} 8s^1$,
$5f^{13} 8p^1$ \\

Fm {\sc ii} &
$5f^{12} 7s^1$,
$5f^{12} 6d^1$,
$5f^{12} 7p^1$,
$5f^{11} 6d^2$,
$5f^{11} 7s^2$,
$5f^{11} 6d^1 7s^1$, & 5,535 & 2,629,353 \\
&
$5f^{11} 6d^1 7p^1$,
$5f^{11} 7s^1 7p^1$,
$5f^{13}$,
$5f^{12} 7d^1$,
$5f^{12} 8s^1$,
$5f^{12} 8p^1$ \\

Fm {\sc iii} &
$5f^{12}$,
$5f^{11} 7s^1$,
$5f^{11} 6d^1$,
$5f^{11} 7p^1$ & 723 & 42,671 \\

Fm {\sc iv} &
$5f^{11}$,
$5f^{10} 7s^1$,
$5f^{10} 6d^1$,
$5f^{10} 7p^1$ & 1,837 & 259,812 \\
\hline

Md {\sc i} &
$5f^{13} 7s^2$,
$5f^{12} 6d^1 7s^2$,
$5f^{12} 6d^1 7s^1 7p^1$,
$5f^{13} 6d^1 7s^1$,
$5f^{13} 7s^1 7p^1$,
$5f^{13} 7s^1 7d^1$,
$5f^{13} 7s^1 8s^1$,
$5f^{13} 7s^1 8p^1$, & 5,614 & 2,089,545 \\
&
$5f^{12} 7s^2 7p^1$,
$5f^{13} 7p^2$,
$5f^{13} 6d^1 7p^1$,
$5f^{13} 6d^1 7d^1$,
$5f^{13} 6d^1 8s^1$,
$5f^{13} 6d^1 8p^1$,
$5f^{12} 6d^2 7s^1$, & & \\
&
$5f^{12} 6d^2 7p^1$,
$5f^{13} 6d^2$,
$5f^{14} 7s^1$,
$5f^{14} 6d^1$,
$5f^{14} 7p^1$,
$5f^{14} 7d^1$,
$5f^{14} 8s^1$,
$5f^{14} 8p^1$ \\

Md {\sc ii} &
$5f^{13} 7s^1$,
$5f^{13} 6d^1$,
$5f^{13} 7p^1$,
$5f^{12} 6d^2$,
$5f^{12} 7s^2$,
$5f^{12} 6d^1 7s^1$, & 1,521 & 216,741 \\
&
$5f^{12} 6d^1 7p^1$,
$5f^{12} 7s^1 7p^1$,
$5f^{13} 7d^1$,
$5f^{13} 8s^1$,
$5f^{13} 8p^1$,
$5f^{14}$ \\

Md {\sc iii} &
$5f^{13}$,
$5f^{12} 7s^1$,
$5f^{12} 6d^1$,
$5f^{12} 7p^1$ & 202 & 3,797 \\

Md {\sc iv} &
$5f^{12}$,
$5f^{11} 7s^1$,
$5f^{11} 6d^1$,
$5f^{11} 7p^1$ & 723 & 42,671 \\
\hline

No {\sc i} &
$5f^{14} 7s^1 7\ell^1$,
$5f^{14} 7s^1 8\ell^1$,
$5f^{14} 7s^1 9\ell^1$,
$5f^{14} 7s^1 10\ell^1$,
$5f^{14} 7s^1 11\ell^1$,
$5f^{14} 7s^1 12\ell^1$,
$5f^{14} 6d^1 7\ell^1$,
$5f^{14} 6d^1 8\ell^1$, & 1,863 & 367,754 \\
&
$5f^{14} 6d^1 9\ell^1$,
$5f^{14} 6d^1 10\ell^1$,
$5f^{14} 6d^1 11\ell^1$,
$5f^{14} 6d^1 12\ell^1$,
$5f^{14} 6d^1 6f^1$,
$5f^{14} 6d^1 6g^1$,
$5f^{13} 6d^1 7s^2$,
$5f^{13} 6d^2 7s^1$, & & \\
&
$5f^{13} 6d^1 7s^1 7p^1$,
$5f^{13} 7s^2 7p^1$,
$5f^{13} 7s^1 7p^2$,
$5f^{13} 6d^2 7p^1$,
$5f^{13} 7s^2 7d^1$,
$5f^{14} 7p^2$,
$5f^{14} 6d^2$,
$5f^{13} 6d^1 7p^2$ \\

No {\sc ii} &
$5f^{14} 7\ell^1$,
$5f^{14} 8\ell^1$,
$5f^{14} 9\ell^1$,
$5f^{14} 6d^1 $,
$5f^{14} 6f^1$,
$5f^{14} 6g^1$, & 292 & 10,037 \\
&
$5f^{13} 6d^2 $,
$5f^{13} 7s^2 $,
$5f^{13} 6d^1 7s^1$,
$5f^{13} 6d^1 7p^1$,
$5f^{13} 7s^1 7p^1$ \\

No {\sc iii} &
$5f^{14}$,
$5f^{13} 7s^1$,
$5f^{13} 6d^1$,
$5f^{13} 7p^1$,
$5f^{13} 7d^1$,
$5f^{13} 7f^1$, & 171 & 3,192 \\
&
$5f^{13} 6f^1$,
$5f^{13} 8s^1$,
$5f^{13} 8p^1$,
$5f^{13} 8d^1$,
$5f^{13} 8f^1$ \\

No {\sc iv} &
$5f^{13}$,
$5f^{12} 7s^1$,
$5f^{12} 6d^1$,
$5f^{12} 7p^1$ & 202 & 3,797 \\
\hline

%\enddata
\end{tabular}
%%%\label{tab:configs}
\end{table*}
%\end{deluxetable*}

%\end{appendices}

%\clearpage 

%%%%%%%%%%%%%%%%%%%%%%%%%%%%%%%%%%%%%%%%%%%%%%%%%%

% Don't change these lines
\bsp    % typesetting comment
\label{lastpage}
\end{document}